\documentclass[10pt]{iopart}
\usepackage{iopams} 
\usepackage{amsfonts}
\usepackage{amssymb}
\usepackage{graphicx}
\graphicspath{{pics/}} 
\usepackage[caption=false]{subfig}

\usepackage{iopext_my}
\usepackage{iopmath_my}
\usepackage{qm}
\usepackage{table_my}

\usepackage{ifthen}
\def\bibfile{My_Library}
\ifthenelse{\equal{\bibfile}{}}{%
  \Def\myprintbibliography{}%
}{%
  \usepackage[numbers,sort&compress]{natbib}
  \def\myprintbibliography{%
    \bibliographystyle{iopart-num}
    \bibliography{\bibfile}%
  }%
}%
\usepackage[%
]{hyperref}

\usepackage[acronym]{glossaries}
\makeglossaries
\newglossaryentry{CreationOp}
{
    name= {%
      $\widehat{a}^{\dagger}_{m}$, $\widehat{b}^{\dagger}_{m}$
    },
    description={%
      The creation operators of the $m$-th input and output mode, respectively
    },
    sort= ab modes
}

\newglossaryentry{LONUnitary}
{
    name= {%
      $\widehat{U}$
    },
    description={%
      A general unitary linear optics transformation%
    },
    sort= U transform
}

\newglossaryentry{MsmntConfig}
{
    name= {%
      $\mathbb{L}$%
    },
    description={%
      A set of Pauli operators that represents a set of Pauli measurement configurations%
    },
    sort= L msmnt config
}

\newglossaryentry{BoundCmplQ}
{
    name= {%
      $\mathcal{B}_{\mathcal{S}}(\mathbb{E})$
    },
    description={%
      The upper bound $\mathcal{B}_{\mathcal{S}}^{(\max)}(\mathbb{E})$ and lower bound $\mathcal{B}_{\mathcal{S}}^{(\min)}(\mathbb{E})$ on a complementary quantity within a Pauli subspace $\mathbb{H}_{\mathbb{E}}$%
    },
    sort= Bounds on compl property
}

\newglossaryentry{GCD}
{
    name= {%
      $\gcd$
    },
    description={%
      The greatest common divisor%
    },
    sort= greatest common divisor
}

\newglossaryentry{PauliOp}
{
    name= {$\widehat{\Lambda}_{i,j}$, $\widehat{\Lambda}_{j}$},
    description={The generalized Pauli operator, i=1 by default for $\widehat{\Lambda}_{j}$},
    sort= Lambda Operator
}

\newglossaryentry{XOp}
{
    name= {$\widehat{X}$},
    description={The mode-shift operator},
    sort= X Operator
}

\newglossaryentry{ZOp}
{
    name= {$\widehat{Z}$},
    description={The phase-shift operator},
    sort= Z Operator
}

\newglossaryentry{ZClockLabel}
{
    name= {$\mu(\boldvec{n})$},
    description={The $\widehat{Z}$-clock label of a Fock state $\boldvec{n}$, which is equal to its total mode index},
    sort= mu Z clock label
}

\newglossaryentry{XiOp}
{
    name= {$\widehat{\Xi}$},
    description={The generalized clock operator},
    sort= Xi Operator
}

\newglossaryentry{FockVec}
{
    name= {$\boldvec{n}$, $\ket{\boldvec{n}}$},
    description={A vector of photon number occupations in modes and its corresponding Fock state},
    sort= N vector
}

\newglossaryentry{phaseUnit}
{
    name= {$w_{M}$, $w$},
    description={The $M$-th root of unity},
    sort= w phase
}

\newglossaryentry{PauliClass}
{
    name= {$\mathbb{E}_{\boldvec{n}}$},
    description={The Pauli class represented by the Fock state $\ket{\boldvec{n}}$},
    sort= E Pauli class
}

\newcommand{\dimE}[1]{d_{\mathbb{E}_{#1}}}
\newglossaryentry{dimE}
{
    name= {$\dimE{\boldvec{n}}$},
    description={The dimension of a Pauli subspace $\mathbb{H}_{\mathbb{E}_{\boldvec{n}}}$},
    sort= Dimension of E
}

\newglossaryentry{PauliSubspace}
{
    name= {$\mathbb{H}_{\mathbb{E}_{\boldvec{n}}}$},
    description={The Pauli subspace spanned by a Pauli class $\mathbb{E}_{\boldvec{n}}$},
    sort= H Pauli subspace
}

\newcommand{\ESt}[3]{\mathbb{E}_{#1, #2}(\Lambda_{#3})}
\newglossaryentry{EState}
{
    name= {$\ket{\ESt{\boldvec{n}}{m}{j}}$},
    description={The $\widehat{\Lambda}_{j}$ eigenstate labeled by $m$ in a Pauli subspace $\mathbb{H}_{\mathbb{E}_{\boldvec{n}}}$},
    sort= E State
}

\newcommand{\eSt}[2]{e_{#1, #2}}
\newglossaryentry{eState}
{
    name= {$\ket{\eSt{\boldvec{n}}{m}}$},
    description={The eigenstates of the generalized clock operator $\widehat{\Xi}$ labeled by $m$, which are also the computational basis in a Pauli subspace $\mathbb{H}_{\mathbb{E}_{\boldvec{n}}}$},
    sort= e State
}

\newcommand{\plProj}[3]{\widehat{\pi}_{#1,#2}(#3)}
\newglossaryentry{PauliProj}
{
    name= {$\plProj{N}{m}{L}$},
    description={The $N$-photon Pauli projector that projects quantum states onto the $N$-photon $m$-labeled eigenstates of an operator $\widehat{L}$},
    sort= Pauli projector
}

\newglossaryentry{Hadamard}
{
    name= {$\widehat{H}_{j}$},
    description={The generalized Hadamard operator},
    sort= Hadamard operator
}

\newcommand{\plQ}[1]{\mathcal{Q}_{#1}}
\newglossaryentry{PauliQauntity}
{
    name= {$\plQ{L}$},
    description={The Pauli quantity $\mathcal{Q}$ in an $\widehat{L}$-Pauli measurement},
    sort= Q Pauli operator
}

\newglossaryentry{PauliDecoherence}
{
    name= {$D(\rho)$},
    description={The decoherence of a quantum state $\widehat{\rho}$ over Pauli subspaces},
    sort= Decoherence
}

\newcommand{\cmplQ}[2]{\mathcal{C}_{#1,#2}}
\newglossaryentry{ComplQuantity}
{
    name= {$\cmplQ{\mathcal{Q}}{\mathbb{L}}$},
    description={The complementary Pauli quantity $\mathcal{Q}$ evaluated in a set of measurement configurations $\mathbb{L}$},
    sort= Complementary Pauli quantity
}

\newglossaryentry{MI}
{
    name= {$\mathcal{I}_{\alpha:\beta}(\rho)$},
    description={Mutual information of $\widehat{\rho}$ in a $\widehat{\alpha}\otimes\widehat{\beta}$-Pauli measurement},
    sort= I Mutual information
}

\newglossaryentry{CmplMI}
{
    name= {$\cmplQ{\mathcal{I}}{\mathbb{L}}(\rho)$},
    description={The complementary mutual information of $\widehat{\rho}$ in a measurement setting of the configurations $\mathbb{L}$},
    sort= Complementary Mutual Information
}

\newglossaryentry{MP}
{
    name= {$\mathcal{F}_{\phi}(\widehat{\alpha},\widehat{\beta}|\widehat{\rho})$},
    description={Mutual predictability of $\widehat{\rho}$ in a $\widehat{\alpha}\otimes\widehat{\beta}$-Pauli measurement },
    sort= F Mutual predictability
}

\newglossaryentry{CmplMP}
{
    name= {$\cmplQ{\mathcal{F}_{\phi}}{\mathbb{L}}(\rho)$},
    description={The complementary mutual predictability of $\widehat{\rho}$ for a target state $\ket{\phi}$ in a measurement setting of the configurations $\mathbb{L}$},
    sort= Complementary Mutual Predictability
}

\newglossaryentry{ConvexProp}
{
    name= {$\mathcal{S}$},
    description={A convex-extendible property},
    sort= S Convex Property
}

\newacronym{mubs}{MUBs}{Mutually unbaised bases}
\newacronym{lon}{LON}{Linear optcics network}
\newacronym{pnrd}{PNRD}{Photon number resolving detection}
\newacronym{cmi}{CMI}{Complementary mutual information}
\newacronym{cmp}{CMP}{Complementary mutual predictability} %
\def\myprintglossary{%
  \clearpage
  \glsfindwidesttoplevelname
  \setglossarystyle{alttree}
  \setglossarypreamble[main]{The notations employed in this paper are listed with their page numbers as follows.\par}
  \setglossarypreamble[acronym]{The acronyms employed in this paper are listed with their page numbers as follows.\par}
  \printglossary[title=Notations]
  \printglossary[type=\acronymtype]
}
\newcommand{\hiddengls}[1]{\glslink*{#1}{}} 

\begin{document}
\title{Complementary properties of multiphoton quantum states in linear optics networks}
\author{
  Jun-Yi Wu$^1$
  , Mio Murao$^{1,2}$
}
\address{
$^1$
Department of Physics, Graduate School of Science, The University of Tokyo, Hongo 7-3-1, Bunkyo-ku, Tokyo 113-0033, Japan
\par
$^{2}$
Trans-scale Quantum Science Institute, The University of Tokyo, Bunkyo-ku,
Tokyo 113-0033, Japan
}
%
\ead{junyiwuphysics@gmail.com}
\begin{abstract}
We have developed a theory for accessing quantum coherences in mutually unbiased bases associated with generalized Pauli operators in multiphoton multimode linear optics networks (LONs). We show a way to construct complementary Pauli measurements in multiphoton LONs and establish a theory for evaluation of their photonic measurement statistics without dealing with the computational complexity of Boson samplings. This theory extends characterization of complementary properties in single-photon LONs to multiphoton LONs employing convex-roof extension. It allows us to detect quantum properties such as entanglement using complementary Pauli measurements, which reveals the physical significance of entanglement between modes in bipartite multiphoton LONs.
\end{abstract}
\maketitle

\section{Introduction}
\label{sec::intro}

Multiphoton multimode linear optics networks (LONs) are the physical platforms for the implementation of possible quantum supremacy in Boson sampling\cite{AaronsonArkhipov2011-CmplxLinOps}.\hiddengls{lon}
Experiments of Boson sampling have been realized and rapidly developed in various linear-optics-network systems\cite{BrodEtAlSciarrinoRev-PhBsnSmp,
BroomeEtAlWhite2013-BsnSmpl,
SpringEtAlWalmsley2013-BsnSmplOnChip,
TillmannEtAlWalther2013-BsnSmpl,
CrespiEtAlSciarrino2013-BsnSmplOnChip,
CarolanEtAlLaing2014-BsnSmpl,
CarolanEtAlLaing2015-UniLinOpt,
LoredoEtAlWhite2017-BsnSmplQDot,
HeEtAlPan2017-TmBinBsnSmpl,
WangEtAlPan2017-BsnSmp,
WangEtAlPan2018-BsnSmpPhLss,
PaesaniEtAlLaing2019-BsnSmpl,
WangEtAlPan2019-BsnSmpN20M60}.
Despite the simulation complexity of Boson sampling,
statistical characteristics can be exploited to benchmark Boson samplers\cite{BentivegnaEtAlSciarrino2014-BayesianBsnSmplVld, LiuEtAlRalph2016-CertfBsnSmpl, Shchesnovich2016-BunchingAssmntBsnSmpl, WalschaersEtAlBuchleitner2016-StatBnchmkBsnSmp, AgrestiEtAlSciarrino2019-PttnRcgnBsnSmp, ViggianielloEtAlSciarrino2018-OptPhIndMltmdNtwk, GiordaniEtAlScirarrino2018-ExpStatSignMBdQIntef}. 
For specific linear optics transformations of permutation symmetric states, one can even predict their zero-probability outputs by the suppression laws\cite{TichyEtAlBuchleitner2010-MultiBS,
CrespiEtAlSciarrino2016-SuppLawFT, ViggianielloEtAlSciarrino2018-ExpSupLawSylvInterf}.
The permutation symmetric states that exhibit the suppression laws are not restricted to Fock states, but also valid for quantum superposition of them\cite{WuHofmann2017-BiEntMltMd, DittelEtAlKeil2018-DestrInterfPermSymMPtclSt, DittelEtAlKeil2018-DestrMPtclInterf}.
It implies that quantum coherences play an important role in photon statistics of multiphoton LONs, if we consider their inputs as general multiphoton states.
From a different perspective, in this paper, we will consider the characterization of physical properties that related to quantum coherences of multiphoton states by evaluation of the photon statistics in LONs.

To characterize quantum coherences between Fock states, one can employ quantum state tomography to reconstruct the full description of a general multiphoton state in the entire multiphoton Hilbert space of LON systems.
The experimental setup of a full quantum state tomography in multiphoton LONs requires either a large number of measurement configurations or a large amount of additional ancillary modes\cite{BanchiKolthammerKim2018-MltphTmgrLnOpt}, both of which are still very challenging for currently available experimental facilities.
In many cases, instead of the full information of a quantum state, one just needs partial information about quantum coherences in measurements of two non-compatible observables.
It is therefore meaningful to consider the possibility of accessing quantum coherences in LONs by a reasonable number of measurement configurations associated with non-compatible observables, which can be meanwhile implemented by a set of experimentally available linear optics transforms without any additional ancillary modes and photons.

In single-photon LONs, which are equivalent to qudit systems, mutually unbiased bases (MUBs)\cite{DurtAtElZyczkowski2010-MUBs} are the optimal bases for obtaining maximal quantum coherences\cite{ChengHall2015-CmplRelQCoh, YaoEtAlSun2016-MaxCoh, HuShenFan2017-MaxCohMUB, StreltsovEtAlBruss2018-MaxCohPurity}.
It implies that complementary measurements, which measure quantum states in MUBs, are appropriate for revealing quantum coherences in qudit systems.
They can serve as coherence quantifiers\cite{Rastegin2017-UnRelQCohMUBs} through the uncertainty relationship of quantum measurements\cite{MaassenUffink1988-EntUnctRel, Sanchez-Ruiz1995-BdEntrUncCmplObsv, WuYuMolmer2009-EntUnRelMUB, WehnerWinter2010-EntrUncRel}.
In multipartite qudit systems, correlations in complementary measurements can be also exploited to detect entanglement\cite{MacconeBrussMacchiavello2015-CmplCrr, HuangEtAlPeruzzo2016-HghDimEntCert, SauerweinAtElKraus2017-MltptCrrMUBs, SpenglerHuberEtAlHiesmayr2012-EntWitViaMUB}, as well as the dimensionality of entanglement\cite{ErkerKrennHuber2017-QtfyHghDmEnt2MUBs, BavarescoEtAlHuber2018-2MUBsCrtfyHghDmEnt}.
In experiments, one can always choose the basis, in which the measurement is the most feasible and efficient, as the computational basis. To access the maximal quantum coherences in such a system, one needs additional measurements of which the measurement bases are mutually unbiased to the computational basis. Such complementary measurements can be implemented with the help of generalized Hadamard transforms, which map the computational basis to the eigenbases of generalized Pauli operators\cite{DurtAtElZyczkowski2010-MUBs}.
These complementary measurements are already feasible in experiments\cite{CarolanEtAlLaing2015-UniLinOpt}.
To open up experimental access to quantum properties associated with quantum coherences in multiphoton LONs, complementary measurements associated with generalized Pauli operators are therefore the desirable keys.

However, in multiphoton LONs, indistinguishability of photons leads to photon bunching in output modes of a LON, which makes the explicit photon statistics of a generalized Hadamard transform $\#P$-hard to determine\cite{AaronsonArkhipov2011-CmplxLinOps}.
This phenomenon tangles the complementarity of Pauli operators.
In this paper, we will tackle this problem to find the complementary structures of generalized Pauli operators and construct complementary measurements in multiphoton LONs.
Our goal is then to establish a theoretical framework for experimental access to complementary properties of multiphoton states in LONs through these complementary measurements.
We will show that complementary properties of convex sets of multiphoton states in LONs can be quantified through convex-roof extensions over the subspaces that are well-defined qudit systems and characterized by cyclicly translational mode shifting.

As an important application of this theoretical framework, we will then derive two approaches for entanglement detection in bipartite multiphoton LON systems employing complementary correlations.
Since photons are indistinguishable identical particles, entanglement between photons is only a well-defined concept after exclusion of the ``entanglement'' arising from particle-label symmetrization in their wavefunctions\cite{
EckertSchliemannEtAlLewenstein2002-QCorrIndistPtcls,
DowlingDohertyWiseman2006-EntIndistPtcl, KilloranCramerPlenio2014-IdPtclEnt, ChinHuh2019-EntIdPtclCoh1QL,
GhirardiMarinattoWeber2002-Ent, GhirardiMarinatto2004-EntCritIdP,
Tichy2011Thesis, TichyEtAlBuchleitner2013-DtctLvlEnt, TichyMintertBuchleitner2013-LimMEntBsnFmn,
ReuschSperlingVogel2015-IdParticleEntWit, GrabowskiKusMarmo2011-EntMltptIndstPcl,
FrancoCompagno2016-EntIdPtclByITNotion, LourencoDebarbaDuzzioni2019-EntIndstPtcl}.
In bosonic systems like LONs, entanglement between modes in the second quantization formalism\cite{LiEtAlLong2001-Ent2PtclSys}, which automatically excludes the ``entanglement'' arising from particle symmetrization, is therefore a legitimate entanglement concept.
In this paper, we therefore assume the perfect indistinguishability of photons in multimode interference of LONs, and consider the entanglement between modes with fixed local photon numbers, which is also called entanglement of ``particles''\cite{WisemanVaccaro2003-IdPtclEnt}.
Although there exist many entanglement detection approaches in bipartite qudit systems which are experimentally feasible, these methods can not be straightforwardly employed in bipartite multiphoton LONs for entanglement between modes.
It means that entanglement between modes in bipartite multiphoton LONs still lacks its physical significance in experiments.
The theoretical framework developed in this paper can extend well-established entanglement detection methods in bipartite qudit systems to bipartite multiphoton LONs, e.g. the entanglement detection approaches in \cite{MacconeBrussMacchiavello2015-CmplCrr,SpenglerHuberEtAlHiesmayr2012-EntWitViaMUB} employing two measures of complementary correlations, which we call complementary mutual information and complementary mutual predictability.
We will extend the threshold of these two complementary correlations for separable states in bipartite qudit systems\cite{MacconeBrussMacchiavello2015-CmplCrr,SpenglerHuberEtAlHiesmayr2012-EntWitViaMUB} to bipartite multiphoton LONs, such that complementary correlations exceeding these thresholds signify entanglement between modes in a bipartite multiphoton LON system.
Our results therefore open up access to the physical significance of entanglement between modes in multiphoton LONs.

This paper is structured as follows.
In Section \ref{sec::cmpl_strct}, we show complementary structures of generalized Pauli operators within the subspaces of multiphoton LONs characterized by a translational mode-shifting operator.
In Section \ref{sec::Pauli_msmnt}, we show the construction of generalized Pauli measurements, which allows us to access complementary properties within the subspaces specified in the previous section.
In Section \ref{sec::cmpl_msmnt}, we show complementary Pauli quantities evaluated in complementary measurements can be exploited to characterize convex sets of quantum states, which leads to measurement uncertainty relationship in multiphoton LONs.
In Section \ref{sec::cmpl_corr_ent}, we demonstrate an application of the theoretical framework established in previous sections in the detection of entanglement between modes in bipartite multiphoton LONs.
Section \ref{sec::conclusion} concludes the paper.

\section{Complementary structures of generalized Pauli operators in linear optics networks}
\label{sec::cmpl_strct}

\begin{figure}%
  \subfloat[]{\includegraphics[width=0.63\textwidth]{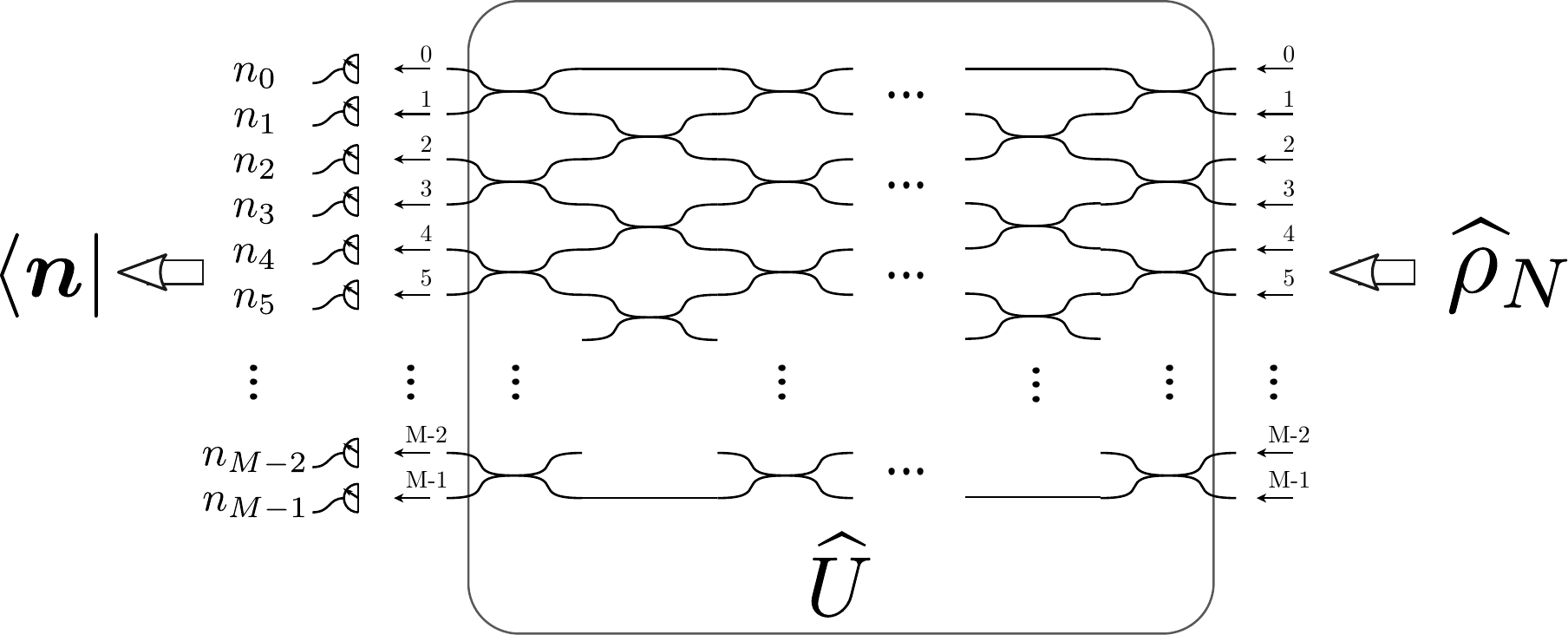}}
  \subfloat[]{\includegraphics[width=0.37\textwidth]{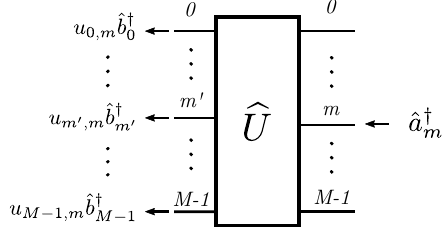}}
  \caption{(a) An $N$-photon state $\widehat{\rho}_{N}$ is transformed by an $M$-mode linear optics network. The transformation is described by a unitary $\widehat{U}$. Photon numbers $\{n_{m}\}_{m}$ are measured by PNRD in the output modes, which corresponds to a Fock state $\ket{\boldvec{n}}=\ket{n_{0},...,n_{M-1}}$.
  (b) An $M$-mode linear optics network transformation $\widehat{U}$ can be described by a unitary matrix $\{u_{m',m}\}_{m',m}$.
  Note that the time flows from right to left as the arrows indicate.
  }
  \label{fig::LONs}%
\end{figure}

\hiddengls{lon}
A  linear optics network (LON) is a multimode interferometer, which is a unitary transform of modes constructed by linear optics elements.
In principle, one can construct any unitary transform of modes using Beam splitters  \cite{ReckEtAlBertani1994-ExpUniOP}.
As shown in Fig. \ref{fig::LONs} (a), each input and output mode of a LON are indexed by $m=0, ..., M-1$.
A state transformed by a linear optics interferometer $\widehat{U}$ is measured by photon number resolving detection (PNRD) at each output mode, which resolves a number of photons $n_{m}$. \hiddengls{pnrd}%
An output event is then denoted by a Fock number vectors $\boldvec{n} = (n_{0},...,n_{M-1})$, which is associated with a projection onto the Fock state $\projector{\boldvec{n}}$. \hiddengls{FockVec}
Due to energy conservation, linear optics does not change the total photon number $|\boldvec{n}|$.
A LON unitary $\widehat{U}$ is therefore diagonal with respect to the subspaces of different total photon numbers.
It is therefore legitimate to describe the mechanism of a LON quantum system independently for quantum states with different photon numbers.
Consider an $N$-photon input state $\widehat{\rho}_{N}$ in the LON shown in Fig. \ref{fig::LONs}, the probability of detecting a photon number vector $\boldvec{n}$ is given by
\begin{equation}
  \mathrm{Pr}\left[\boldvec{n}\;|\;\widehat{U}\widehat{\rho}_{N}\widehat{U}^{\dagger}\right] = \Braket{\boldvec{n}|\widehat{U}\widehat{\rho}_{N}\widehat{U}^{\dagger}|\boldvec{n}}.
\end{equation}
Note that if the input is a Fock state, Fig. \ref{fig::LONs} (a) is a Boson sampling scenario.
Here we consider a more general scheme which allows an input state to be a superposition of Fock states.

In LONs, each input and output mode can be represented by photon creation operators $\widehat{a}_{m}^{\dagger}$ and $\widehat{b}_{m'}^{\dagger}$, respectively.
As shown in Fig. \ref{fig::LONs} (b), under the assumption of perfect indistinguishability of photons in interferometers, a general LON transformation $\widehat{U}$ between input and output modes can be represented by a unitary matrix $\{u_{m'm}\}_{m',m}$ in the Heisenberg picture of second quantization formalism, \hiddengls{CreationOp}\hiddengls{LONUnitary}
\begin{equation}
\label{eq::md_transf_mtx}
  \widehat{U}\widehat{a}_{m}^{\dagger}\widehat{U}^{\dagger}
  =
  \sum_{m'} u_{m'm}\widehat{b}^{\dagger}_{m'}.
\end{equation}
From such a transformation of modes, one can then derive the unique transition matrix $\{\braket{\boldvec{n}'|\widehat{U}|\boldvec{n}}\}_{\boldvec{n}',\boldvec{n}}$ of $\widehat{U}$ in the whole Fock space, which is believed to be a $\#P$-hard problem\cite{AaronsonArkhipov2011-CmplxLinOps}.
As an alternative of the $\#P$-hard transition matrix $\{\braket{\boldvec{n}'|\widehat{U}|\boldvec{n}}\}_{\boldvec{n}',\boldvec{n}}$, Eq. \eqref{eq::md_transf_mtx} can serve as an efficient definition of an $M$-mode transformation $\widehat{U}$ for all photon-number Fock states in LONs\cite{CamposEtAlTeich1989-BSSU2nPHStat, TichyEtAlBuchleitner2012-MPInteference, WuHofmann2017-BiEntMltMd, DittelEtAlKeil2018-DestrInterfPermSymMPtclSt}\footnote{%
There is an alternative representation of $\widehat{U}$ called the $N$-fold symmetric tensor product of $\{u_{m'm}\}_{m',m}$\cite{HayashiBook-GrpReprQThry}, which is defined from the perspective of first quantization formalism. Since we study the complementary properties of LONs from the second quantization perspective, we adopt the definition in Eq. \eqref{eq::md_transf_mtx} for a LON transformation $\widehat{U}$.}.
In single-photon LONs, the mode-transformation matrix $\{u_{m'm}\}_{m',m}$ determines the transformation between single-photon Fock states,
\begin{equation}
\label{eq::md_transf_mtx_1}
  \braket{\cdots01_{m'}0\cdots|\widehat{U}|\cdots01_{m}0\cdots} = u_{m'm}.
\end{equation}
Eq. \eqref{eq::md_transf_mtx} therefore allows us to extend a single-photon transition matrix $\{\braket{\cdots1_{m'}\cdots|\widehat{U}|\cdots1_{m}\cdots}\}_{m',m}$ to its corresponding multiphoton transition matrix $\{\braket{\boldvec{n}'|\widehat{U}|\boldvec{n}}\}_{\boldvec{n}',\boldvec{n}}$.

%

\bigskip

In  an $M$-mode single-photon LON system, which is equivalent to an $M$-dimensional qudit system, two operators that have mutually unbiased eigenbases are complementary for accessing maximal quantum coherences\cite{ChengHall2015-CmplRelQCoh, YaoEtAlSun2016-MaxCoh, HuShenFan2017-MaxCohMUB, StreltsovEtAlBruss2018-MaxCohPurity}.
In experiments, measurements associated with two complementary operators can be constructed as follows.
One first chooses the most feasible measurement basis as its computational basis, e.g. the Fock state basis associated with direct PNRD measurements on the input modes of LONs.
Then one employs a LON transformation to map the computational basis to a MUB.
For the implementation of such complementary measurements, Hadamard transforms are the legitimate candidates\cite{DurtAtElZyczkowski2010-MUBs}, which are already feasible in experiments\cite{CarolanEtAlLaing2015-UniLinOpt}.
Note that a Hadamard-transformed basis is the eigenbasis of a corresponding generalized Pauli operator.
We therefore focus on the complementarity between generalized Pauli operators in the rest of this section.

\begin{figure}
  \centering
  \subfloat[]{\includegraphics[width = 0.36\columnwidth]{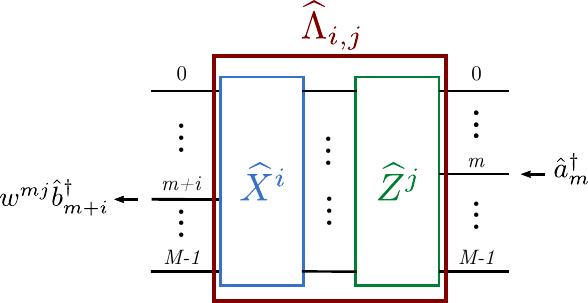}}
  \hspace{0.01\textwidth}
  \subfloat[]{\includegraphics[width = 0.3\columnwidth]{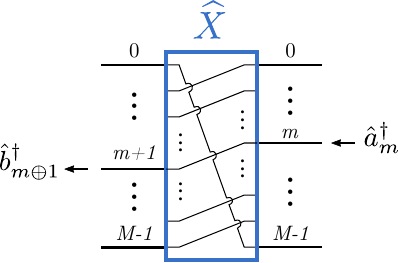}}
  \hspace{0.01\textwidth}
  \subfloat[]{\includegraphics[width = 0.3\columnwidth]{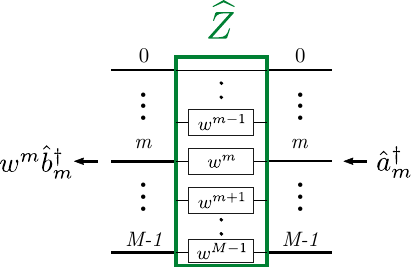}}
  \caption{(a) A Pauli operator $\widehat{\Lambda}_{i,j} = \widehat{X}^{i}\widehat{Z}^{j}$ constructed by a mode-shift operator $\widehat{X}$ and a phase-shift operator $\widehat{Z}$. (b) LON of the mode-shift operator $\widehat{X}$. (c) LON of the phase-shift operator $\widehat{Z}$.
  Note that the time in (a-c) flows from right to left as the arrows indicate.
  }
\label{fig::Pauli_op}
\end{figure}

A \emph{generalized Pauli operator} $\widehat{\Lambda}_{i,j}$ is a combination of a mode-shift operator $\widehat{X}$ and a phase-shift operator $\widehat{Z}$ (see Fig. \ref{fig::Pauli_op}), which are called the shift and clock operator, respectively, in qudit systems, \hiddengls{PauliOp}
\begin{equation}
  \widehat{\Lambda}_{i,j}:=\widehat{X}_{M}^{i}\widehat{Z}_{M}^{j}.
\end{equation}
The mode-shift operator $\widehat{X}$ shifts a mode to its next neighboring mode translationally and cyclicly, while the phase-shift operator $\widehat{Z}$ adds phases to each mode, \hiddengls{XOp}\hiddengls{ZOp}
\begin{equation}
  \widehat{X}\widehat{a}_{m}^{\dagger}\widehat{X}^{\dagger} = \widehat{b}_{m\oplus1}^{\dagger}
  \;\;\;\;\text{and}\;\;\;\;
  \widehat{Z}\widehat{a}_{m}^{\dagger}\widehat{Z}^{\dagger} = w^{m}\widehat{b}_{m}^{\dagger},
\end{equation}
where $w = \exp(\imI2\pi/M)$ is a phase given by the $M$-th root of unity and $m\oplus1 = (m+1)_{\pmod M}$ is the $M$-modulus sum. \hiddengls{phaseUnit}
Since the Pauli operator $\{\widehat{\Lambda}_{i,ij}\}_{i}$ have the same eigenbasis as the Pauli operator $\widehat{\Lambda}_{1,j}$, they can be evaluated in the same measurement associated with $\widehat{\Lambda}_{1, j}$.
As a consequence, most of the Pauli-operator eigenspaces can be characterized by the $\widehat{\Lambda}_{1,j}$.
In the rest of this section, we therefore focus on the complementary structures in the $\widehat{\Lambda}_{1,j}$ eigenspaces.
For conciseness, we shorten the notation for the Pauli operator $\widehat{\Lambda}_{1,j}$ by $\widehat{\Lambda}_{j}$.

In an $M$-mode multiphoton LON systems, the operator $\widehat{X}$ shifts a Fock state $\ket{\boldvec{n}}$ translationally and cyclicly,
\begin{equation}
\label{eq::X_shift}
  \widehat{X}\ket{\boldvec{n}} = \ket{n_{M-1},n_{0},...,n_{M-2}},
\end{equation}
while $\widehat{Z}$ adds a phase shift $\mu(\boldvec{n})$ which is equal to the \emph{total mode index} of the Fock state $\ket{\boldvec{n}}$,
\begin{equation}\hiddengls{ZClockLabel}
\label{eq::Z_shift}
  \widehat{Z}\ket{\boldvec{n}} = w^{\mu(\boldvec{n})}\ket{\boldvec{n}}
  \;\;\text{  with  }\;\;
  \mu(\boldvec{n}) = \sum_{m}n_{m}\,m.
\end{equation}
The effect of a generalized Pauli operator performed on a Fock state is a combination of the mode shift and phase shift,
\begin{equation}
  \widehat{\Lambda}_{j}\ket{\boldvec{n}}
  =
  w^{j\mu(\boldvec{n})}\ket{n_{M-1},n_{0},...,n_{M-2}}.
\end{equation}
After $M$-times $\widehat{\Lambda}_{j}$ operations, a Fock state $\ket{\boldvec{n}}$ will be periodically shifted back to its original.
Such a periodic operation connects and groups multiphoton Fock states in different orbits, which we call \emph{Pauli classes}.
\begin{definition}[Pauli classes and subspaces]
\label{def::Pauli_class}
  A \emph{Pauli class} $\mathbb{E}_{\boldvec{n}}$ in a linear optics network is a set of Fock states, whose elements are generated by the mode-shift operator performed on the representative Fock state $\ket{\boldvec{n}}$, \hiddengls{PauliClass}
  \begin{equation}
  \label{eq::Pauli_class}
    \mathbb{E}_{\boldvec{n}} := \left\{\widehat{X}^{k}\ket{\boldvec{n}}: k=0,...,\dimE{\boldvec{n}}-1 \right\},
  \end{equation}
  where $\dimE{\boldvec{n}}$ is the cardinality of the Pauli class. \hiddengls{dimE}
  The $\dimE{\boldvec{n}}$-dimensional Hilbert subspace $\mathbb{H}_{\mathbb{E}_{\boldvec{n}}}$ spanned by a Pauli class $\mathbb{E}_{\boldvec{n}}$ is called a \emph{Pauli subspace}, \hiddengls{PauliSubspace}
  \begin{equation}
    \mathbb{H}_{\mathbb{E}_{\boldvec{n}}} := \spn(\mathbb{E}_{\boldvec{n}}).
  \end{equation}
\end{definition}%

Since the operation of a Pauli operator $\widehat{\Lambda}_{j}$ performed on a multiphoton LON can be described independently within each Pauli class, one can decompose $\widehat{\Lambda}_{j}$ into diagonal blocks $\widehat{\Lambda}_{j}^{(\mathbb{E})}$, $\widehat{\Lambda}_{j} = \sum_{\mathbb{E}}\widehat{\Lambda}_{j}^{(\mathbb{E})}$,
where each block $\widehat{\Lambda}_{j}^{(\mathbb{E})}$ is an irreducible representation of $\widehat{\Lambda}_{j}$ within a Pauli subspace $\mathbb{H}_{\mathbb{E}}$.
In a Pauli subspace $\widehat{\mathbb{H}}_{\mathbb{E}_{\boldvec{n}}}$, eigenstates of $\widehat{\Lambda}_{j}$ are constructed by \hiddengls{EState}
\begin{equation}
\label{eq::En_Pauli_eigenstate}
  \ket{\ESt{\boldvec{n}}{m}{j}}
  :=
  \frac{1}{\sqrt{\dimE{\boldvec{n}}}}\sum_{k=0}^{\dimE{\boldvec{n}}-1} w^{-(\frac{1}{2}(M-1)j|\boldvec{n}| + m)k}\widehat{\Lambda}_{j}^{k} \ket{\boldvec{n}}
  \;\;
  \text{ with }
  \;\;
  m \in \left\{0, \frac{M}{\dimE{\boldvec{n}}}, ..., (\dimE{\boldvec{n}}-1)\frac{M}{\dimE{\boldvec{n}}}\right\}.
\end{equation}
The eigenstate $\ket{\ESt{\boldvec{n}}{m}{j}}$ satisfies the eigenequation
\begin{equation}
\label{eq::eigeneq_Lambda_j}
  \widehat{\Lambda}_{j} \ket{\ESt{\boldvec{n}}{m}{j}}
  =
  w^{\frac{1}{2}(M-1)j|\boldvec{n}|+m} \ket{\ESt{\boldvec{n}}{m}{j}}.
\end{equation}
As a result, the Pauli operator $\widehat{\Lambda}_{j}$ is a sum of all $\widehat{\Lambda}_{j}^{(\mathbb{E}_{\boldvec{n}})}$ constructed within Pauli subspaces,
\begin{equation}
  \widehat{\Lambda}_{j} = \sum_{\mathbb{E}_{\boldvec{n}}}\widehat{\Lambda}_{j}^{(\mathbb{E}_{\boldvec{n}})}
  \;\;\;\text{ with }\;\;\;
  \widehat{\Lambda}_{j}^{(\mathbb{E}_{\boldvec{n}})}
  =
  w^{\frac{1}{2}(M-1)j|\boldvec{n}|}\sum_{m}w^{m}\Projector{\mathbb{E}_{\boldvec{n}, m}(\Lambda_{j})}.
\end{equation}%
According to Eq. \eqref{eq::En_Pauli_eigenstate}, within a Pauli subspace $\mathbb{H}_{\mathbb{E}_{\boldvec{n}}}$, the computational Fock basis $\mathbb{E}_{\boldvec{n}}$ given in Definition \ref{def::Pauli_class} is mutually unbiased with a $\widehat{\Lambda}_{j}$-Pauli eigenbasis $\{\ket{\ESt{\boldvec{n}}{m}{j}}\}_{m}$ for any $j=0, ..., M-1$.

\bigskip

\begin{figure}
  \centering
  \subfloat[]{\includegraphics[width=0.33\textwidth]{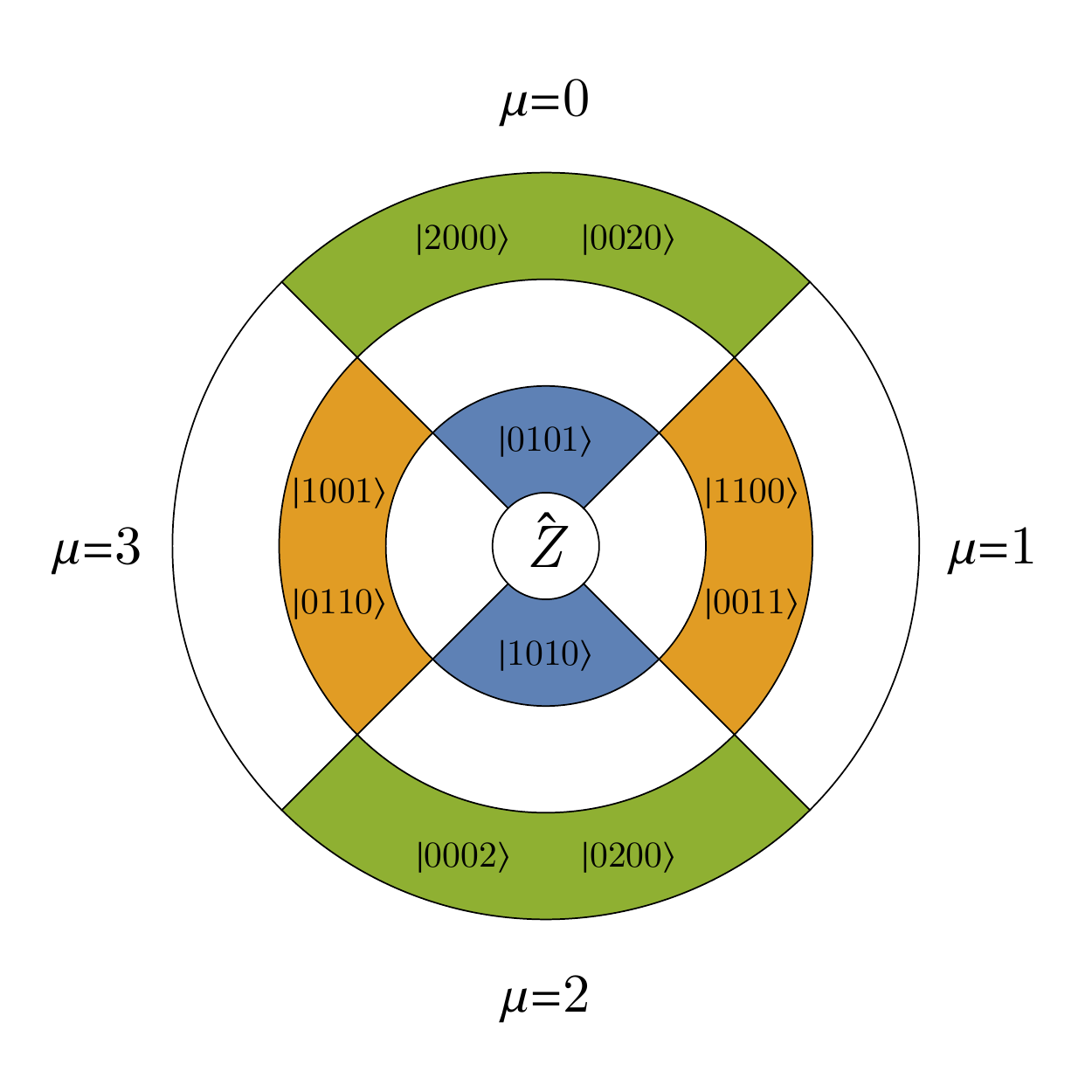}}  \subfloat[]{\includegraphics[width=0.33\textwidth]{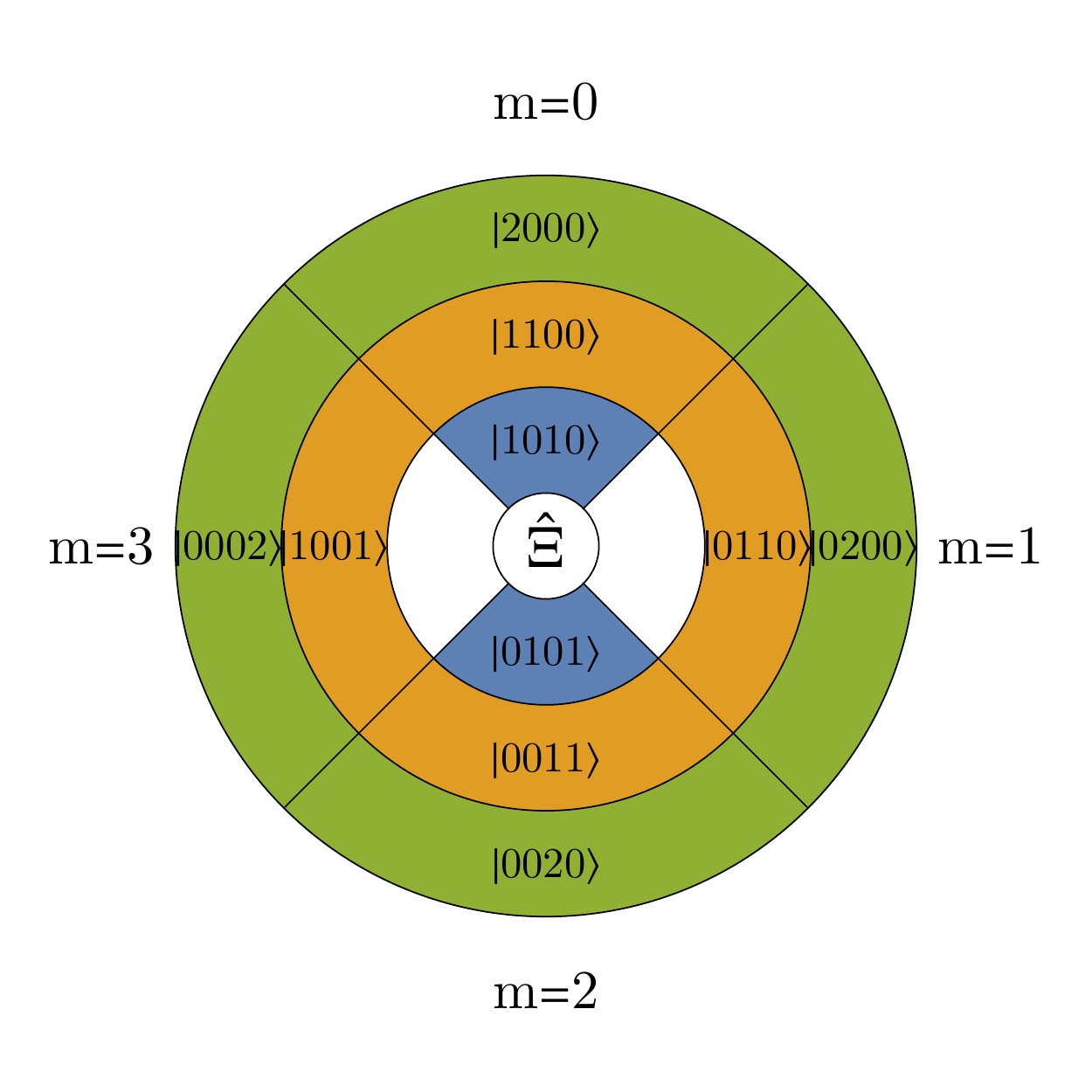}}
  \subfloat[]{\includegraphics[width=0.33\textwidth]{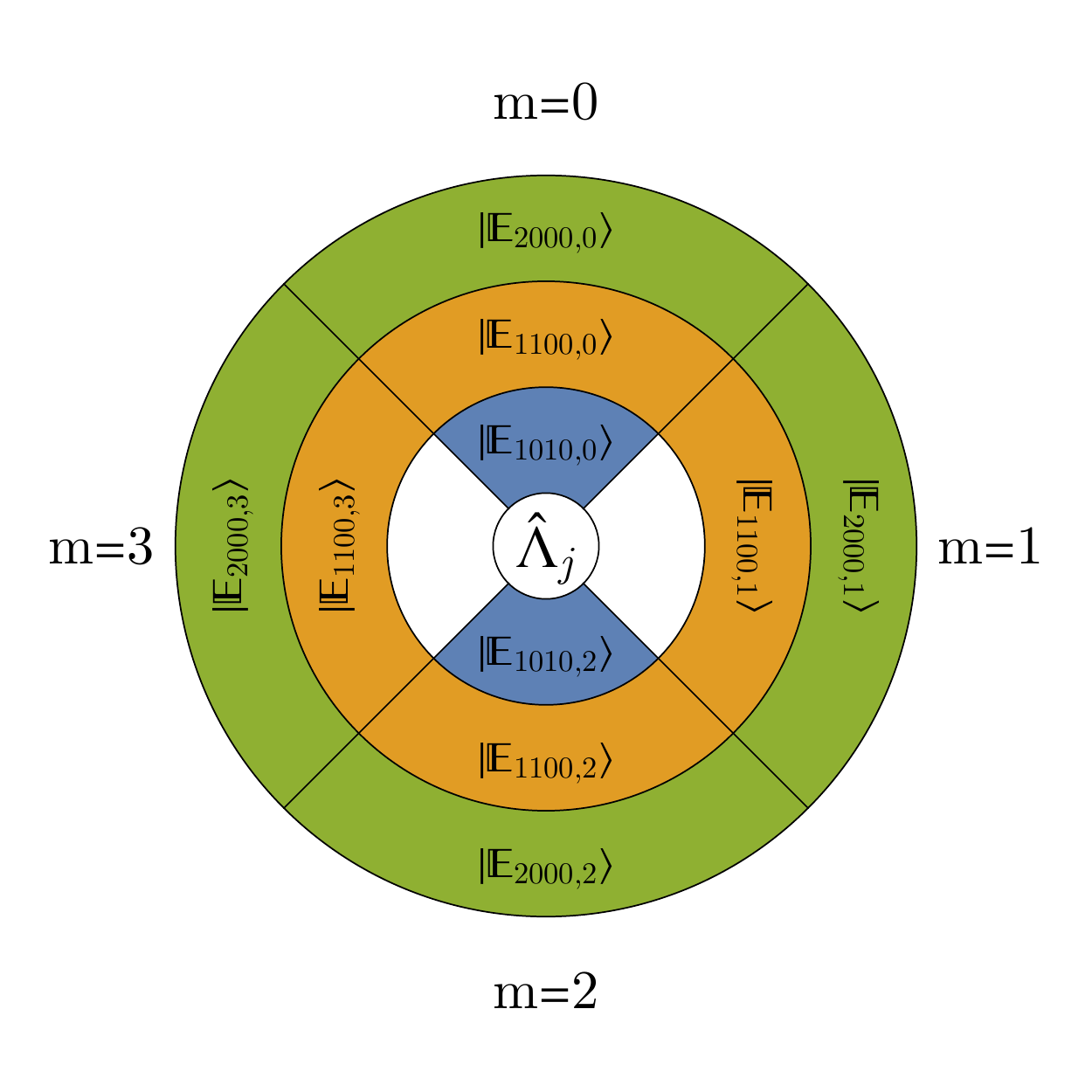}}
  \caption{Clock diagrams of generalized Pauli operators in $2$-photon $4$-mode LONs.
    (a) The clock diagram of $\widehat{Z}$. Fock states $\ket{\boldvec{n}}$ are grouped and sorted by their $\widehat{Z}$-Pauli eigenvalues $w^{\mu(\boldvec{n})}$. They are degenerated in their $\widehat{Z}$-clock labels $\mu$.
    (b) The clock diagram of $\widehat{\Xi}$ defined in Eq. \eqref{eq::Xi} with $\sigma=\id$. The operator $\widehat{\Xi}$ is non-degenerated in each Pauli subspace.
    (c) The clock diagram of $\widehat{\Lambda}_{j}$ in its eigenspace. Within each Pauli subspace, a label $m$ refers to an eigenstate uniquely.
  }
  \label{fig::clock}
\end{figure}

If the eigenbases of two operators are MUBs in a Pauli subspace $\mathbb{H}_{\mathbb{E}_{\boldvec{n}}}$, we say these two operators are \emph{complementary within the Pauli subspace $\mathbb{H}_{\mathbb{E}_{\boldvec{n}}}$}.
A pair of complementary operators should therefore define non-degenerated eigenstates within each Pauli subspace.
Since eigenstates of $\widehat{Z}$ are degenerated within particular Pauli subspaces, it is not appropriate to represent a physical property in the computational basis that is complementary to the $\widehat{\Lambda}_{j}$ operator.
The degeneracy of $\widehat{Z}$ operator can be seen from the clock-like diagram of the $2$-photon $4$-mode $\widehat{Z}$ eigenspace as shown in Fig. \ref{fig::clock} (a).
In this diagram, Fock states $\ket{\boldvec{n}}$ are grouped by the phases $\mu(\boldvec{n})$ of their $\widehat{Z}$-operator eigenvalues given in Eq. \eqref{eq::Z_shift}, which we call the \emph{$\widehat{Z}$-clock labels}.\hiddengls{ZClockLabel}
For an operator complementary to $\widehat{\Lambda}_{j}$, we need to construct it with non-degenerate labeling in each Pauli subspace as follows,\hiddengls{XiOp}
\begin{equation}
\label{eq::Xi}
  \widehat{\Xi} := \sum_{\mathbb{E}_{\boldvec{n}}}\widehat{\Xi}^{(\mathbb{E}_{\boldvec{n}})}
  \;\;\text{ with }\;\;
  \widehat{\Xi}^{(\mathbb{E}_{\boldvec{n}})}
  :=
  \sum_{m=0, \frac{M}{d_{\mathbb{E}_{\boldvec{n}}}}, ... , (d_{\mathbb{E}_{\boldvec{n}}}-1)\frac{M}{d_{\mathbb{E}_{\boldvec{n}}}}} w^{m}
  \projector{\eSt{\boldvec{n}}{m}},
\end{equation}
where $\widehat{\Xi}^{(\mathbb{E}_{\boldvec{n}})}$ is a clock operator in $\mathbb{H}_{\mathbb{E}_{\boldvec{n}}}$ with $\{\ket{\eSt{\boldvec{n}}{m}}\}_{m}$ being the computational basis of $\mathbb{H}_{\mathbb{E}_{\boldvec{n}}}$ associated with the eigenvalues $w^{m}$ and labeled as follows, \hiddengls{eState}
\begin{equation}
\label{eq::eState_def}
  \ket{\eSt{\boldvec{n}}{m}} := \widehat{X}^{
    \sigma\left(m \frac{\dimE{\boldvec{n}}}{M}\right)
  }\ket{\boldvec{n}}
  \;\;\text{ with }\;\;
  m\in\left\{0,\frac{M}{\dimE{\boldvec{n}}},..., (\dimE{\boldvec{n}}-1)\frac{M}{\dimE{\boldvec{n}}}\right\}.
\end{equation}
Here, $\sigma$ is an arbitrary permutation in the set $\{0,...,\dimE{\boldvec{n}}-1\}$.
Fig. \ref{fig::clock} (b) shows the non-degeneracy of $\widehat{\Xi}$ within $2$-photon $4$-mode Pauli subspaces for the permutation $\sigma = \id$.
The permutation $\sigma$ works as a relabeling of the eigenbasis of $\widehat{\Xi}$.
The computational basis state $\ket{e_{\boldvec{n},m}}$ in the Pauli subclass $\mathbb{E}_{\boldvec{n}}$ depends on the labeling of $m$, which is determined by the permutation $\sigma$.
No matter which permutation one takes, the eigenbasis of $\widehat{\Xi}$ is unchanged and mutually unbiased to the eigenbasis of $\widehat{\Lambda}_{j}$.
Compare Fig. \ref{fig::clock} (b) and (c), one can see that the operator $\widehat{\Xi}$ and $\widehat{\Lambda}_{j}$ define non-degenerated eigenstates that are mutually unbiased with each other in each Pauli subspace,
while the operator $\widehat{Z}$ has degenerated eigenstates in the Pauli subspaces $\mathbb{H}_{\mathbb{E}_{1100}}$ and $\mathbb{H}_{\mathbb{E}_{2000}}$.
This degeneracy leads to the ambiguity of the eigenbasis of $\widehat{Z}$ within $\mathbb{H}_{\mathbb{E}_{1100}}$ and $\mathbb{H}_{\mathbb{E}_{2000}}$, which decreases the degree of complementarity between $\widehat{Z}$ and $\widehat{\Lambda}_{j}$.
In this case, for the study of complementary properties of multiphoton states in LONs, it is therefore appropriate to refer to the operator $\widehat{\Xi}$ instead of the phase-shift operator $\widehat{Z}$.
Note that in the case $\gcd(|\boldvec{n}|,M)=1$, which guarantees non-degeneracy of $\widehat{Z}$, the operator $\widehat{\Xi}$ can be constructed as the phase-shift operator $\widehat{Z}$ with the permutation that satisfies $\sigma(m)|\boldvec{n}| \oplus \mu(\boldvec{n}) = m$.

\bigskip

Besides the operator pairs $\{\widehat{\Xi}, \widehat{\Lambda}_{j}\}$, two Pauli operators $\{\widehat{\Lambda}_{j},\widehat{\Lambda}_{l}\}$ can also be complementary.
However, their complementarity within a Pauli subspace $\mathbb{H}_{\mathbb{E}_{\boldvec{n}}}$ is not guaranteed.
The MUB structures of two Pauli operator $\{\widehat{\Lambda}_{j}, \widehat{\Lambda}_{l}\}$ in a Pauli subspace $\mathbb{H}_{\mathbb{E}_{\boldvec{n}}}$ depends on the degeneracy of the $\widehat{Z}^{l-j}$-Pauli operator in $\mathbb{H}_{\mathbb{E}_{\boldvec{n}}}$.
\begin{theorem}[MUBs within Pauli subspaces]
\label{theorem:MUBs_in_Pauli_subspace}
  Two Pauli operators $\{\widehat{\Lambda}_{j}, \widehat{\Lambda}_{l}\}$ are complementary in a Pauli subspace $\mathbb{H}_{\mathbb{E}_{\boldvec{n}}}$, if  and only if the eigenbasis of $\widehat{Z}^{l-j}$ is not degenerated in $\mathbb{H}_{\mathbb{E}_{\boldvec{n}}}$, which is equivalent to the condition that $\gcd((l-j)|\boldvec{n}|\dimE{\boldvec{n}}/M, \dimE{\boldvec{n}}) = 1$.
\begin{proof}
See Appendix.
\end{proof}
\end{theorem}%
\noindent This theorem implies that two Pauli operators $\widehat{\Lambda}_{j}$ and $\widehat{\Lambda}_{l}$ can be complementary within a Pauli subspace $\mathbb{H}_{\mathbb{E}_{\boldvec{n}}}$, while non-complementary within the other subspace $\mathbb{H}_{\mathbb{E}_{\boldvec{n}'}}$.
The complementarity in Pauli subspaces can be directly seen from the clock diagram of the operator $\widehat{Z}^{l-j}$.
For example, in a $4$-mode linear optics network, the clock diagram in Fig. \ref{fig::clock} (a) shows that the operator $\widehat{Z}$ is non-degenerate only in the Pauli subspace $\mathbb{E}_{1010}$.
According to Theorem \ref{theorem:MUBs_in_Pauli_subspace}, the operators $\{\widehat{\Lambda}_{j},\widehat{\Lambda}_{j+1}\}$ are complementary within the $2$-photon Pauli subspace $\mathbb{H}_{1010}$, but non-complementary within the subspaces $\mathbb{H}_{1100}$ and $\mathbb{H}_{2000}$.
It is therefore necessary to refer to complementary Pauli operators in LONs with reference to Pauli subspaces.
Examples of the complementary structures of Pauli operators in $4$-mode LONs are demonstrated in Table \ref{tab::eg_E_compl}.
In $3$-photon Pauli subspaces $\mathbb{H}_{\mathbb{E}_{\boldvec{n}}}$, the operator $\widehat{Z}$ is non-degenerate in any Pauli subspace, which leads to the complementarity of the operators $\{\widehat{\Lambda}_{j},\widehat{\Lambda}_{j+1}\}$ in $3$-photon $4$-mode LONs.

\begin{table}
  \centering
  \begin{tabular}
    [c]{|c|c|c|c|c|c|c|c|}
    \hline
      & Pauli class
      & $\{\Lambda_{0},\Lambda_{1}\}$ & $\{\Lambda_{0},\Lambda_{2}\}$ & $\{\Lambda_{0},\Lambda_{3}\}$
      & $\{\Lambda_{1},\Lambda_{2}\}$ & $\{\Lambda_{1},\Lambda_{3}\}$
      & $\{\Lambda_{2},\Lambda_{3}\}$
    \\ \hline
      1 photon & $\mathbb{E}_{1000}$
      & \checkyes & \checkno & \checkyes
      & \checkyes & \checkno
      & \checkyes
    \\ \hline
      \multirow{3}{*}{2 photons} & $\mathbb{E}_{2000}$
      & \checkno & \checkno & \checkno
      & \checkno & \checkno
      & \checkno
    \\ \cline{2-8}
    & $\mathbb{E}_{1100}$
      & \checkno & \checkno & \checkno
      & \checkno & \checkno
      & \checkno
    \\ \cline{2-8}
      & $\mathbb{E}_{1010}$
      & \checkyes & \checkno & \checkyes
      & \checkyes & \checkno
      & \checkyes
    \\ \hline
      \multirow{5}{*}{3 photons} & $\mathbb{E}_{1110}$
      & \checkyes & \checkno & \checkyes
      & \checkyes & \checkno
      & \checkyes
    \\ \cline{2-8}
      & $\mathbb{E}_{2100}$
      & \checkyes & \checkno & \checkyes
      & \checkyes & \checkno
      & \checkyes
    \\ \cline{2-8}
      & $\mathbb{E}_{2010}$
      & \checkyes & \checkno & \checkyes
      & \checkyes & \checkno
      & \checkyes
    \\ \cline{2-8}
      & $\mathbb{E}_{2001}$
      & \checkyes & \checkno & \checkyes
      & \checkyes & \checkno
      & \checkyes
    \\ \cline{2-8}
      & $\mathbb{E}_{3000}$
      & \checkyes & \checkno & \checkyes
      & \checkyes & \checkno
      & \checkyes
    \\ \hline
      4 photons & all Pauli classes
      & \checkno & \checkno & \checkno
      & \checkno & \checkno
      & \checkno
    \\ \hline
    \vdots & \vdots  & \vdots  & \vdots  & \vdots & \vdots  & \vdots  & \vdots
    \\ \hline
  \end{tabular}
  \caption{The complementary structure of Pauli operators $\widehat{\Lambda}_{j}$ within different Pauli subspaces $\mathbb{H}_{\mathbb{E}_{\boldvec{n}}}$ in a $4$-mode system. The symbols {\checkyes} indicate $\mathbb{E}_{\boldvec{n}}$-complementary of corresponding Pauli operator pairs $\{\widehat{\Lambda}_{j},\widehat{\Lambda}_{l}\}$, while {\checkno} indicate non-complementary in $\mathbb{H}_{\mathbb{E}_{\boldvec{n}}}$}
  \label{tab::eg_E_compl}
\end{table}

Theorem \ref{theorem:MUBs_in_Pauli_subspace} tells us how to construct a set of complementary operators for a Pauli subspace.
In general, an $N$-photon state can be a superposition of quantum states in different Pauli subspaces.
It is therefore worthwhile to construct a set of operators $\mathbb{L}$ that are complementary in all $N$-photon Pauli subspaces.
According to Theorem \ref{theorem:MUBs_in_Pauli_subspace}, complementary operators in $N$-photon LON systems can be selected from $\{\widehat{\Xi}, \widehat{\Lambda}_{0},...,\widehat{\Lambda}_{M-1}\}$ as follows.
\begin{corollary}[Complementary operators in $N$-photon LONs]
\label{coro::MUBs_for_Mmd_Nph}
  In an $N$-photon LON system,
  \begin{enumerate}
    \item the operators $\{\widehat{\Xi}, \widehat{\Lambda}_{j}\}$ are complementary in all $N$-photon Pauli subspaces, for any $j\in\{0,...,M-1\}$;
    \item the operators $\{\widehat{\Xi}, \widehat{\Lambda}_{j}, \widehat{\Lambda}_{l}\}$ are complementary in all $N$-photon Pauli subspaces, if $\gcd(N,M)=\gcd(|l-j|,M)=1$;
    \item the operators $\{\widehat{\Xi}, \widehat{\Lambda}_{0},...,\widehat{\Lambda}_{p_{1}-1}\}$ with $p_{1}$ being the minimum nontrivial prime divisor of $M$ are complementary in all $N$-photon Pauli subspaces, if $\gcd(N,M)=1$;
    \item  the operators $\{\widehat{\Xi}, \widehat{\Lambda}_{0}, ..., \widehat{\Lambda}_{M-1}\}$ are complementary in all $N$-photon Pauli subspaces, if $M$ is prime and the total photon number $N$ is not a multiple of $M$.
  \end{enumerate}
\end{corollary}
\noindent For a prime $M$, one can therefore construct a complete set of complementary operators in $N$-photon LONs.
If $M$ is a prime power $p^{k}$, one has to decompose the $p^{k}$-mode LON into a $k$-level $p$-branch tree-style LON and construct MUBs in each $p$-mode subsystem followed by its extension to higher levels.
In qudit systems, it is shown that a complete set of complementary operators for $M = p^{k}$ exists\cite{DurtAtElZyczkowski2010-MUBs}.
However, the complete set of MUBs in multiphoton LONs with $M = p^{k}$ is not straightforwardly extendible from the qudit system due to the photonic bunching effects.
Since characterization of the complete set of MUBs in multiphoton LONs is out of the scope of this paper, we leave this question open.

Other than Pauli operators, general unitary operators in multiphoton LONs also have eigenstates, which are superpositions of the computational-basis states. Their quantum coherence can also be studied within their irreducible subspaces\cite{HayashiBook-GrpReprQThry}, which are in general larger than the Pauli subspaces.
The theoretical aspects of complementary properties within these larger irreducible subspaces are also important for the fundamental understanding of quantum coherences in multiphoton LON systems.
However, in this paper, we focus on the characterization of complementary properties in multiphoton LONs in an experimentally feasible way.
We therefore study the complementary structures of the Pauli operators in multiphoton LONs, as their eigenbases are mutually unbiased to the computational basis within their irreducible subspaces.
It means that together with the measurement in the computational basis, one can employ just one additional measurement associated with a Pauli operator $\widehat{\Lambda}_{j}$ to access maximum quantum coherence between Fock states in multiphoton LONs within Pauli subspaces $\mathbb{H}_{\mathbb{E}_{\boldvec{n}}}$.
This is a preferable property for experimental implementation.
In the next section, we will study the Pauli measurements that allow us to access quantum coherence between Fock states in multiphoton LONs within these Pauli subspaces.

\section{Complementary Pauli measurements in linear optics networks}
\label{sec::Pauli_msmnt}

Measurements in the MUBs associated with a set of complementary Pauli operators constructed in Corollary \ref{coro::MUBs_for_Mmd_Nph} can be exploited to evaluate complementary properties of multiphoton states in LONs.
A trivial measurement is in the computational basis, which is associated with the operator $\widehat{\Xi}$ specified in Eq. \eqref{eq::Xi}.
For $N$-photon states, the operator $\widehat{\Xi}$ can be decomposed into the sum of projectors $\plProj{N}{m}{\Xi}$ that project onto eigenspaces with eigenvalues $w^{m}$ labeled by $m$ as \hiddengls{PauliProj}
\begin{equation}
\label{eq::Xi_op}
  \widehat{\Xi}
  =
  \sum_{m=0}^{M-1} w^{m}\widehat{\pi}_{N, m}(\Xi)
  \;\;\text{ with }\;\;
  \plProj{N}{m}{\Xi} := \sum_{\mathbb{E}_{\boldvec{n}}:|\boldvec{n}|=N}\projector{\eSt{\boldvec{n}}{m}},
\end{equation}
where $\ket{e_{\boldvec{n},m}}$ are the computational basis states in the Pauli subspace $\mathbb{E}_{\boldvec{n}}$ labeled by $m$, which are defined in Eq. \eqref{eq::eState_def}.
The expectation value of $\widehat{\Xi}$ can be then evaluated in the projective measurement $\{\plProj{N}{m}{\Xi}\}_{m}$, which we call a \emph{$\widehat{\Xi}$-Pauli measurement}.
Since the assignment of the eigenvalues $\omega^{m}$ depends on the relabeling permutation $\sigma$ as shown in Eq. \eqref{eq::eState_def}, the expectation value $\braket{\widehat{\Xi}}$ also depends on the choice of $\sigma$. However, no matter which permutation $\sigma$ one chooses, the measurement $\{\widehat{\pi}_{N,m}(\Xi)\}_{m}$ associated with $\widehat{\Xi}$ is always the measurement in the computational basis, which does not depend on the relabeling of measurement basis states by $\sigma$.
The relabeling permutation $\sigma$ is therefore a degree of freedom for the construction of $\widehat{\Xi}$ that can be deployed according to demands even after the implementation of the measurement in the computational basis.
If the mode number $M$ is prime, an instinct construction of $\widehat{\Xi}$ is the operator $\widehat{Z}$.

A Pauli operator $\widehat{\Lambda}_{j}$ can be also decomposed as a sum of eigenvalue projectors $\widehat{\pi}_{N,m}(\Lambda_{j})$,
\begin{equation}
  \widehat{\Lambda}_{j}
  =
  \sum_{m} w^{\frac{1}{2}(M-1)jN+m} \widehat{\pi}_{N,m}(\Lambda_{j}),
\end{equation}
where the projector $\widehat{\pi}_{N,m}(\Lambda_{j})$ is called an $N$-photon \emph{$\widehat{\Lambda}_{j}$-Pauli projector} for the label $m$, and explicitly defined by \hiddengls{PauliProj}
\begin{equation}
\label{eq::Pauli_proj_def}
  \widehat{\pi}_{N,m}(\Lambda_{j})
  :  =
  \sum_{\mathbb{E}_{\boldvec{\nu}}: |\boldvec{\nu}|=N}
  \projector{\mathbb{E}_{\boldvec{\nu},m}(\Lambda_{j})}.
\end{equation}
In $N$-photon LONs, a \emph{$\widehat{\Lambda}_{j}$-Pauli measurement} in the eigenspace of $\widehat{\Lambda}_{j}$ is then a projective measurement represented by the $\widehat{\Lambda}_{j}$-Pauli projectors $\{\plProj{N}{m}{\Lambda_{j}}\}_{m}$.

\begin{figure}
  \centering
  \subfloat[]{\includegraphics[width = 0.4\columnwidth]{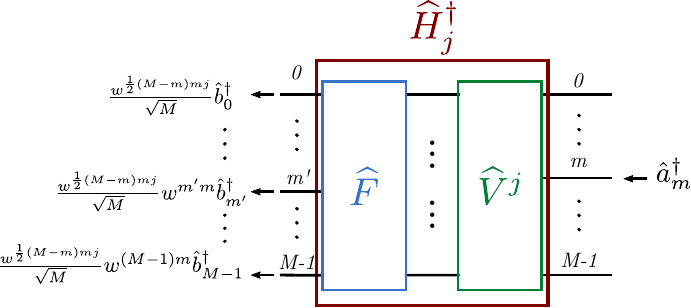}}
  \subfloat[]{\includegraphics[width = 0.29\columnwidth]{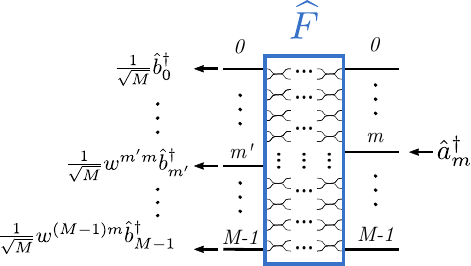}}
  \hspace{0.01\textwidth}
  \subfloat[]{\includegraphics[width = 0.29\columnwidth]{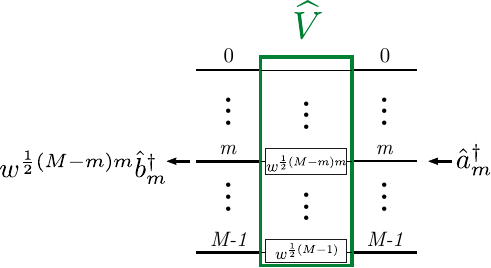}}
  \caption{(a) The LON of a generalized inverse Hadamard operator $\widehat{H}_{j}^{\dagger}$ can be decomposed into the standard discrete Fourier transform $\widehat{F}$ and a phase-shift transform $\widehat{V}$. (b) The LON of the standard discrete Fourier transforms $\widehat{F}$. (c) The LON of the phase shift transforms $\widehat{V}$ which are necessary for $\widehat{H}_{j}^{\dagger}$.
  Note that the time in (a-c) flows from right to left as the arrows indicate.
  }
\label{fig::H_op}
\end{figure}%

For $\widehat{\Lambda}_{j}$-Pauli measurements, one needs the corresponding inverse Hadamard transform $\widehat{H}_{j}^{\dagger}$ to transform a $\widehat{\Lambda}_{j}$-Pauli eigenbasis to the computational Fock-state basis, such that one can employ photon number resolving detection in the outputs of $\widehat{H}_{j}^{\dagger}$ to measure input states in the $\widehat{\Lambda}_{j}$-Pauli eigenbasis.
As shown in Eq. \eqref{eq::md_transf_mtx} and \eqref{eq::md_transf_mtx_1}, an inverse Hadamard transform of modes is determined by its transformation of single-photon states.
In single-photon LONs, a generalized Hadamard transform $\widehat{H}_{j}$ maps the computational basis states $\{\ket{\cdots01_{m}0\cdots}\}_{m}$ to the eigenstates $\{\ket{\mathbb{E}_{10\cdots,m}(\Lambda_{j})}\}_{m}$ of a $\widehat{\Lambda}_{j}$-Pauli operator,
\begin{equation}
  \widehat{H}_{j}^{(\mathbb{E}_{10\cdots})}
  =\sum_{m}\ket{\mathbb{E}_{10\cdots, m}(\Lambda_{j})}\bra{\cdots01_{m}0\cdots}.
\end{equation}
The matrix $\{\braket{\cdots01_{m}0\cdots|\mathbb{E}_{10\cdots, m'}(\Lambda_{j})}\}_{m,m'}$ of the overlaps between the single-photon computational basis states and the single-photon $\widehat{\Lambda}_{j}$-eigenbasis states is a complex Hadamard matrix\cite{DurtAtElZyczkowski2010-MUBs}.
According to Eq. \eqref{eq::En_Pauli_eigenstate}, the inverse Hadamard transform $\widehat{H}_{j}^{\dagger}$ is then described by \hiddengls{Hadamard}
\begin{align}
  \widehat{H}_{j}^{\dagger}\widehat{a}_{m}^{\dagger}\widehat{H}_{j}%
  = &
  \sum_{m'}
  \braket{\ESt{10\cdots}{m'}{j}|\cdots01_{m}0\cdots}\widehat{b}^{\dagger}_{m'}
  \nonumber \\
  =&
  \frac{w^{\frac{1}{2}(M-m)mj}}{\sqrt{M}}\sum_{m'} w^{m' m }\widehat{b}^{\dagger}_{m'}.
\end{align}
As shown in Fig. \ref{fig::H_op}, the LONs of Hadamard operators can be decomposed into a combination of the standard discrete Fourier transform and a phase shift $\widehat{V}$ as
\begin{equation}
  \widehat{H}_{j}^{\dagger}=\widehat{F}\widehat{V}^{j}
  \;\;\text{ with }\;\;
  \widehat{F}\widehat{a}^{\dagger}_{m}\widehat{F}^{\dagger}  =
  \frac{1}{\sqrt{M}}\sum_{m'}w^{m'm}\widehat{a}^{\dagger}_{m'}
  \;\;\text{ and }\;\;
  \widehat{V}\widehat{a}^{\dagger}_{m}\widehat{V}^{\dagger} = w^{\frac{1}{2}(M-m)m}\widehat{a}_{m}^{\dagger}.
\end{equation}

\begin{figure}
  \centering
  \subfloat[]{\includegraphics[width=0.57\textwidth]{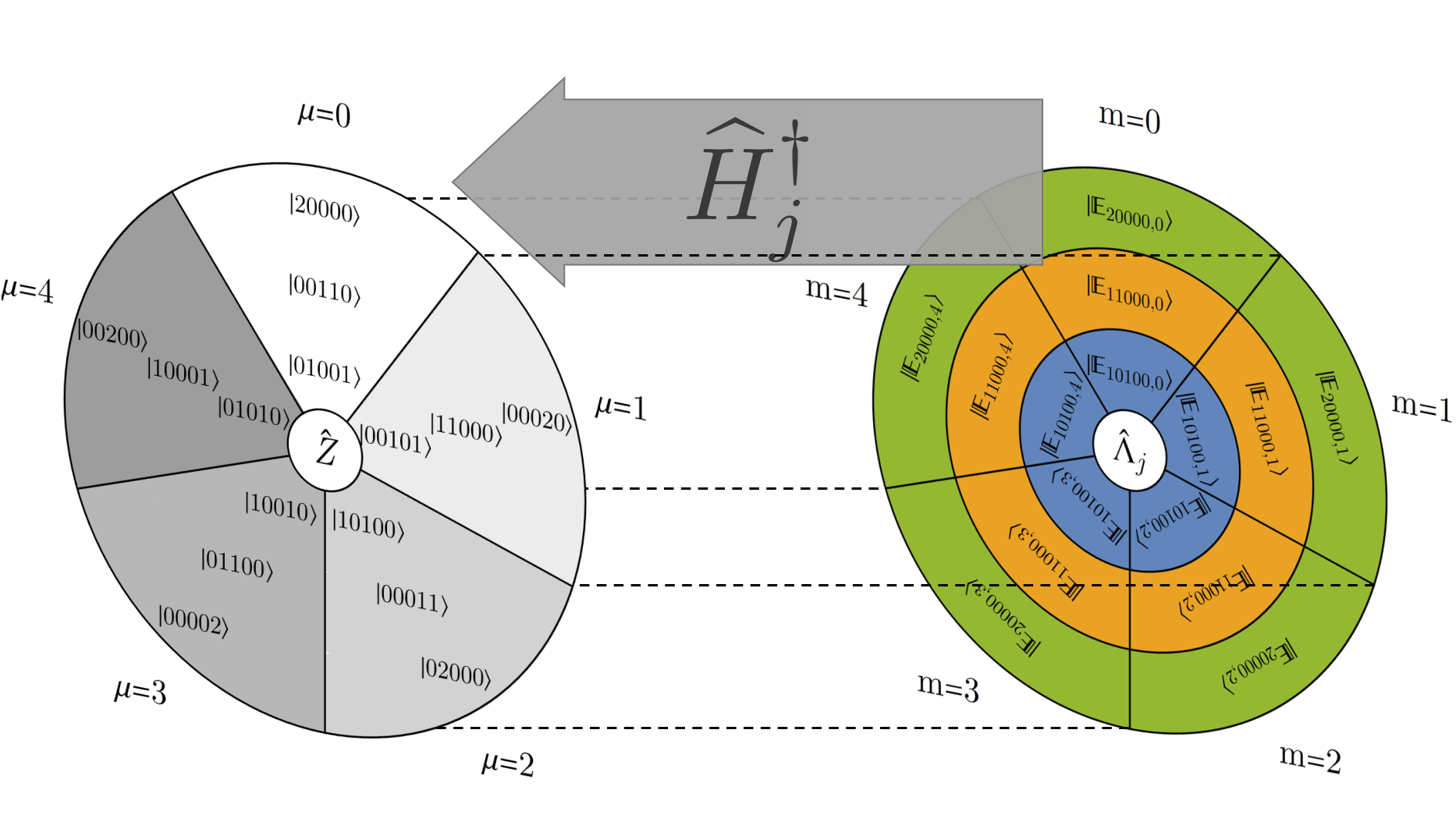}}
  \hspace{0.02\textwidth}
  \subfloat[]{\includegraphics[width=0.4\textwidth]{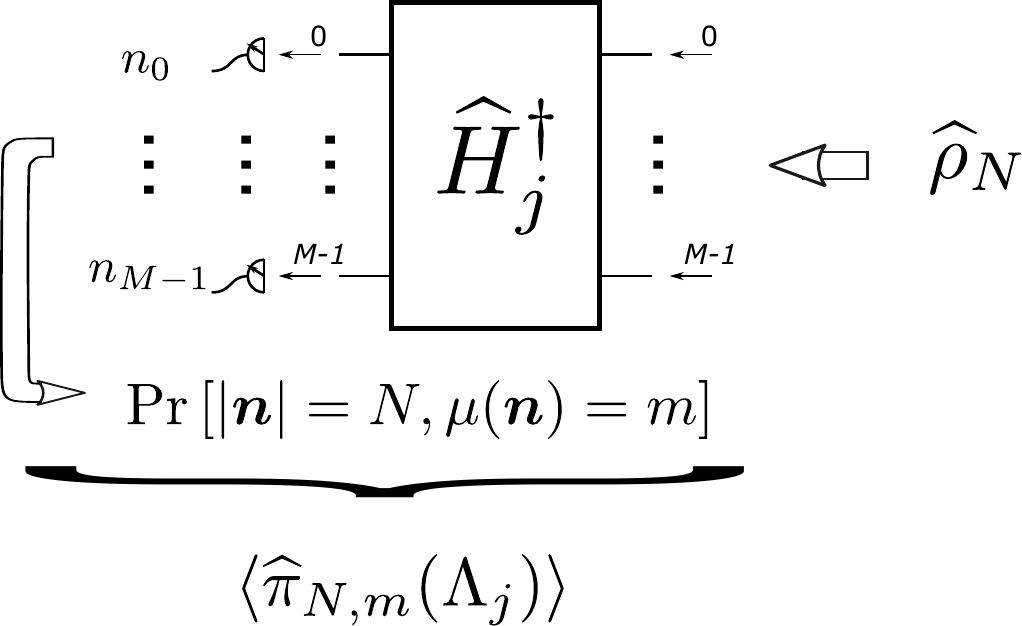}}
  \caption{Pauli measurements in LONs.
    (a) A $\widehat{H}^{\dagger}_{j}$-Hadamard transformation of the $\widehat{\Lambda}_{j}$-Pauli
    eigenspaces to the $\widehat{Z}$-Pauli eigenspaces in a $2$-photon $5$-mode LON.
    (b) A $\widehat{\Lambda}_{j}$-Pauli measurement $\{\widehat{\pi}_{N,m}(\Lambda_{j})\}_{m}$ in an $N$-photon LON. The $N$-photon $\widehat{\Lambda}_{j}$-Pauli projectors    $\{\widehat{\pi}_{N,m}(\Lambda_{j})\}_{m}$ are evaluated by counting the probability of photon number detection events that satisfy $\mu(\boldvec{n})=m$.
  }
  \label{fig::Pauli_msmnt}
\end{figure}

For a single photon, the Pauli operator $\widehat{\Lambda}_{j}$ performed on inputs of $\widehat{H}^{\dagger}_{j}$ is equivalent to the phase-shift operator $\widehat{Z}$ performed on outputs of $\widehat{H}^{\dagger}_{j}$ up to a phase,
\begin{equation}
\label{eq::HL_ZH}
  \widehat{H}_{j}^{\dagger}\widehat{\Lambda}_{j}
  =
  w^{\frac{1}{2}(M-1)j}
  \widehat{Z} \widehat{H}_{j}^{\dagger}
   .
\end{equation}
In the outputs of a $\widehat{H}_{j}^{\dagger}$ transform of an $N$-photon input, the additional phase $w^{(M-1)j/2}$ is added to each photon and leads to a total phase shift $w^{(M-1)j|\boldvec{n}|/2}$.
Applying this relation to an $\widehat{\Lambda}_{j}$ eigenstate $\ket{\ESt{\boldvec{n}}{m}{j}}$ and according to the eigenequation of $\widehat{\Lambda}_{j}$ given in Eq. \eqref{eq::eigeneq_Lambda_j}, the additional phase shift will be eliminated, which leads to the following eigenequation,
\begin{equation}
\label{eq::eigeneq_H_trans_E_St}
  \widehat{Z}\widehat{H}^{\dagger}_{j}\ket{\ESt{\boldvec{n}}{m}{j}}
  =
  w^{m}\widehat{H}^{\dagger}_{j}\ket{\ESt{\boldvec{n}}{m}{j}}.
\end{equation}
It means that the $\widehat{H}_{j}^{\dagger}$ transforms a $\widehat{\Lambda}_{j}$ eigenstate $\ket{\ESt{\boldvec{n}}{m}{j}}$ to a $\widehat{Z}$ eigenstate with the eigenvalue $w^{m}$.
As a result, the only possible outputs of the transformation $\widehat{H}_{j}^{\dagger}\ket{\ESt{\boldvec{n}}{m}{j}}$ are the $\widehat{Z}$ eigenstates with the eigenvalue $w^{m}$, which are the Fock states $\ket{\boldvec{\nu}}$ with the $\widehat{Z}$-clock label $\mu(\boldvec{\nu})=m$.
This is a suppression law of inverse Hadamard transforms, which is a special case of the suppression law of general permutation invariant states \cite{DittelEtAlKeil2018-DestrInterfPermSymMPtclSt, DittelEtAlKeil2018-DestrMPtclInterf}.
Eq. \eqref{eq::eigeneq_H_trans_E_St} shows that the eigenspaces of $\widehat{\Lambda}_{j}$ are transformed to the eigenspaces of $\widehat{Z}$ by the inverse Hadamard $\widehat{H}_{j}^{\dagger}$, which means that a $\widehat{\Lambda}_{j}$-Pauli projector is equivalent to a $\widehat{H}_{j}$-transformed $\widehat{Z}$-Pauli projector,
\begin{equation}
\label{eq::Pauli_proj_Hj_out}
  \widehat{\pi}_{N,m}(\Lambda_{j})
  = \widehat{H}_{j}\plProj{N}{m}{Z}\widehat{H}_{j}^{\dagger}
  \;\;\text{ with }\;\;
  \plProj{N}{m}{Z} := \sum_{|\boldvec{n}|=N, \mu(\boldvec{n})=m}
  \projector{\boldvec{n}}.
\end{equation}
From the example in a $2$-photon $5$-mode system shown in Fig. \ref{fig::Pauli_msmnt} (a), one can see that the Hadamard transform $\widehat{H}_{j}^{\dagger}$ maps each $\widehat{\Lambda}_{j}$ eigenspace to its corresponding $\widehat{Z}$ eigenspace without changing the clock labels.
Note that from an output event $\boldvec{n}$ of a Hadamard transform $\widehat{H}_{j}^{\dagger}$, one can not distinguish the Pauli subspaces of inputs.
We can only distinguish the eigenspaces of the Pauli operator $\widehat{\Lambda}_{j}$ associated with different labels $m$ by taking all possible outputs satisfying $\mu(\boldvec{n})=m$ into account.
As a result, one can implement a $\widehat{\Lambda}_{j}$-Pauli measurement through PNRD on the output modes of the corresponding inverse Hadamard $\widehat{H}_{j}^{\dagger}$ to obtain the measurement statistics  $\{\braket{\widehat{\pi}_{N,m}(\Lambda_{j})}\}_{m}$ according to the following theorem.
\begin{theorem}[Pauli measurement]
\label{theorem::Pauli_msmnt}
  Given a quantum state $\widehat{\rho}$, its expectation value of a $\widehat{\Lambda}_{j}$-Pauli projector $\widehat{\pi}_{N,m}(\Lambda_{j})$ can be evaluated by simply counting the probability of detecting photon number occupations $\boldvec{n}$ satisfying $\mu(\boldvec{n})=m$ in the output modes of a $\widehat{H}_{j}^{\dagger}$ transform
  \begin{equation}
  \label{eq::Pauli_proj_msmnt}
    \braket{\widehat{\pi}_{N, m}(\Lambda_{j})}
    =
    \mathrm{Pr}\left[
      \left(|\boldvec{n}|=N, \mu(\boldvec{n})=m\right)\;\vert\;\widehat{H}^{\dagger}_{j}\widehat{\rho}\widehat{H}_{j}
    \right].
  \end{equation}%
\end{theorem}%

A schematic $\widehat{\Lambda}_{j}$-Pauli measurement is shown in Fig. \ref{fig::Pauli_msmnt} (b). This theorem shows that the computational complexity of Boson sampling can be lifted up in a $\widehat{\Lambda}_{j}$-Pauli measurement, if one sums up the sampling probability distribution $\{\mathrm{Pr}[\boldvec{n}|\widehat{H}_{j}^{\dagger}\widehat{\rho}\widehat{H}_{j}]\}_{\boldvec{n}}$ to a collective one $\{\mathrm{Pr}[\mu(\boldvec{n})=m|\widehat{H}_{j}^{\dagger}\widehat{\rho}\widehat{H}_{j}]\}_{m}$.
Although the exact $\widehat{H}_{j}^{\dagger}$ transform of a state is $\#P$-hard to calculate, Theorem \ref{theorem::Pauli_msmnt} allows us to theoretically predict and experimentally verify its collective probability distribution $\{\mathrm{Pr}\left[\mu(\boldvec{n})=m\right]\}_{m}$ in Pauli measurements.
Together with Theorem \ref{theorem:MUBs_in_Pauli_subspace} and Corollary \ref{coro::MUBs_for_Mmd_Nph}, one can now construct complementary Pauli measurements in multiphoton LONs.

\bigskip

\begin{figure}
  \centering
  \subfloat[]{\includegraphics[width=0.33\textwidth]{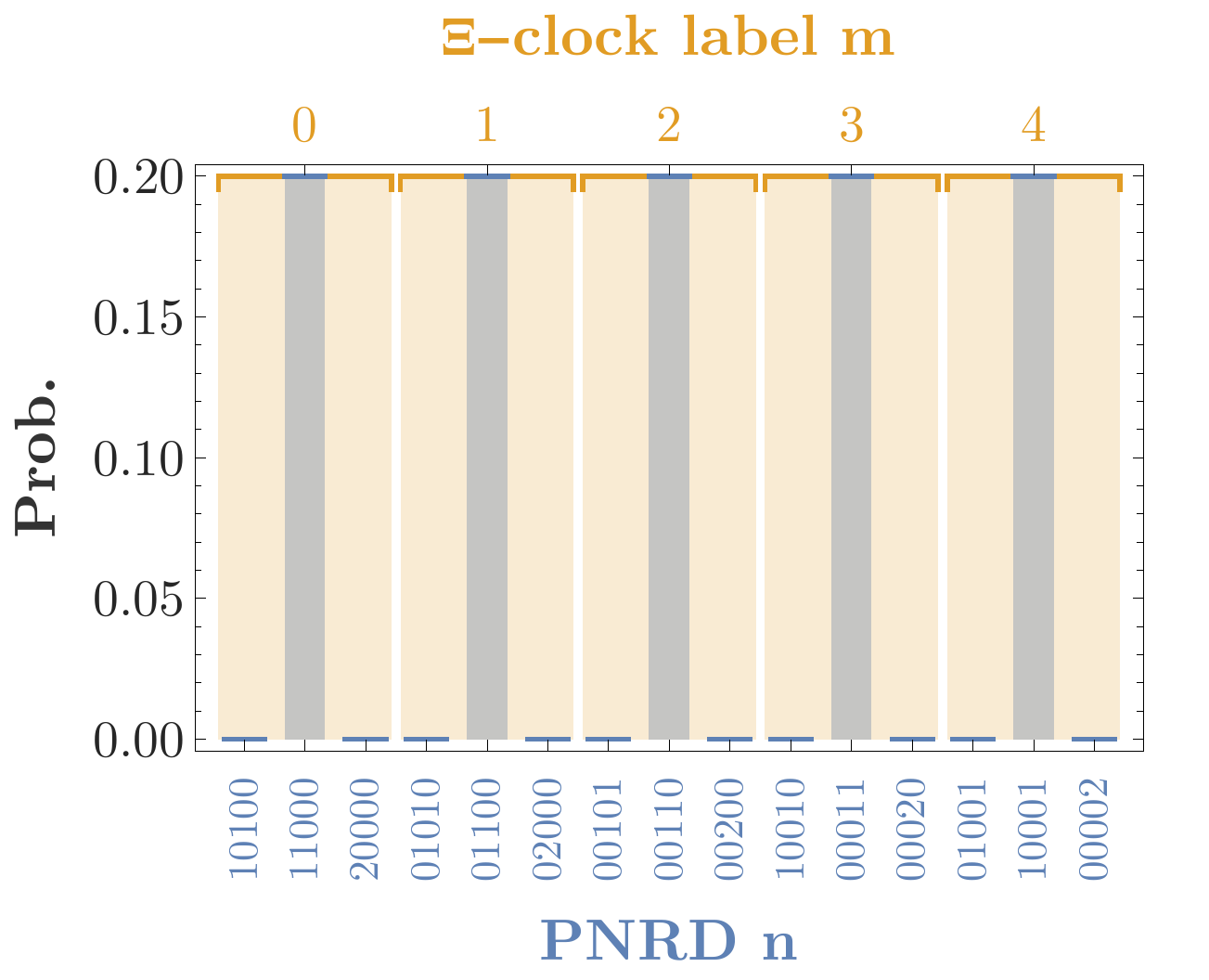}}
  \subfloat[]{\includegraphics[width=0.33\textwidth]{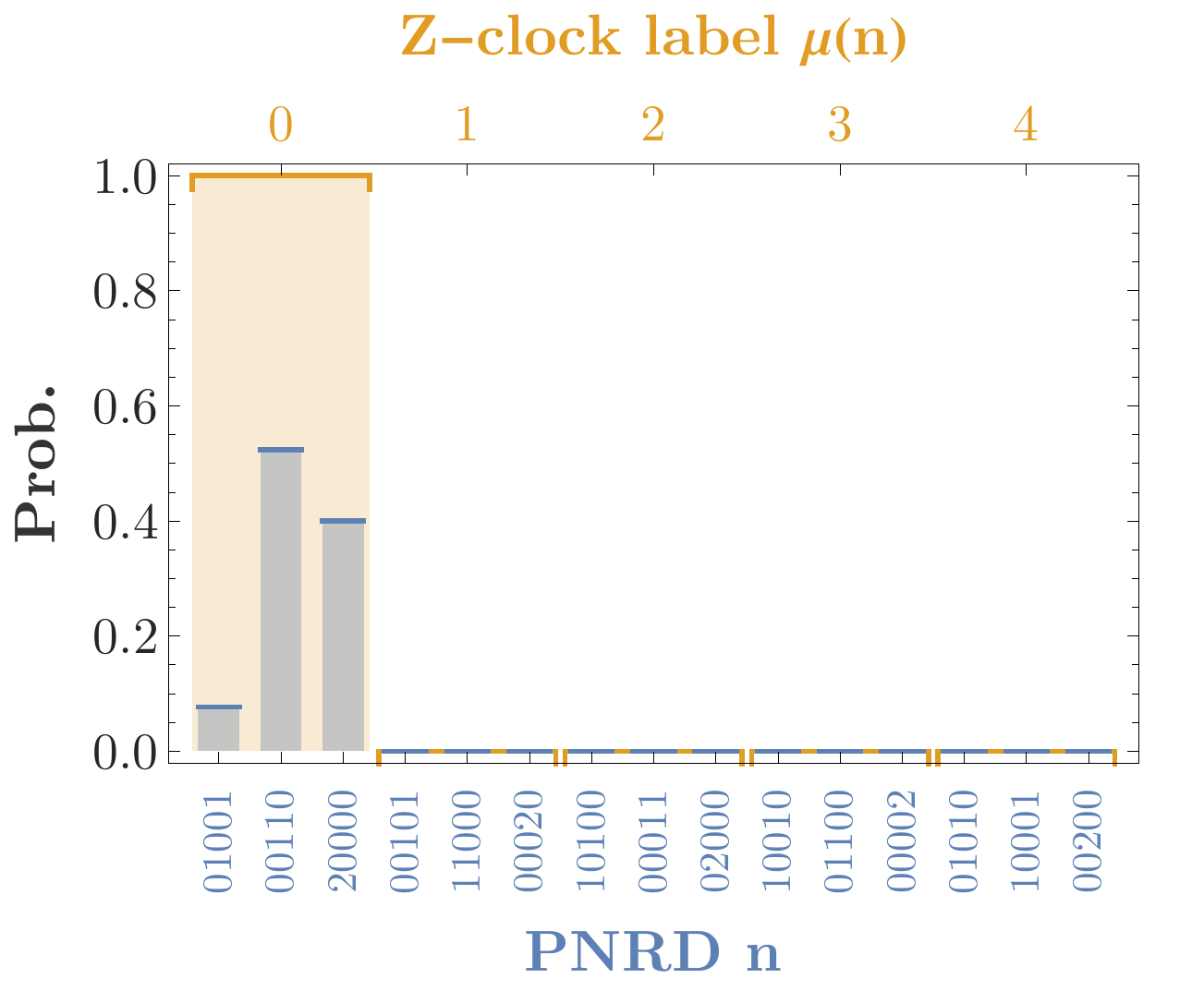}}
  \subfloat[]{\includegraphics[width=0.33\textwidth]{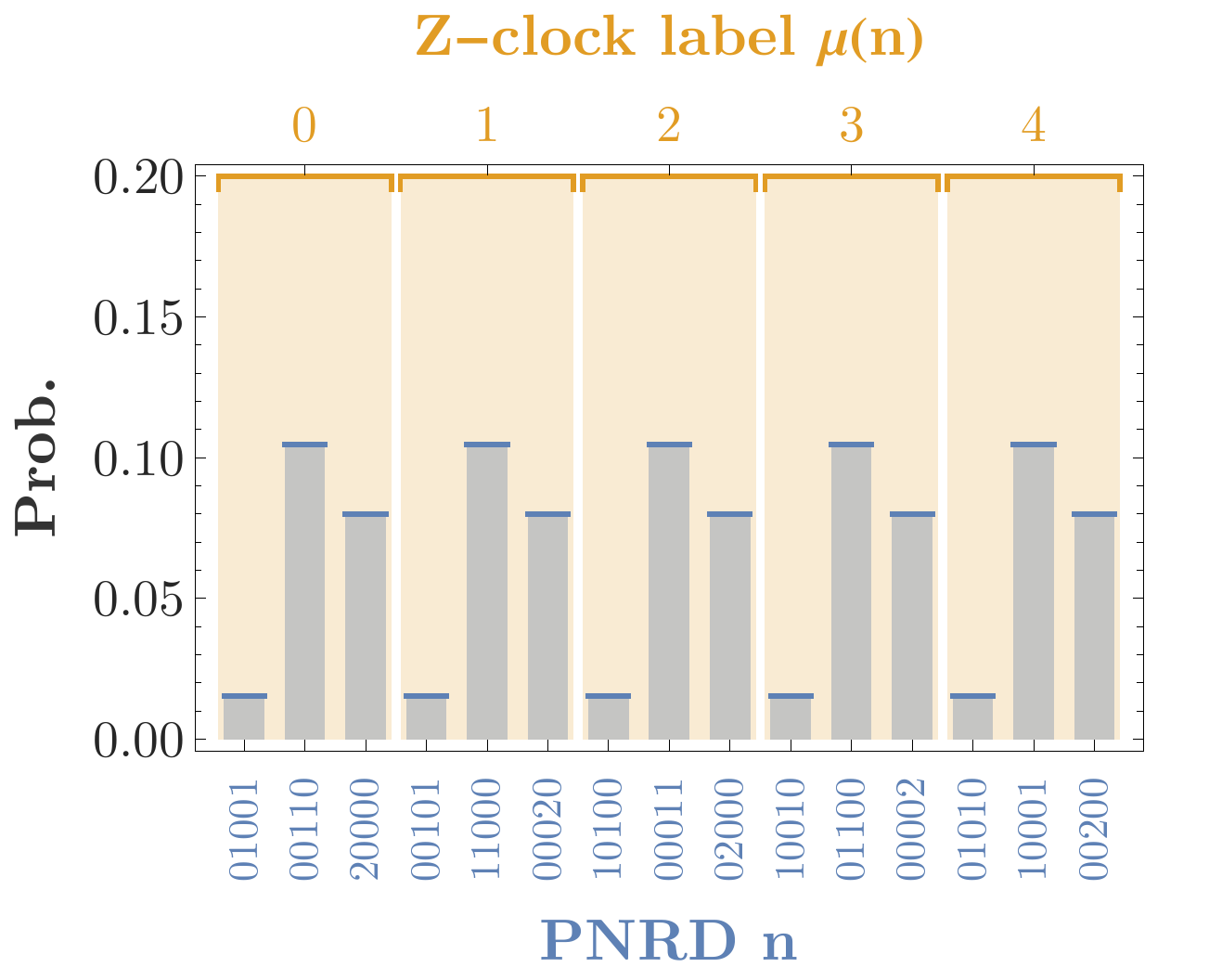}}
  \caption{Pauli measurements of the state $\ket{\ESt{11000}{0}{0}}$. PNRD statistics $\{\mathrm{Pr}(\boldvec{n})\}_{\boldvec{n}}$ are plotted in blue, while the Pauli measurement statistics $\{\braket{\widehat{\pi}_{N,m}(\cdot)}\}_{m}$ are plotted in orange.
    (a) The measurement in the computational basis associated with $\widehat{\Xi}$.
    (b) The $\widehat{\Lambda}_{0}$-Pauli measurement.
    (c) The $\widehat{\Lambda}_{j}$-Pauli measurement with $j=\{1,...,4\}$.
  }
  \label{fig::Pauli_msmnt_eigenSt}
\end{figure}

In Fig. \ref{fig::Pauli_msmnt_eigenSt}, Pauli measurement statistics of a $2$-photon $5$-mode $\widehat{\Lambda}_{0}$ eigenstate $\ket{\ESt{11000}{0}{0}}$ is demonstrated. Fig. \ref{fig::Pauli_msmnt_eigenSt} (a) shows the measurement statistics in the computational basis, in which the output are grouped by their $\Xi$-clock labels $m$ defined in Eq. \eqref{eq::eState_def} with $\sigma=\id$, where $\widehat{\Xi}=\widehat{\Xi}^{(20000)}+\widehat{\Xi}^{(11000)}+\widehat{\Xi}^{(10100)}$ is defined within the Pauli subspaces $\mathbb{E}_{20000}$, $\mathbb{E}_{11000}$ and $\mathbb{E}_{10100}$.
Fig. \ref{fig::Pauli_msmnt_eigenSt} (b) is the $\widehat{\Lambda}_{0}$-Pauli measurement.
It shows that the possible outputs of the $\widehat{H}_{0}^{\dagger}$ transformation of $\ket{\ESt{11000}{0}{0}}$ are constrained by the condition that their $Z$-clock labels are $\mu(\boldvec{n})=0$.
Fig. \ref{fig::Pauli_msmnt_eigenSt} (c) is the statistics obtained by the other $\widehat{\Lambda}_{j}$-Pauli measurements with $j=\{1,...,4\}$, which are identical for any $j\neq0$.
According to Corollary \ref{coro::MUBs_for_Mmd_Nph}, the operator $\widehat{\Lambda}_{j}$ with $j\neq0$ is complementary to $\widehat{\Lambda}_{0}$ within the all Pauli subspaces, the expectation value of a $\widehat{\Lambda}_{j}$-Pauli projector of the $\widehat{\Lambda}_{0}$ eigenstate $\ket{\ESt{11000}{0}{0}}$ is therefore uniformly equal to $\braket{\plProj{N}{m}{\Lambda_{j}}}=1/5$.
As a result of Theorem \ref{theorem::Pauli_msmnt}, the expectation value of a $\widehat{\Lambda}_{j}$-Pauli projector $\braket{\plProj{N}{m}{\Lambda_{j}}}$ is equal to the collective probability $\mathrm{Pr}[(\mu(\boldvec{n})=m)\,|\,\widehat{H}_{j}^{\dagger}]$.
The complementarity of $\{\widehat{\Lambda}_{0},....,\widehat{\Lambda}_{4}\}$ can be therefore verified in their corresponding Pauli measurement statistics $\mathrm{Pr}[\mu\,|\,\widehat{H}_{j}^{\dagger}]$, which are plotted in orange bars in Fig. \ref{fig::Pauli_msmnt_eigenSt}.

\bigskip

\section{Convex roof extension of complementary properties over Pauli subspaces}
\label{sec::cmpl_msmnt}

Now we are ready to implement measurements in the computational basis and the $\widehat{\Lambda}_{j}$-Pauli eigenbasis.
As shown in Theorem \ref{theorem:MUBs_in_Pauli_subspace} and Corollary \ref{coro::MUBs_for_Mmd_Nph}, complementary measurements in $N$-photon LON systems can be configured in the eigenspace of operators $\widehat{L}$ selected from $\{\widehat{\Xi},\widehat{\Lambda}_{0},...,\widehat{\Lambda}_{M-1}\}$.
Let $\mathbb{L}\subseteq\{\widehat{\Xi},\widehat{\Lambda}_{0},...,\widehat{\Lambda}_{M-1}\}$ be a set of selected complementary Pauli operators in $N$-photon LON systems.\hiddengls{MsmntConfig}
In an $\widehat{L}$-Pauli measurement configuration with $\widehat{L}\in\mathbb{L}$, one can quantify a certain physical property of a quantum state by a corresponding assessment function $\mathcal{Q}$ that maps the $M$-dimensional $\widehat{L}$-Pauli measurement statistics $\{\braket{\plProj{N}{m}{L}}\}_{m}$ to a real-value quantity,
\begin{equation}\hiddengls{PauliQauntity}
  \plQ{L}:\Re^{M} \rightarrow \Re,
  \;\;
  \plQ{L}(\rho) := \plQ{}[\braket{\widehat{\pi}_{N,0}(L)},...,\braket{\plProj{N}{M-1}{L}}].
\end{equation}
We call such a quantity an \emph{$\widehat{L}$-Pauli quantity}.
The average of a Pauli quantity $\plQ{L}$ obtained in the complementary Pauli measurements configured by the set $\mathbb{L}$ can be exploited to quantify complementary properties of quantum states.
We call such a quantity a \emph{complementary Pauli quantity} in the measurement configurations $\mathbb{L}$ and define it as follows.
\begin{definition}[Complementary Pauli quantity]
  Let $\mathbb{L} \subseteq \{\widehat{\Xi}, \widehat{\Lambda}_{0}, ..., \widehat{\Lambda}_{M-1}\}$ be a set of complementary Pauli operators in $N$-photon LON systems.
  A \emph{complementary Pauli quantity} $\cmplQ{\plQ{}}{\mathbb{L}}$ is the average of Pauli quantities $\plQ{L}$ evaluated in the measurement configurations  $\widehat{L}\in\mathbb{L}$,
  \begin{equation}\hiddengls{ComplQuantity}
    \cmplQ{\plQ{}}{\mathbb{L}}(\rho) :=
    \frac{1}{|\mathbb{L}|} \sum_{\widehat{L}\in\mathbb{L}}
    \plQ{}[\braket{\widehat{\pi}_{N,0}(L)},...,\braket{\plProj{N}{m}{L}}],
  \end{equation}
  where $|\mathbb{L}|$ denotes the cardinality of the operator set $\mathbb{L}$.
\end{definition}

Since all Pauli projectors $\widehat{\pi}_{N,m}(\cdot)$ are block-diagonal with respect to Pauli subspaces by definition (see Eq. \eqref{eq::Xi_op} and \eqref{eq::Pauli_proj_def}), they are invariant under the projections onto Pauli subspaces,
\begin{equation}
\label{eq::diag_Pauli_proj}
  \sum_{\mathbb{E}_{\boldvec{n}}}\widehat{\Pi}_{\mathbb{E}_{\boldvec{n}}}\,
  \widehat{\pi}_{N,m}(\cdot)\,
  \widehat{\Pi}_{\mathbb{E}_{\boldvec{n}}}
  =
  \widehat{\pi}_{N,m}(\cdot)
  \;\;\text{ with }\;\;
  \widehat{\Pi}_{\mathbb{E}_{\boldvec{n}}} =
  \sum_{\boldvec{\nu}\in\mathbb{E}_{\boldvec{n}}}\projector{\boldvec{\nu}}.
\end{equation}
For any operator $\widehat{L}\in\{\widehat{\Xi},\widehat{\Lambda}_{0},...,\widehat{\Lambda}_{M-1}\}$, its
corresponding $\widehat{L}$-Pauli measurement statistics $\{\braket{\plProj{N}{m}{L}}\}_{m}$ of a quantum state $\widehat{\rho}$ is therefore invariant under the decoherence $D(\rho)$ among Pauli subspaces, where the Pauli-subspace decoherence $D(\rho)$ of a state is defined as
\begin{equation}\hiddengls{PauliDecoherence}
  D(\rho) :=
  \sum_{\mathbb{E}_{\boldvec{n}}}
  \widehat{\Pi}_{\mathbb{E}_{\boldvec{n}}}
  \;\widehat{\rho}\;
  \widehat{\Pi}_{\mathbb{E}_{\boldvec{n}}}.
\end{equation}%
As a consequence, Pauli quantities that evaluated from Pauli measurement statistics $\{\braket{\plProj{N}{m}{\cdot}}\}_{m}$ are also invariant under this Pauli-subspace decoherence.
\begin{corollary}[Pauli-subspace decoherence invariance]
\label{coro::Pauli_subspace_decoherence}
  A Pauli quantity and its corresponding complementary Pauli quantity are invariant under the Pauli-subspace decoherence,
  \begin{equation}
    \plQ{L}(\rho) = \plQ{L}(D(\rho))
    \;\;\text{ and }\;\;
    \cmplQ{\mathcal{Q}}{\mathbb{L}}(\rho) = \cmplQ{\mathcal{Q}}{\mathbb{L}}(D(\rho)).
  \end{equation}
\end{corollary}

Let $\Psi_{\mathcal{S}}$ be a set of all the pure states that possess certain property $\mathcal{S}$.
If the convex combination of two states $\ket{\psi_{1,2}}\in\Psi_{\mathcal{S}}$ also possesses the property $\mathcal{S}$, then we say the property $\mathcal{S}$ is \emph{convex-extendible}, e.g. separability, entanglement dimensionality, and so on.\hiddengls{ConvexProp}
It is clear that the set of all quantum states with a convex-extendible property is convex.
In qudit systems, hyperplanes that separate the convex set of $\mathcal{S}$-property quantum states $\widehat{\rho}_{\mathcal{S}}$ from some non-$\mathcal{S}$-property quantum states $\widehat{\rho}_{\overline{\mathcal{S}}}$ can be exploited to characterize the property $\mathcal{S}$.
If the corresponding quantity of a quantum state $\widehat{\rho}$ exceeds the bounds of the hyperplanes tangent to the $\mathcal{S}$-property convex set, one can conclude the
non-$\mathcal{S}$ property of $\widehat{\rho}$.
If these hyperplanes are defined by a quantity which can be measured in experiments, the property $\overline{\mathcal{S}}$  complement to $\mathcal{S}$ is then physically detectible.

In LON systems, quantum states can be characterized in hyperplanes $\{\widehat{\rho}: \cmplQ{\plQ{}}{\mathbb{L}}(\widehat{\rho})=q\}$ defined by a complimentary Pauli quantity $\cmplQ{\plQ{}}{\mathbb{L}}$.
Since a complementary Pauli quantity is physically accessible by definition, it provides the physical significance of the property $\overline{\mathcal{S}}$ complement to a convex-extendible property $\mathcal{S}$.
As a Pauli subspace $\mathbb{H}_{\mathbb{E}_{\boldvec{n}}}$ is a well-defined $\dimE{\boldvec{n}}$-dimensional qudit system, the $\cmplQ{\plQ{}}{\mathbb{L}}$-hyperplane boundaries on a $\mathcal{S}$-property set within the Pauli subspace $\mathbb{H}_{\mathbb{E}_{\boldvec{n}}}$ can be determined by well-established theories in qudit systems, \hiddengls{BoundCmplQ}
\begin{align}
\label{eq::Bs_subspace}
  \mathcal{B}_{\mathcal{S}}^{(\text{max})}(\mathbb{E}_{\boldvec{n}})
  := &
  \max_{\ket{\psi}
  \in\Psi_{\mathcal{S}}\cap\mathbb{H}_{\mathbb{E}_{\boldvec{n}}}} \cmplQ{\mathcal{Q}}{\mathbb{L}}(\ket{\psi}),
  \nonumber\\
  \mathcal{B}_{\mathcal{S}}^{(\text{min})}(\mathbb{E}_{\boldvec{n}})
  := &
  \min_{\ket{\psi}
  \in\Psi_{\mathcal{S}}\cap\mathbb{H}_{\mathbb{E}_{\boldvec{n}}}} \cmplQ{\mathcal{Q}}{\mathbb{L}}(\ket{\psi}).
\end{align}
According to Corollary \ref{coro::Pauli_subspace_decoherence}, a complementary Pauli quantity $\cmplQ{\plQ{}}{\mathbb{L}}(\ket{\psi})$ of a pure state $\ket{\psi}$ in LONs is given by its Pauli-subspace decoherence, which is a convex combination of pure states $\ket{\psi_{\mathbb{E}}}$ over Pauli subspaces $\mathbb{H}_{\mathbb{E}}$,
\begin{equation}
  D(\ket{\psi})
  =
  \sum_{\mathbb{E}}p_{\mathbb{E}} \projector{\psi_{\mathbb{E}}}.
\end{equation}
In the case that $\cmplQ{\plQ{}}{\mathbb{L}}$ is convex or concave, the $\cmplQ{\plQ{}}{\mathbb{L}}$-hyperplane boundaries on the $\mathcal{S}$-property set in $N$-photon LONs can be then extended from the bounds determined in Eq. \eqref{eq::Bs_subspace} through a convex-roof extension over all Pauli subspaces.
\begin{theorem}[Complementary Pauli quantities of a convex set]
\label{theorem:convex_prop_witness}
  Let $\mathcal{S}$ be a convex-extendible property,
  and $\Psi_{\mathcal{S}}:=\{\ket{\psi}:\ket{\psi}\text{ is }\mathcal{S}\}$ be the set of all pure states with the property $\mathcal{S}$.
  If $\cmplQ{\mathcal{Q}}{\mathbb{L}}$ is a convex (concave) complementary Pauli quantity, then   $\cmplQ{\mathcal{Q}}{\mathbb{L}}$ of an state $\widehat{\rho}_{\mathcal{S}}$ with the property $\mathcal{S}$ is bounded by
  \begin{align}
  \label{eq::cnvx_rf_bnd_cmplQ}
    \cmplQ{\mathcal{Q}}{\mathbb{L}}(\widehat{\rho}_{\mathcal{S}})
    \le
    \sum_{\mathbb{E}}p_{\mathbb{E}}(\rho_{\mathcal{S}})\mathcal{B}_{\mathcal{S}}^{(\mathrm{max})}(\mathbb{E}),
    &
    \;\;\text{for convex $\cmplQ{\mathcal{Q}}{\mathbb{L}}$,}
    \nonumber \\
    \cmplQ{\mathcal{Q}}{\mathbb{L}}(\widehat{\rho}_{\mathcal{S}})
    \ge
    \sum_{\mathbb{E}}p_{\mathbb{E}}(\rho_{\mathcal{S}})\mathcal{B}_{\mathcal{S}}^{(\mathrm{min})}(\mathbb{E}),
    &
    \;\;\text{for concave $\cmplQ{\mathcal{Q}}{\mathbb{L}}$,}
  \end{align}
  where $p_{\mathbb{E}}(\rho) = \sum_{\ket{\boldvec{n}}\in\mathbb{E}}\braket{\boldvec{n} |\widehat{\rho}|\boldvec{n}}$ is the probability of measurement outcomes belonging to a Pauli subspace $\mathbb{H}_{\mathbb{E}}$ in the computational basis.
  If a state $\widehat{\rho}$ violates these inequalities, then the state $\widehat{\rho}$ does not possess the property $\mathcal{S}$.
\par
\begin{proof}
  According to Corollary \ref{coro::Pauli_subspace_decoherence}, a Pauli quantity is invariant under the decoherence among Pauli subspaces, i.e. $\cmplQ{\mathcal{Q}}{\mathbb{L}}(\rho_{\mathcal{S}}) = \cmplQ{\mathcal{Q}}{\mathbb{L}}\left( \sum_{\mathbb{E}}p_{\mathbb{E}}\widehat{R}_{\mathbb{E}}(\rho_{\mathcal{S}})\right)$, where $\widehat{R}_{\mathbb{E}}(\rho_{\mathcal{S}}) = \widehat{\Pi}_{\mathbb{E}}\widehat{\rho}\widehat{\Pi}_{\mathbb{E}}/p_{\mathbb{E}}$ is the state projected onto the subspace $\mathbb{H}_{\mathbb{E}}$.
  If $\cmplQ{\mathcal{Q}}{\mathbb{L}}$ is convex or concave, it holds then $\cmplQ{\mathcal{Q}}{\mathbb{L}}(\rho_{\mathcal{S}}) \le \sum_{\mathbb{E}} p_{\mathbb{E}} \cmplQ{\mathcal{Q}}{\mathbb{L}}\left(R_{\mathbb{E}}(\rho_{\mathcal{S}})\right)$ or $\cmplQ{\mathcal{Q}}{\mathbb{L}}(\rho_{\mathcal{S}}) \ge \sum_{\mathbb{E}} p_{\mathbb{E}} \cmplQ{\mathcal{Q}}{\mathbb{L}}\left(R_{\mathbb{E}}(\rho_{\mathcal{S}})\right)$.
  The upper (lower) bound $\mathcal{B}_{\mathcal{S}}(\mathbb{E})$ on $\cmplQ{\mathcal{Q}}{\mathbb{L}}(R_{\mathbb{E}}(\rho_{\mathcal{S}}))$ is then determined by the maximum (minimum) $\cmplQ{\mathcal{Q}}{\mathbb{L}}$ for the $\mathcal{S}$-property pure states in the Pauli subspace $\mathbb{H}_{\mathbb{E}}$, which is defined in Eq. \eqref{eq::Bs_subspace}.
  As a result, Eq. \eqref{eq::cnvx_rf_bnd_cmplQ} follows.
\end{proof}
\end{theorem}%
\noindent This theorem allows us to extend well-established hyperplane boundaries on a convex set in qudit system to multiphoton LON systems through convex-roof extension over Pauli subspaces.
Since the weight $p_{\mathbb{E}}(\rho)$ of a state $\widehat{\rho}$ in a Pauli-subspace $\mathbb{H}_{\mathbb{E}}$ can be measured in the computational basis, the boundaries given in Theorem \ref{theorem:convex_prop_witness} can be determined adapted to input states.
As a result, one can reveal the physical significance of the complement of a convex-extendible property in multiphoton LONs by detecting a complementary Pauli quantity exceeding the bounds determined in Theorem \ref{theorem:convex_prop_witness} in a set of complementary measurements.
Since a complementary Pauli quantity takes an average over complementary measurements, the hyperplanes defined by $\cmplQ{\mathcal{Q}}{\mathbb{L}}$ becomes finer, if more Pauli measurements are included in the complementary measurement configurations $\mathbb{L}$, which means more non-$\mathcal{S}$-property states can be detected.

\bigskip

As an example, the Shannon entropy $\mathcal{H}$ is a concave quantity, which can be exploited to quantify randomness of measurement statistics in qudit systems.
Since the property of being a quantum state is by definition convex-extendible, the whole set of quantum states is a convex set.
There exists therefore a lower bound $\mathcal{B}_{\text{quan.}}$ on the average of Shannon entropies of complementary measurement statistics, which implies the uncertainty relation of complementary measurements in qudit systems\cite{MaassenUffink1988-EntUnctRel, Sanchez-Ruiz1995-BdEntrUncCmplObsv, WuYuMolmer2009-EntUnRelMUB, WehnerWinter2010-EntrUncRel, Rastegin2017-UnRelQCohMUBs}.
Such an uncertainty relationship can be extended to multiphoton LON systems according to Theorem \ref{theorem:convex_prop_witness}.
\begin{corollary}[Uncertainty relations in LONs]
\label{coro::un_rel}
  Let $\widehat{\rho}_{N}$ be an $N$-photon state in an $M$-mode LON system.
  \begin{enumerate}
    \item If $\gcd(N,M)\neq 1$, one can construct a pair of complementary Pauli operators $\mathbb{L}=\{\widehat{\Xi},\widehat{\Lambda}_{j}\}$ with $j=0,...,M-1$.
        The lower bound on the corresponding complementary Shannon entropy of $\widehat{\rho}_{N}$ is
        \begin{equation}
        \label{eq::CSE_1}
          \cmplQ{\mathcal{H}}{\mathbb{L}}(\rho_{N})
          \ge
          \frac{1}{2}\sum_{\mathbb{E}_{\boldvec{n}}} p_{\mathbb{E}_{\boldvec{n}}}(\rho_{N}) \log\dimE{\boldvec{n}}.
      \end{equation}%
    \item If $\gcd(N,M)=1$ and $\mathbb{L}$ is a set of complementary Pauli operators constructed  according to Corollary \ref{coro::MUBs_for_Mmd_Nph}, the lower bound on the corresponding complementary Shannon entropy is
        \begin{equation}
        \label{eq::CSE_2}
          \cmplQ{\mathcal{H}}{\mathbb{L}}(\rho_{N})
          \ge
          \left\{
            \begin{array}{ll}
              \frac{1}{2}\log(M), & |\mathbb{L}|\le\sqrt{M}+1; \\
              -\log\frac{|\mathbb{L}|+M-1}{|\mathbb{L}|M}, & |\mathbb{L}|>\sqrt{M}+1.
            \end{array}
          \right.
        \end{equation}%
  \end{enumerate}
\begin{proof}
See Appendix.
\end{proof}
\end{corollary}

\begin{figure}
  \centering
  \subfloat[]{\includegraphics[width=0.33\textwidth]{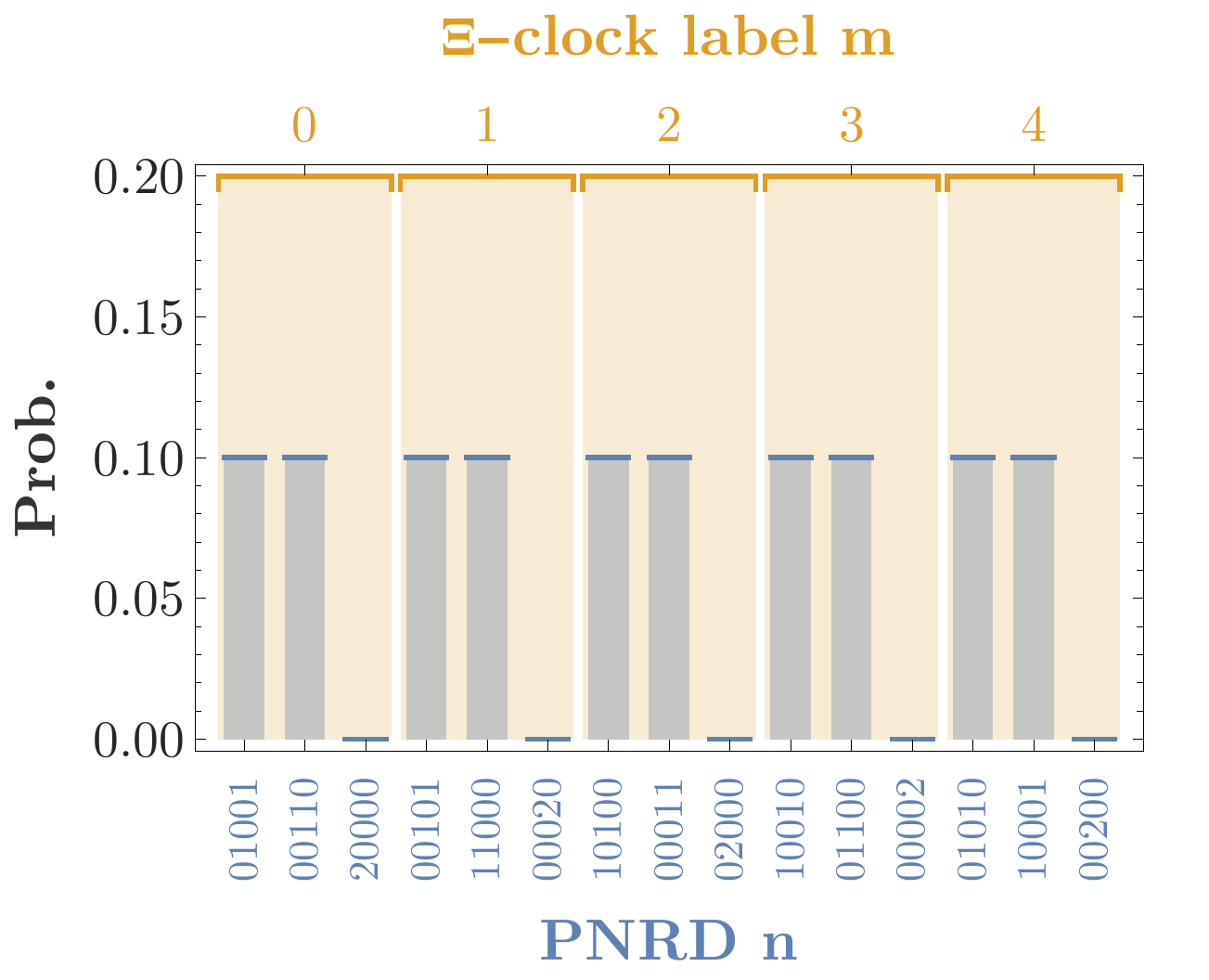}}
  \subfloat[]{\includegraphics[width=0.33\textwidth]{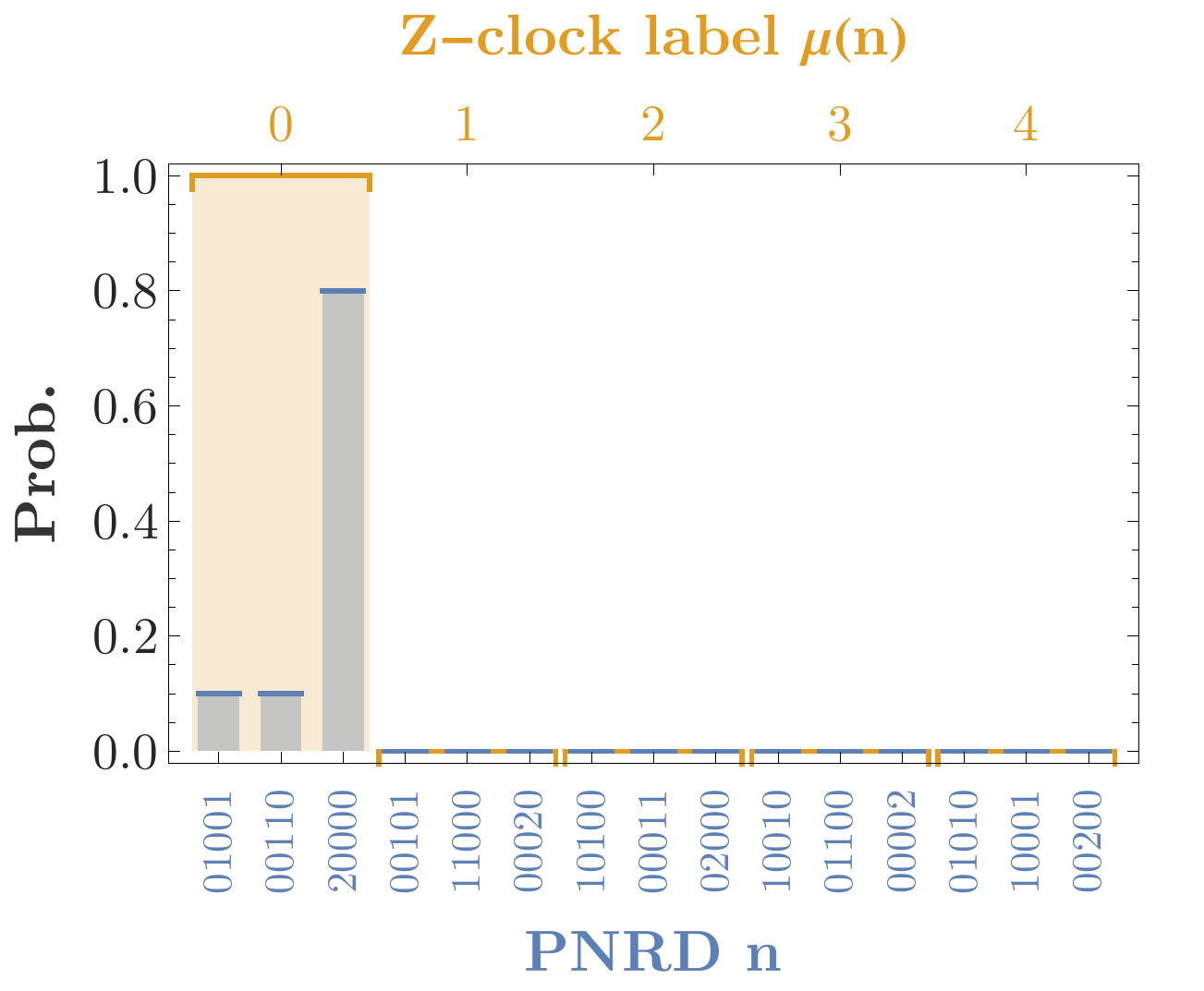}}
  \subfloat[]{\includegraphics[width=0.33\textwidth]{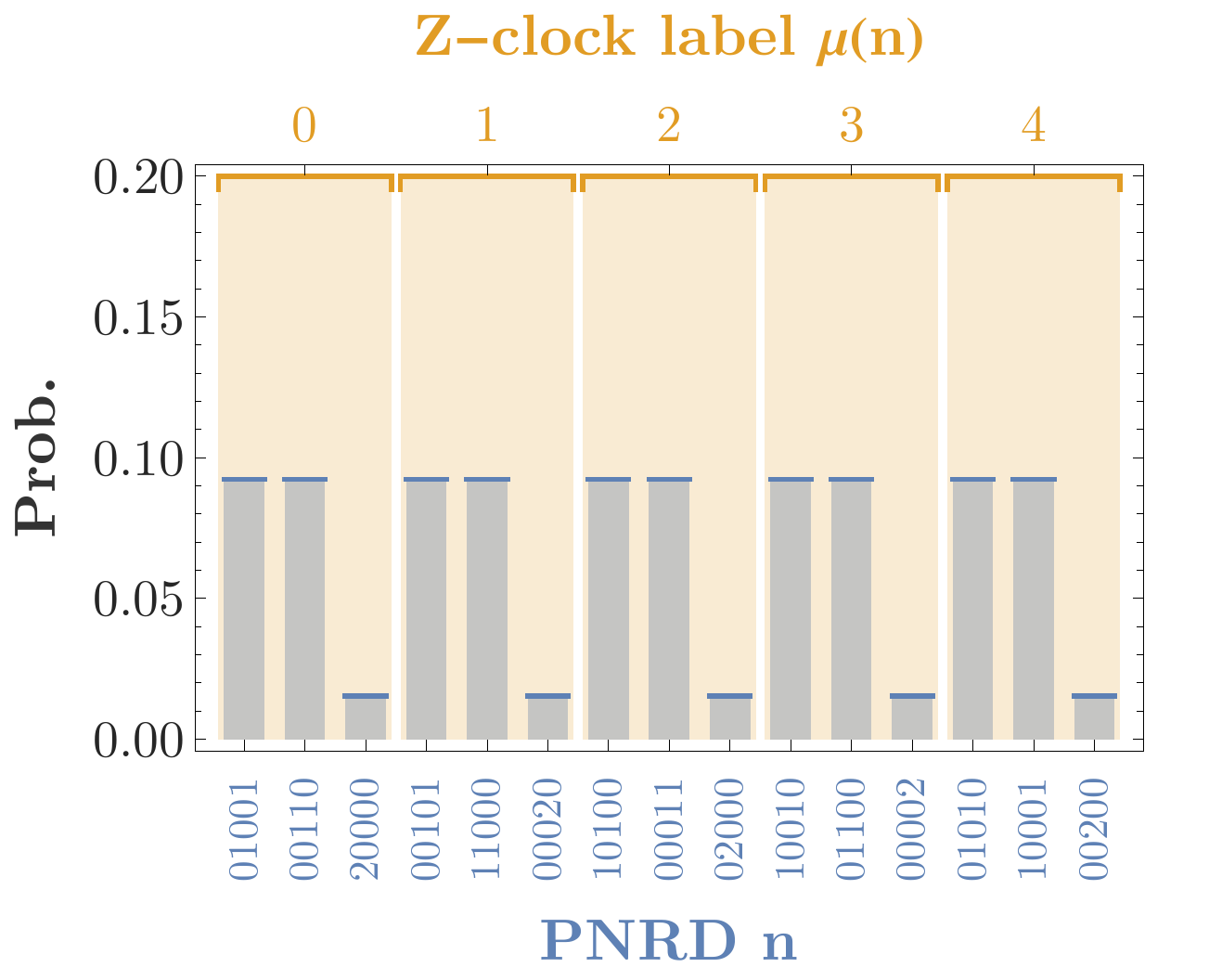}}
  \\
  \subfloat[]{\includegraphics[width=0.33\textwidth]{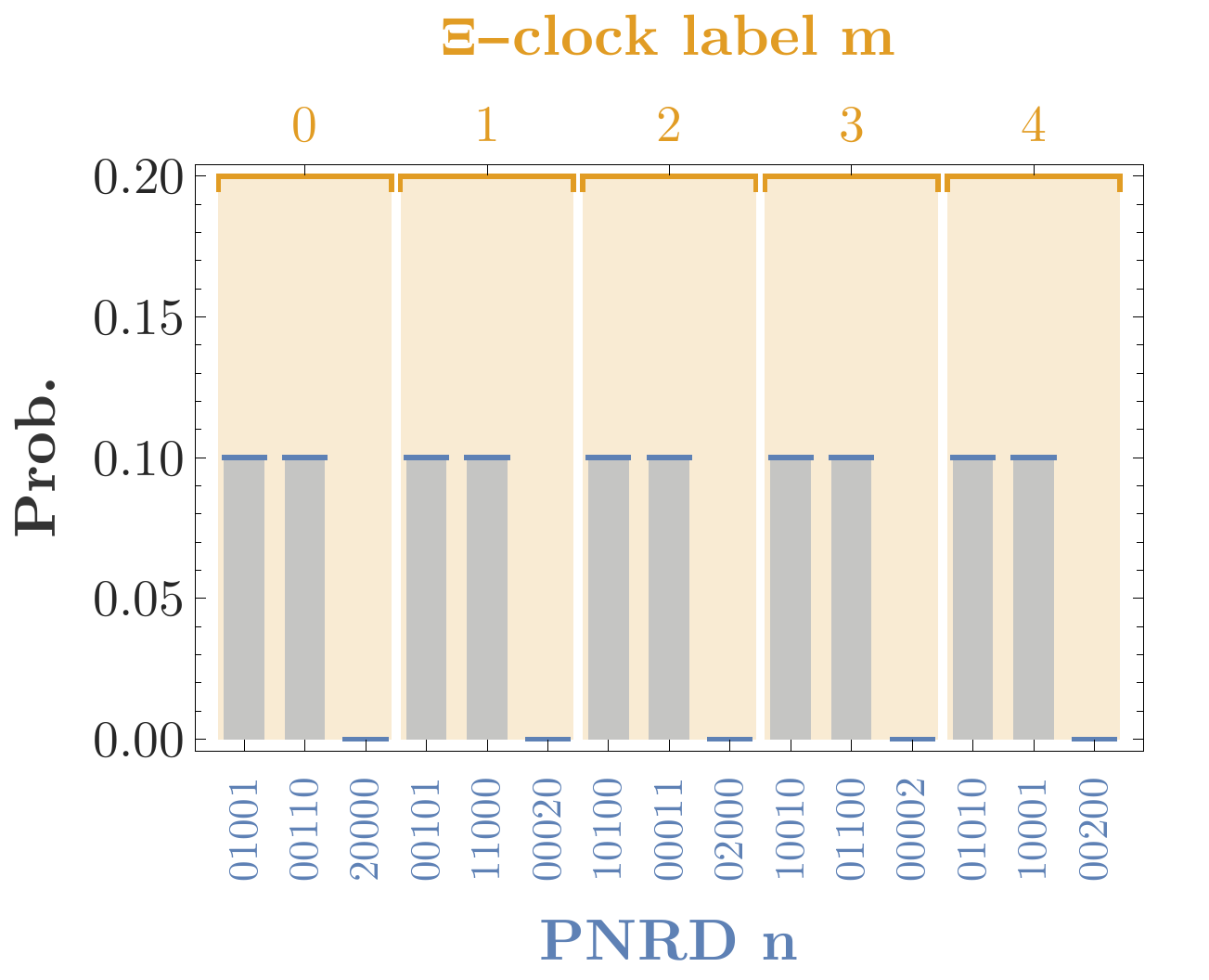}}
  \subfloat[]{\includegraphics[width=0.33\textwidth]{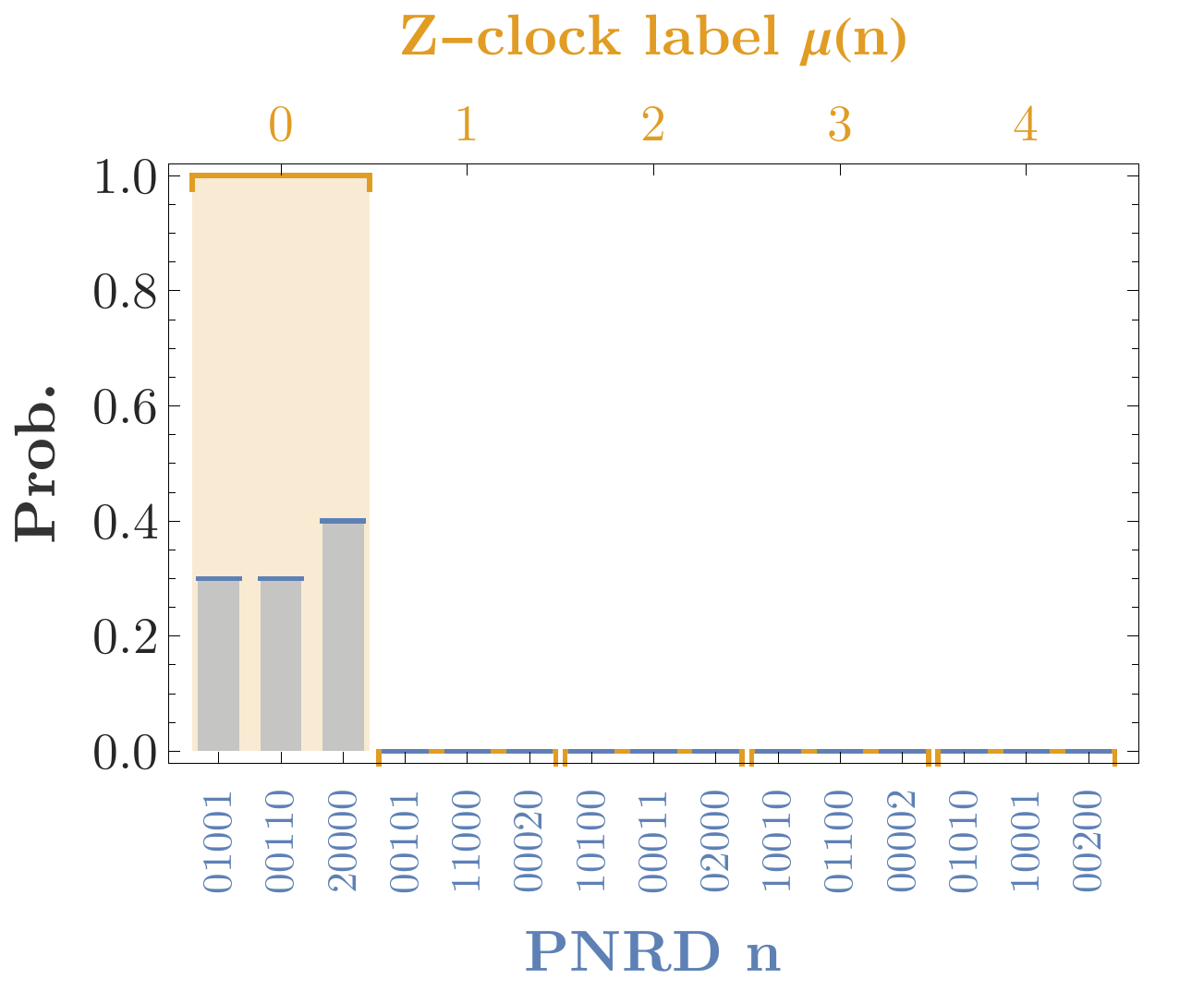}}
  \subfloat[]{\includegraphics[width=0.33\textwidth]{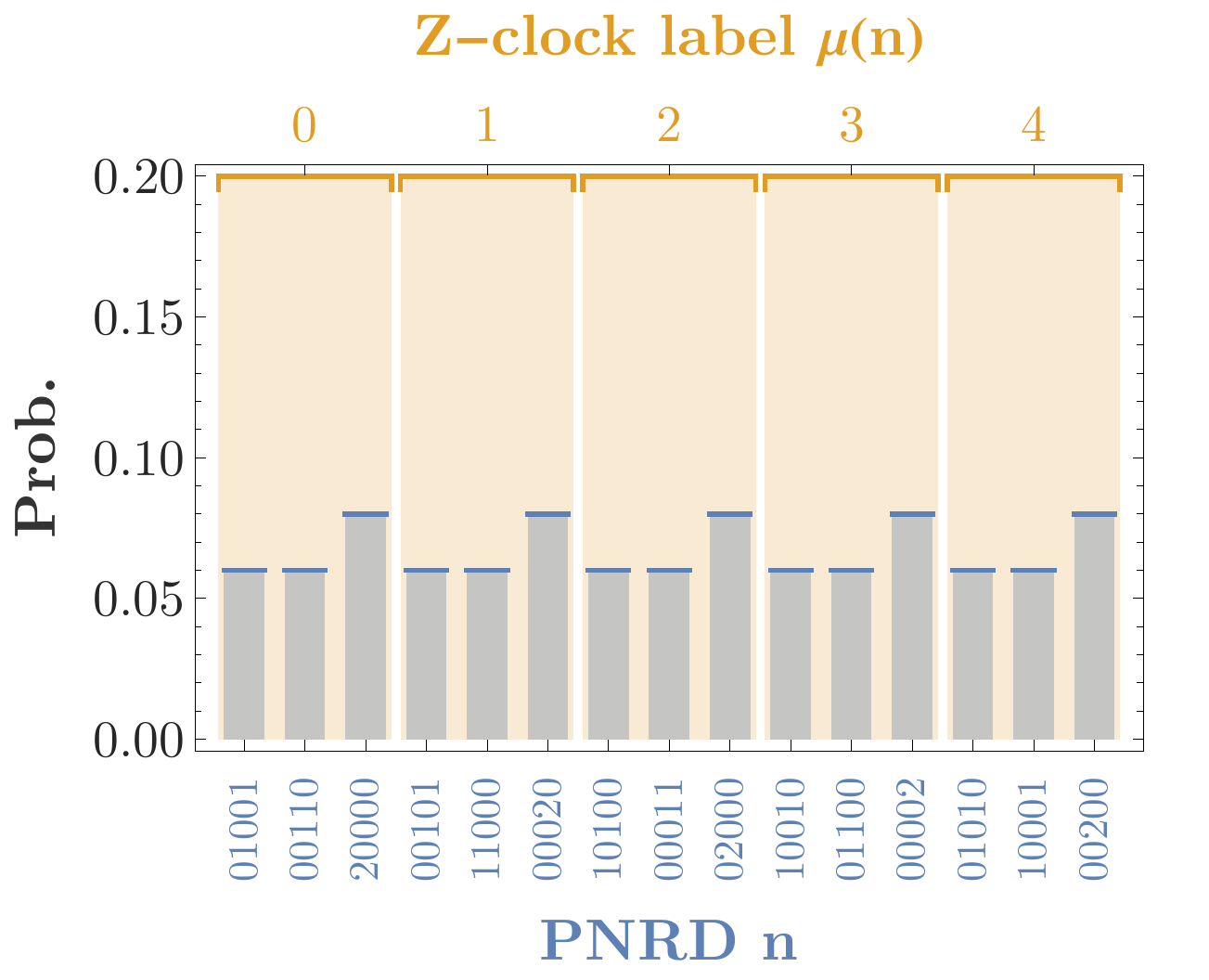}}
  \caption{Pauli measurements of a $\widehat{\Lambda}_{0}$ eigenstate $\ket{\psi_{0}}$ given in Eq. \eqref{eq::psi_0} and its Pauli-subsapce decoherence $D(\ket{\psi_{0}})$ given in Eq. \eqref{eq::psi_0_Decoh}. The Pauli measurement statistics $\{\braket{\widehat{\pi}_{N,m}(\cdot)}\}_{m}$ are plotted in orange.
    (a) The $\widehat{\Xi}$-Pauli measurement of $\ket{\psi_{0}}$.
    (b) The $\widehat{\Lambda}_{0}$-Pauli measurement of $\ket{\psi_{0}}$.
    (c) The $\widehat{\Lambda}_{j}$-Pauli measurement of $\ket{\psi_{0}}$ with $j\in\{0,...,4\}$.
    (a) The $\widehat{\Xi}$-Pauli measurement of $\ket{\psi_{0}}$.
    (d) The $\widehat{\Xi}$-Pauli measurement of $D(\ket{\psi_{0}})$.
    (e) The $\widehat{\Lambda}_{0}$-Pauli measurement of $D(\ket{\psi_{0}})$.
    (f) The $\widehat{\Lambda}_{j}$-Pauli measurement of $D(\ket{\psi_{0}})$ with $j\in\{0,...,4\}$.
  }
  \label{fig::Pauli_msmnt_UnRel}
\end{figure}

The lower bound determined in Corollary \ref{coro::un_rel} is tight for particular states, e.g. for a $\Lambda_{0}$ eigenstate $\ket{\psi_{0}}$ given by
\begin{equation}
\label{eq::psi_0}
  \ket{\psi_{0}}
  =
  \frac{1}{\sqrt{2}}(\ket{\ESt{11000}{0}{0}}+\ket{\ESt{10100}{0}{0}}).
\end{equation}
The statistics of complementary Pauli measurements of this state is shown in Fig. \ref{fig::Pauli_msmnt_UnRel} (a)-(c). Its complementary Shannon entropy $\cmplQ{\mathcal{H}}{\mathbb{L}}$ for $\mathbb{L}=\{\widehat{\Xi}, \widehat{\Lambda}_{j}\}$ is $\cmplQ{\mathcal{H}}{\mathbb{L}}(\ket{\psi_{0}}) = \log(5)/2$, which reaches the lower bound given in Eq. \eqref{eq::CSE_1}.
If one measures $\ket{\psi_{0}}$ in all complementary Pauli measurements $\mathbb{L}=\{\widehat{\Xi}, \widehat{\Lambda}_{0}, ..., \widehat{\Lambda}_{4}\}$, its complementary Shannon entropy is then $\cmplQ{\mathcal{H}}{\mathbb{L}}(\ket{\psi_{0}}) = 5\log(5)/6$, which is greater than the lower bound $\log(3)$ determined by Eq. \eqref{eq::CSE_2}.
According to Corollary \ref{coro::Pauli_subspace_decoherence}, the Pauli measurement statistics of the state $\ket{\psi_{0}}$ is invariant under the Pauli-subspace decoherence,
\begin{equation}
\label{eq::psi_0_Decoh}
  D(\ket{\psi_{0}})
  =
  \frac{1}{2}\left(\projector{\ESt{11000}{0}{0}} + \projector{\ESt{10100}{0}{0}}\right).
\end{equation}
The Pauli measurement statistics of $D(\ket{\psi_{0}})$ is given in Fig. \ref{fig::Pauli_msmnt_UnRel} (d)-(f).
Compare the probability distributions in (a-c) and (d-f), one can see that although the explicit photon number statistics $\mathrm{Pr}(\boldvec{n})$ of $\ket{\psi_{0}}$ and $D(\ket{\psi_{0}})$ differ from each other, their collective probability $\mathrm{Pr}(\mu)$ is invariant under the Pauli-subspace decoherence.
As a result, the complementary Shannon entropy $\cmplQ{\mathcal{H}}{\mathbb{L}}(D(\ket{\psi_{0}}))$ is equal to $\cmplQ{\mathcal{H}}{\mathbb{L}}(\ket{\psi_{0}})$.

\section{Complementary correlations of entanglement in bipartite LONs}
\label{sec::cmpl_corr_ent}

In multipartite qudit systems, complementary correlations have been widely employed to characterize  separability and entanglement dimensionality in theory and experiments \cite{SpenglerHuberEtAlHiesmayr2012-EntWitViaMUB, MacconeBrussMacchiavello2015-CmplCrr, HuangEtAlPeruzzo2016-HghDimEntCert, SauerweinAtElKraus2017-MltptCrrMUBs, ErkerKrennHuber2017-QtfyHghDmEnt2MUBs,BavarescoEtAlHuber2018-2MUBsCrtfyHghDmEnt}.
As a straightforward application of Theorem \ref{theorem:convex_prop_witness}, one can extend the entanglement criteria that employ complementary correlations in bipartite qudit systems to bipartite multiphoton LON systems, so that we can evaluate entanglement between modes of multiphoton states in LONs theoretically and reveal its physical significance experimentally.

In a bipartite qudit system $\mathbb{H}_{A}\otimes\mathbb{H}_{B}$, one can construct complementary operators by two separable Pauli operators $\widehat{\alpha}_{1}\otimes\widehat{\beta}_{1}$ and $\widehat{\alpha}_{2}\otimes\widehat{\beta}_{2}$, where $\{\widehat{\alpha}_{1},\widehat{\alpha}_{2}\}$ and $\{\widehat{\beta}_{1},\widehat{\beta}_{2}\}$ are complementary Pauli operators in each local systems $A$ and $B$, respectively.
A maximally entangled state can be perfectly correlated both in the $\widehat{\alpha}_{1}\otimes\widehat{\beta}_{1}$-Pauli and $\widehat{\alpha}_{2}\otimes\widehat{\beta}_{2}$-Pauli measurements at the same time.
In each local $\widehat{\alpha}_{l}\otimes\widehat{\beta}_{l}$-Pauli measurements, correlations can be evaluated by certain correlation measures, e.g. mutual information,  mutual predictability, Pearson correlation coefficient and so on.
The simultaneous correlations in a set of complementary measurements can be evaluated by the average of these correlation measures, which are called complementary correlations.
The upper bounds on complementary correlations for separable states specify the hyperplanes that divide the convex set of separable states from particular entangled states.
It therefore allows us to detect bipartite entanglement by evaluating complementary correlations exceeding these bounds\cite{MacconeBrussMacchiavello2015-CmplCrr, SpenglerHuberEtAlHiesmayr2012-EntWitViaMUB}.

Here, we consider bipartite multiphoton LON systems with the same number of modes $M_{A}=M_{B}=M$.
For multiphoton states with $N_{A}$ and $N_{B}$ photons in each local system, complementary operators can be constructed locally with separable operators \hiddengls{MsmntConfig}
\begin{equation}
\label{eq::sep_Pauli_ops}
  \mathbb{L} = \{\widehat{\alpha}_{1}\otimes\widehat{\beta}_{1}, ..., \widehat{\alpha}_{l}\otimes\widehat{\beta}_{l}, ...\},
\end{equation}
where $\{\widehat{\alpha}_{l}\}_{l}$ and $\{\widehat{\beta}_{l}\}_{l}$ are complementary Pauli operators in the $N_{A}$-photon and $N_{B}$-photon local system, respectively, which are constructed according to Corollary \ref{coro::MUBs_for_Mmd_Nph}.
We call $\mathbb{L}$ a set of \emph{complementary separable Pauli operators} in $(N_{A},N_{B})$-photon $(M,M)$-mode LON systems.
In this section, we will derive experimentally accessible criteria for entanglement between modes in bipartite multiphoton LON systems using complementary mutual information and complementary mutual predictability.

\bigskip

\subsection{Complementary mutual information (CMI)}
\label{sec::CMI}
For an entangled state that has correlations in the Pauli measurements configured by a set of complementary separable Pauli operators $\mathbb{L}$ given in Eq. \eqref{eq::sep_Pauli_ops}, complementary mutual information (CMI) is a good quantity for entanglement detection.\hiddengls{cmi}
It takes the average of the mutual information
in all $\widehat{\alpha}_{l}\otimes\widehat{\beta}_{l}$-Pauli measurements
\begin{equation}\hiddengls{CmplMI}
  \cmplQ{\mathcal{I}}{\mathbb{L}}(\rho):=
  \frac{1}{|\mathbb{L}|}\sum_{\widehat{\alpha}\otimes\widehat{\beta}\in\mathbb{L}}
  \mathcal{I}_{\alpha:\beta}(\rho),
\end{equation}
where $\mathcal{I}_{\alpha:\beta}(\rho)$ is mutual information of the joint probability distribution $\{\Braket{\plProj{N_{A}}{m_{A}}{\alpha}\otimes\plProj{N_{B}}{m_{B}}{\beta}}\}_{m_{A},m_{B}}$ of a bipartite state $\widehat{\rho}$ measured in an $\widehat{\alpha}\otimes\widehat{\beta}$-Pauli measurements. \hiddengls{MI}
One can detect entanglement between modes, if the CMI of a multiphoton state in LONs exceeds the upper bound for separable states, which can be derived from Theorem \ref{theorem:convex_prop_witness} as follows.
\begin{corollary}[Complementary mutual information in LONs]
\label{coro::cmpl_MI_LONs}
Complementary mutual information of separable states is upper bounded as follows
\begin{enumerate}
  \item In the case that $\gcd(N_{A},M)\neq1$ or $\gcd(N_{B},M)\neq1$, one can construct $\mathbb{L}$ with $\widehat{\alpha}_{l}\in\{\widehat{\Xi},\widehat{\Lambda}_{j_{A}}\}$ and $\widehat{\beta}_{l}\in\{\widehat{\Xi},\widehat{\Lambda}_{j_{B}}\}$. The corresponding CMI of separable states is
      \begin{equation}
      \label{eq::CMI_1}
        \cmplQ{\mathcal{I}}{\mathbb{L}}(\rho)
        \underset{\rm{sep.}}{\le}
        \log(M) - \frac{1}{2}\sum_{\mathbb{E}_{A},\mathbb{E}_{B}}
        p_{\mathbb{E}_{A},\mathbb{E}_{B}}(\rho)\log(d_{\mathbb{E}_{A},\mathbb{E}_{B}}),
      \end{equation}
      where $d_{\mathbb{E}_{A}, \mathbb{E}_{B}}:=\min(d_{\mathbb{E}_{A}}, d_{\mathbb{E}_{B}})$ is the minimum dimension of the local Pauli subspace $\mathbb{H}_{\mathbb{E}_{A}}$ and $\mathbb{H}_{\mathbb{E}_{B}}$.
  \item In the case that $\gcd(N_{A},M)=\gcd(N_{B},M)=1$, one can construct $\{\widehat{\alpha}_{l}\}_{l}$ and $\{\widehat{\beta}_{l}\}_{l}$ according to Corollary \ref{coro::MUBs_for_Mmd_Nph}. The corresponding CMI of separable states is
      \begin{equation}
      \label{eq::CMI_2}
        \cmplQ{\mathcal{I}}{\mathbb{L}}(\rho)
        \underset{\rm{sep.}}{\le}
        \left\{
          \begin{array}{ll}
            \frac{1}{2}\log(M), & |\mathbb{L}|\le\sqrt{M}+1; \\
            \log(\frac{|\mathbb{L}|+M-1}{|\mathbb{L}|}), & |\mathbb{L}|>\sqrt{M}+1.
          \end{array}
        \right.
      \end{equation}
\end{enumerate}
\par
\begin{proof}
See Appendix.
\end{proof}
\end{corollary}
\noindent Note that the upper bound $\log(M)/2$ in Eq. \eqref{eq::CMI_2} is tight for a measurement setting with two complementary configurations $|\mathbb{L}|=2$.

\bigskip

\begin{figure}
  \centering
  \subfloat[]{\includegraphics[width=0.4\textwidth]{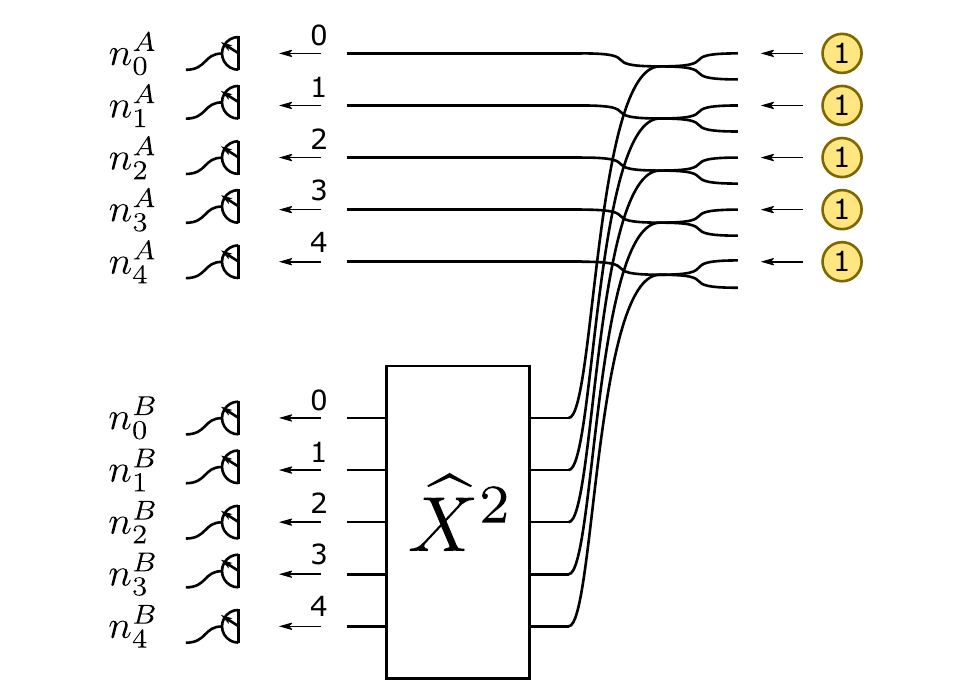}}
  \hspace{0.1\textwidth}
  \subfloat[]{\includegraphics[width=0.4\textwidth]{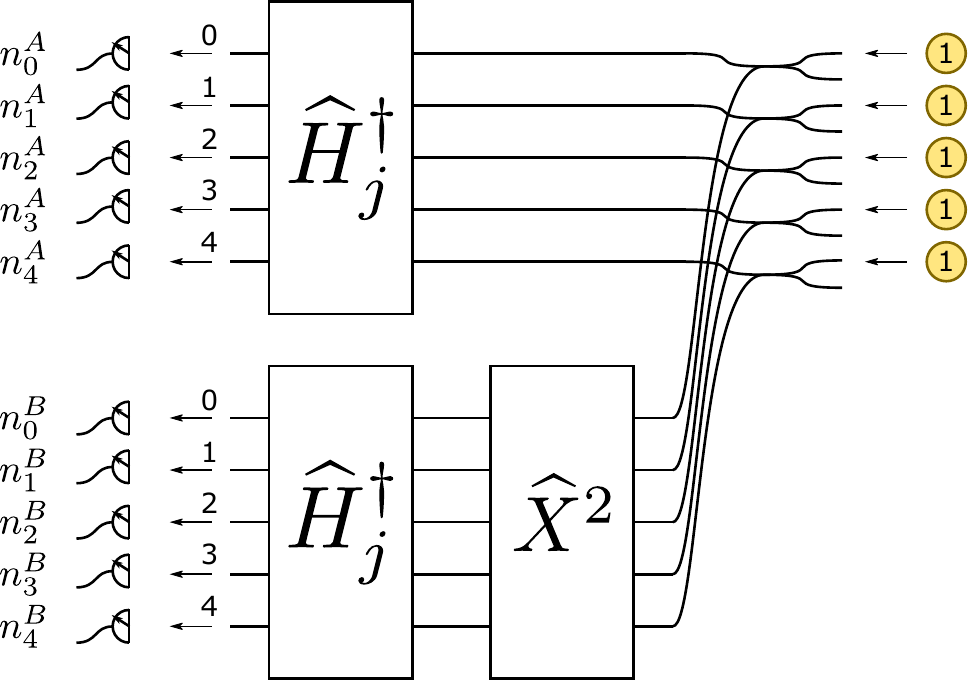}}
  \caption{%
    Generation and measurements of bipartite entangled states in $(N_{A},N_{B})$-photon $(5,5)$-mode LONs according to the approach proposed in \cite{WuHofmann2017-BiEntMltMd}.
    The yellow circles are single-photon sources.
    With post-selection on the outputs with $(N_{A},N_{B}) = (3,2)$, one can obtain the corresponding Pauli measurement statistics of the target state $\ket{\phi_{3_{A},2_{B}}}$ given in Eq. \eqref{eq::eg_phi_3A2B}.
    (a) The measurement in the computational basis.
    (b) The $(\widehat{\Lambda}_{j}\otimes\widehat{\Lambda}_{j})$-Pauli measurement.
  }%
  \label{fig::ent_gen_dect}
\end{figure}

In the following example, we demonstrate this entanglement criterion in a $(5,5)$-mode bipartite LON system.
According to Corollary \ref{coro::MUBs_for_Mmd_Nph}, one can construct complementary separable Pauli operators $\mathbb{L}$ with $\widehat{\alpha}_{l}, \widehat{\beta}_{l}\in\{\widehat{\Xi},\widehat{\Lambda}_{0},...,\widehat{\Lambda}_{M-1}\}$.
Since the $\widehat{Z}$ operator is non-degenerate in all $5$-mode Pauli subspaces, we can construct the $\widehat{\Xi}$ operator as $\widehat{\Xi} = \widehat{Z}$.
For entanglement detection of quantum states that have correlations in $\widehat{Z}\otimes
\widehat{Z}$ and $\widehat{\Lambda}_{j}\otimes\widehat{\Lambda}_{j}$ eigenbases, one can construct measurement configurations $\mathbb{L}$ as follows,
\begin{equation}
\label{eq::cmpl_msmnt_config_phi_3A2B}
  \mathbb{L} \subseteq \{\widehat{Z}\otimes\widehat{Z}, \widehat{\Lambda}_{0}\otimes\widehat{\Lambda}_{0}, ..., \widehat{\Lambda}_{4}\otimes\widehat{\Lambda}_{4}\}.
\end{equation}
An entangled state, which is an eigenstate of every Pauli operator $\widehat{\Lambda}_{j}\otimes\widehat{\Lambda}_{j}$, has perfect correlations in all measurement configurations $\widehat{L}\in\mathbb{L}$.
An example of such entangled states with $(3_{A},2_{B})$ photons can be generated using beam splitters and single photon sources \cite{WuHofmann2017-BiEntMltMd},
\begin{equation}
\label{eq::eg_phi_3A2B}
  \ket{\phi_{3_{A},2_{B}}}
  =
  \frac{1}{\sqrt{10}}
  \sum_{m=0}^{4}\widehat{X}^{m}\otimes\widehat{X}^{m}
  \left(\ket{11100}\ket{11000}+\ket{11010}\ket{01001}\right).
\end{equation}
The state $\ket{\phi_{3_{A},2_{B}}}$ satisfies the following eigenequations
\begin{equation}
\label{eq::stab_phi}
  \widehat{Z}\otimes\widehat{Z} \ket{\phi_{3_{A},2_{B}}}
  =
  w^{-1}\ket{\phi_{3_{A},2_{B}}}
  \;\;\text{ and }\;\;
  \widehat{\Lambda}_{j}\otimes\widehat{\Lambda}_{j} \ket{\phi_{3_{A},2_{B}}}
  =
  w^{-j}\ket{\phi_{3_{A},2_{B}}}.
\end{equation}
According to Theorem \ref{theorem::Pauli_msmnt}, an $(\widehat{\alpha}_{l}\otimes\widehat{\beta}_{l})$-Pauli measurement of $\ket{\phi_{3_{A},2_{B}}}$ has perfect correlations in the $\widehat{Z}$-clock labels of local photon-number-occupation-vector outputs,
\begin{equation}
\label{eq::corr_MI_muAmuB}
  \left\{
    \begin{array}{ll}
      \mu(\boldvec{n}_{A}) + \mu(\boldvec{n}_{B}) = M-1, & \hbox{in $\widehat{Z}\otimes\widehat{Z}$ measurement;} \\
      \mu(\boldvec{n}_{A}) + \mu(\boldvec{n}_{B}) = M-j, & \hbox{in $\widehat{\Lambda}_{j}\otimes\widehat{\Lambda}_{j}$ measurement.}
    \end{array}
  \right.
\end{equation}

\begin{figure}
  \centering
  \subfloat[]{\includegraphics[width=0.5\textwidth]{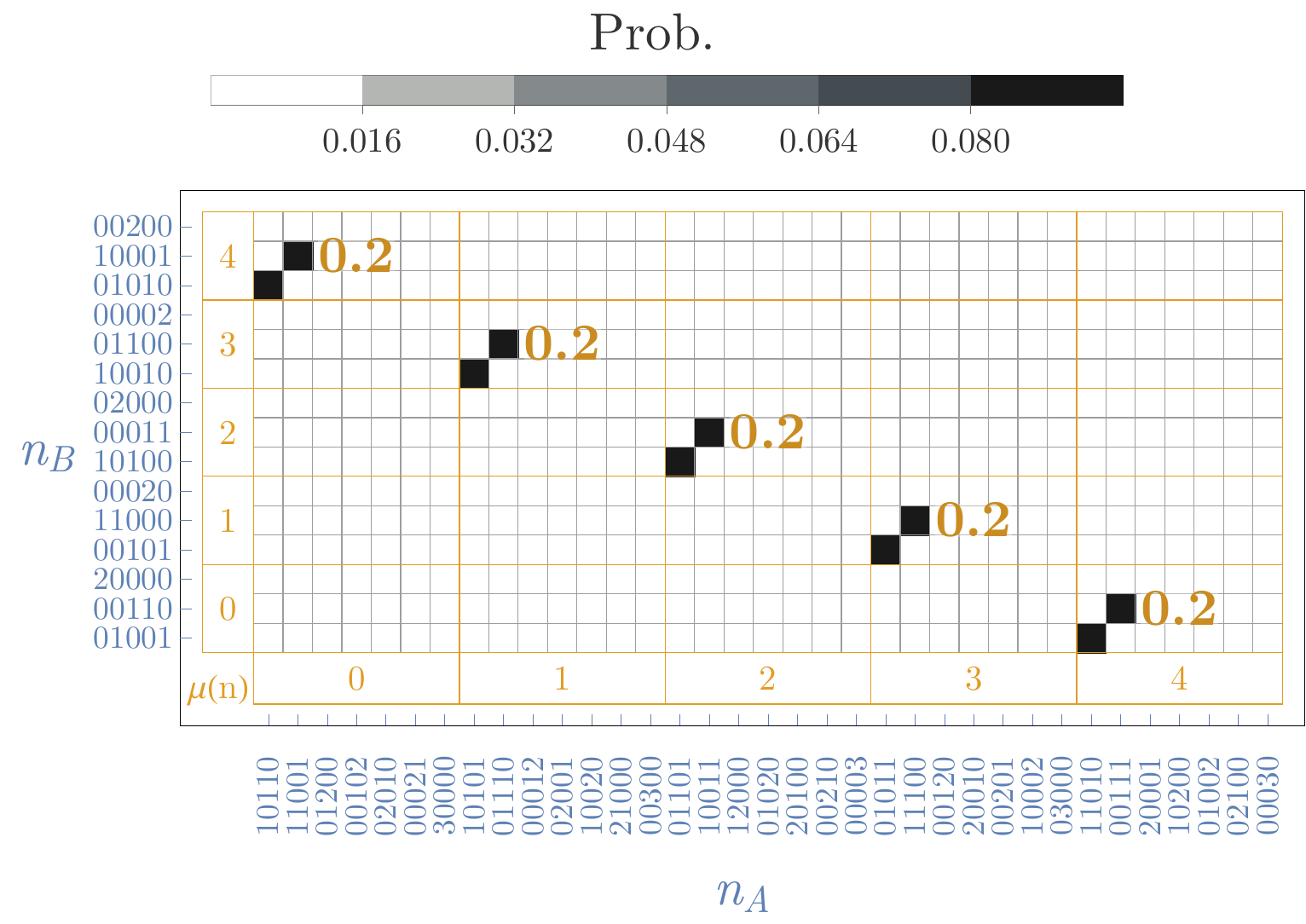}}
  \subfloat[]{\includegraphics[width=0.5\textwidth]{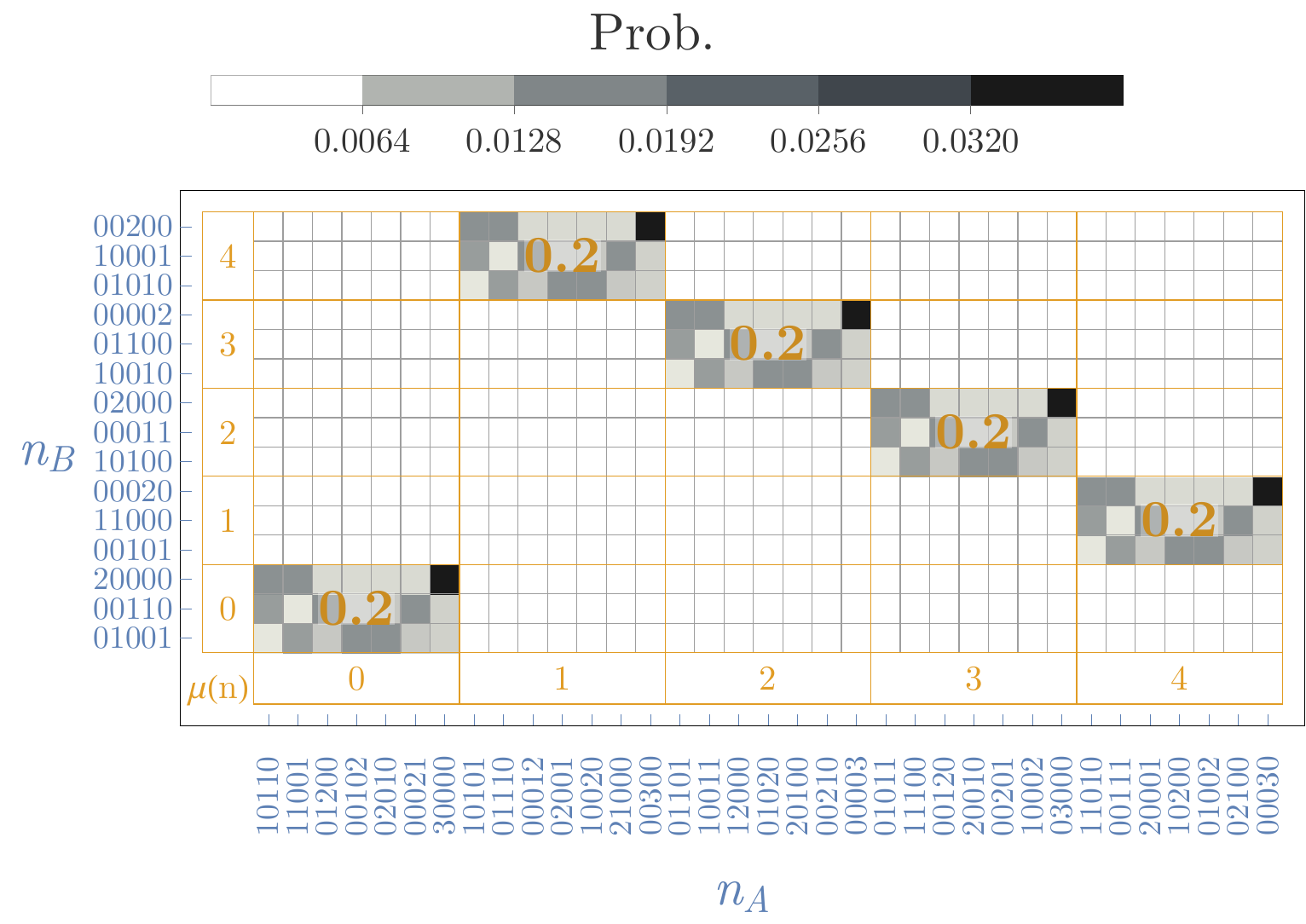}}
  \\
  \subfloat[]{\includegraphics[width=0.25\textwidth]{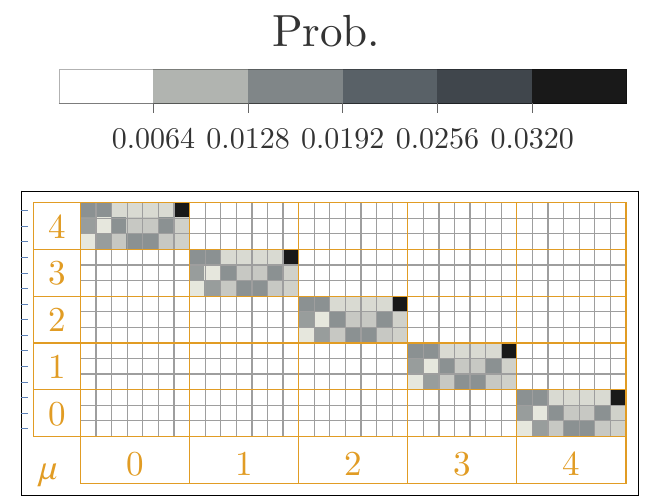}}
  \subfloat[]{\includegraphics[width=0.25\textwidth]{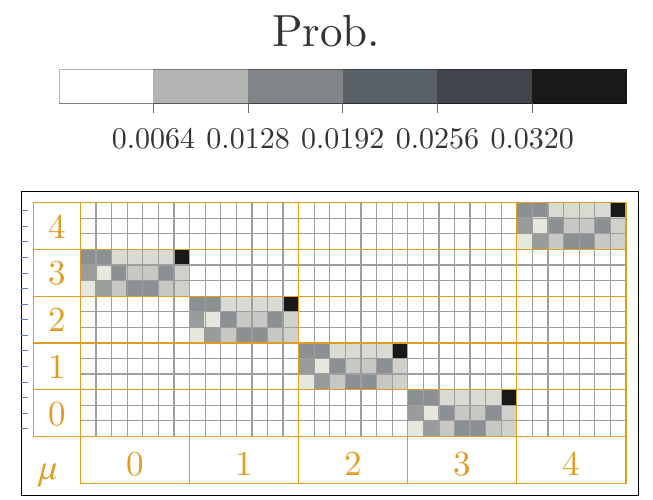}}
  \subfloat[]{\includegraphics[width=0.25\textwidth]{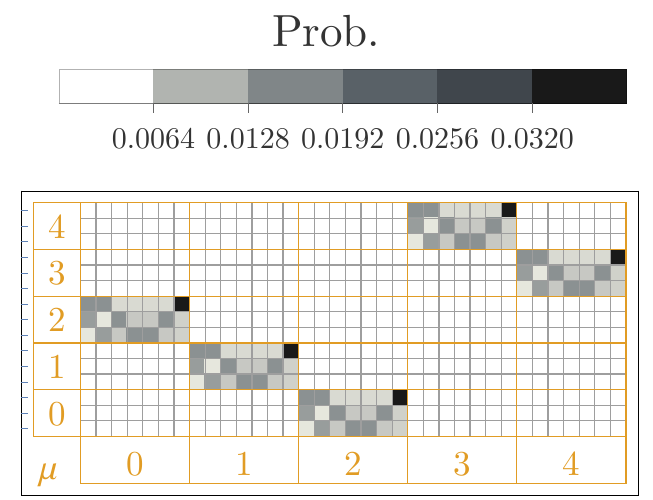}}
  \subfloat[]{\includegraphics[width=0.25\textwidth]{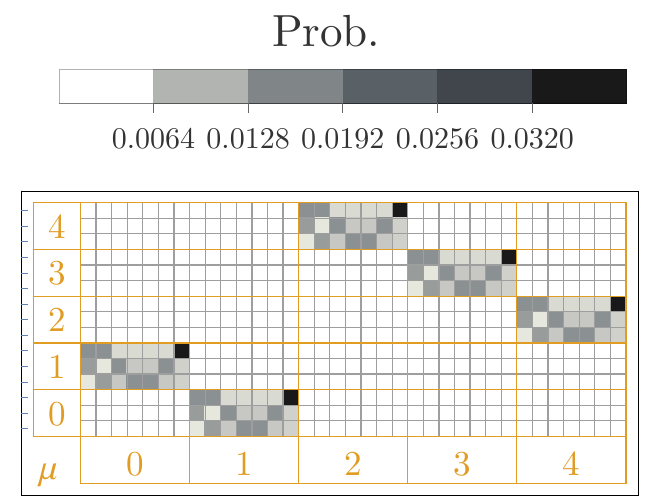}}
  \caption{Complementary Pauli measurements of $\ket{\phi_{3_{A},2_{B}}}$. Local PNRD outputs $\boldvec{n}_{A,B}$ are sorted by their $\widehat{Z}$-clock labels $\mu(\boldvec{n})$ and divided into blocks by orange lines. The collective probabilities $\mathrm{Pr}(\mu_{A},\mu_{B})$ are indicated in each clock-label block.
    (a) The local PNRD measurement in the computational basis.
    (b) The $\widehat{\Lambda}_{0}\otimes\widehat{\Lambda}_{0}$-Pauli measurement.
    (c) The $\widehat{\Lambda}_{1}\otimes\widehat{\Lambda}_{1}$-Pauli measurement.
    (d) The $\widehat{\Lambda}_{2}\otimes\widehat{\Lambda}_{2}$-Pauli measurement.
    (e) The $\widehat{\Lambda}_{3}\otimes\widehat{\Lambda}_{3}$-Pauli measurement.
    (f) The $\widehat{\Lambda}_{4}\otimes\widehat{\Lambda}_{4}$-Pauli measurement.
  }
  \label{fig::WHSt_cmpl_msmnt}
\end{figure}

A schematic experimental implementation of entanglement generation and measurements for $\ket{\phi_{3_{A},2_{B}}}$ following the approach in \cite{WuHofmann2017-BiEntMltMd} is shown in Fig. \ref{fig::ent_gen_dect}, where five single-photon inputs are distributed into two local systems by beam splitters with the modes of the local system $B$ being permuted by $\widehat{X}^{2}$.
One can obtain the corresponding $(\widehat{\alpha}_{l}\otimes\widehat{\beta}_{l})$-Pauli measurement statistics of $\ket{\phi_{3_{A},2_{B}}}$ by post-selection on the local photon numbers $(N_{A},N_{B})= (3,2)$, which is shown in Fig. \ref{fig::WHSt_cmpl_msmnt}.
The local photon number vectors $\boldvec{n}_{A,B}$ are sorted by their $\widehat{Z}$-clock labels $\mu(\boldvec{n})$.
One can see that the measurement outcomes $(\boldvec{n}_{A},\boldvec{n}_{B})$ are perfectly correlated in $(\mu_{A}, \mu_{B})$ blocks in each measurement configuration as given in Eq. \eqref{eq::corr_MI_muAmuB}.
Fig. \ref{fig::WHSt_cmpl_msmnt} (a) is the trivial measurement in the computational basis.
The collective probability of $(\mu_{A},\mu_{B})$ is $\mathrm{Pr}(\mu_{A},\mu_{B}) = 0.2\,\delta_{\mu_{A}}^{4-\mu_{B}}$.
The mutual information in this measurement is therefore $\mathcal{I}_{Z:Z}=\log(5)$.
Fig. \ref{fig::WHSt_cmpl_msmnt} (b) is the $\widehat{\Lambda}_{0}\otimes\widehat{\Lambda}_{0}$-Pauli measurement.
The collective probability of $(\mu_{A},\mu_{B})$ is $\mathrm{Pr}(\mu_{A},\mu_{B}) = 0.2\,\delta_{\mu_{A}}^{-\mu_{B}}$, and hence $\mathcal{I}_{\Lambda_{0}:\Lambda_{0}}=\log(5)$.
Fig. \ref{fig::WHSt_cmpl_msmnt} (c)-(f) show the measurement statistics in the configuration $\Lambda_{1}\otimes\Lambda_{1}, ..., \Lambda_{4}\otimes\Lambda_{4}$, respectively.
The $(\mu_{A},\mu_{B})$ probabilities in each non-zero block are all $0.2$.
In each measurement configuration $\widehat{\alpha}_{l}\otimes\widehat{\beta}_{l}\in\mathbb{L}$ one therefore obtains a mutual information $\mathcal{I}_{\alpha_{l}:\beta_{l}} = \log(5)$.
As a result, the complementary mutual information of the state $\ket{\phi_{3_{A},2_{B}}}$ in the complementary measurement configurations $\mathbb{L}$ is
\begin{equation}
  \cmplQ{\mathcal{I}}{\mathbb{L}}(\ket{\phi_{3_{A},2_{A}}}) = \log(5).
\end{equation}
If we implement all the six complementary measurements, the upper bound on $\cmplQ{\mathcal{I}}{\mathbb{L}}$ for separable states determined in Corollary \ref{coro::cmpl_MI_LONs} is $\log(5/3)$, which is much smaller than the CMI of the entangled state $\ket{\phi_{3_{A},2_{A}}}$.

\subsection{Complementary mutual predictability (CMP)\hiddengls{cmp}}
\label{sec::CMP}
If a state $\widehat{\rho}$ is close to a target entangled state $\ket{\phi_{N_{A},N_{B}}}$, which is an eigenstate of all complementary Pauli separable operators $\widehat{\alpha}_{l}\otimes\widehat{\beta}_{l}$ with eigenvalues $w^{\tilde{\mu}_{l}}$
\begin{equation}
  \widehat{\alpha}_{l}\otimes\widehat{\beta}_{l}\ket{\phi_{N_{A},N_{B}}}
  =w^{\tilde{\mu}_{l}}\ket{\phi_{N_{A},N_{B}}},
\end{equation}
mutual predictability \cite{SpenglerHuberEtAlHiesmayr2012-EntWitViaMUB} can be exploited to quantify the specific complementary correlations close to the target entangled state $\ket{\phi_{N_{A},N_{B}}}$.
The mutual predictability $\mathcal{F}_{\phi}(\widehat{\alpha}_{l},\widehat{\beta}_{l})$ of a quantum state $\widehat{\rho}$ for a target entangled state $\ket{\phi_{N_{A},N_{B}}}$ is the probability of measuring the expected correlated outputs specified by $\mu_{A}+\mu_{B}=\tilde{\mu}$,
\begin{equation}\hiddengls{MP}
  \mathcal{F}_{\phi}(\widehat{\alpha}_{l},\widehat{\beta}_{l}|\widehat{\rho})
  :=
  \sum_{\mu_{A}+\mu_{B}=\tilde{\mu}_{l}}
  \mathrm{Pr}_{\alpha_{l}, \beta_{l}}\left(\mu_{A},\mu_{B}|\widehat{\rho}\right),
\end{equation}
where $\mathrm{Pr}_{\alpha_{l}, \beta_{l}}\left(\mu_{A},\mu_{B}|\widehat{\rho}\right)$ is the probability of $(\mu_{A},\mu_{B})$ outputs in the $\widehat{\alpha_{l}}\otimes\widehat{\beta_{l}}$-Pauli measurement of $\widehat{\rho}$.
According to Theorem \ref{theorem::Pauli_msmnt}, it is equivalent to the expectation value of Pauli projectors that project onto the specific correlations $m_{A}+m_{B}=\tilde{\mu}$,
\begin{equation}\hiddengls{MP}
  \mathcal{F}_{\phi}(\widehat{\alpha}_{l},\widehat{\beta}_{l}|\widehat{\rho})
  =
  \sum_{m_{A}+m_{B} = \tilde{\mu}_{l}}
  \tr\left(
  \widehat{\pi}_{N_{A},m_{A}}(\alpha_{l})\otimes\widehat{\pi}_{N_{B},m_{B}}(\beta_{l})\;\widehat{\rho}
  \right).
\end{equation}
Mutual predictability $\mathcal{F}_{\phi}(\widehat{\alpha},\widehat{\beta}|\widehat{\rho})$ quantifies the closeness of a testing state $\widehat{\rho}$ to a target state $\ket{\phi}$ in an  $\widehat{\alpha}\otimes\widehat{\beta}$-Pauli measurement.
Complementary mutual predictability (CMP) for a target state $\ket{\phi_{N_{A},N_{B}}}$ therefore quantifies the closeness of $\widehat{\rho}$ to $\ket{\phi_{N_{A},N_{B}}}$ by taking the average of mutual predictability in the complementary Pauli measurements configurations $\mathbb{L}$,
\begin{equation}\hiddengls{CmplMP}
  \cmplQ{\mathcal{F}_{\phi}}{\mathbb{L}}(\rho) :=
  \frac{1}{|\mathbb{L}|}
  \sum_{\widehat{\alpha}\otimes\widehat{\beta}\in\mathbb{L}}
  \mathcal{F}_{\phi}(\widehat{\alpha},\widehat{\beta}|\widehat{\rho}).
\end{equation}
If a state $\widehat{\rho}$ is close enough to the target entangled state $\ket{\phi_{N_{A},N_{B}}}$ such that its CMP is above the threshold for separable states, then one can confirm the entanglement of $\widehat{\rho}$.
The threshold for entanglement determination can be derived analogous to Corollary \ref{coro::cmpl_MI_LONs} by the convex-roof extension over Pauli subspaces according to Theorem \ref{theorem:convex_prop_witness}.
\begin{corollary}[Complementary mutual predictability in LONs]
\label{coro::cmpl_MP_LONs}%
  Let $\ket{\phi_{N_{A},N_{B}}}$ be an entangled state, which is an eigenstate of complementary separable Pauli operators $\mathbb{L}=\{\widehat{\alpha}_{l}\otimes\widehat{\beta}_{l}\}_{l}$ with $\widehat{\alpha}_{l},\widehat{\beta}_{l}\in\{\widehat{\Xi},\widehat{\Lambda}_{0},...,\widehat{\Lambda}_{M-1}\}$.
  \begin{enumerate}
    \item In the case that $\gcd(N_{A},M)\neq1$ or $\gcd(N_{B},M)\neq1$, $\mathbb{L}$ can be constructed by $\widehat{\alpha}_{l}\in\{\widehat{\Xi},\widehat{\Lambda}_{j_{A}}\}$ and $\widehat{\beta}_{l}\in\{\widehat{\Xi},\widehat{\Lambda}_{j_{B}}\}$.
        The corresponding CMP of separable states is upper bounded by
        \begin{equation}
        \label{eq::CMP_sep_bnd_LEq2}
          \cmplQ{\mathcal{F}_{\phi}}{\mathbb{L}}(\rho)
          \underset{\rm{sep.}}{\le}
          \frac{1}{2} + \frac{1}{2}\sum_{\mathbb{E}_{A},\mathbb{E}_{B}}     p_{\mathbb{E}_{A},\mathbb{E}_{B}}(\rho)\frac{1}{d_{\mathbb{E}_{A},\mathbb{E}_{B}}},
      \end{equation}
      where $d_{\mathbb{E}_{A}, \mathbb{E}_{B}}:=\min(d_{\mathbb{E}_{A}}, d_{\mathbb{E}_{B}})$ is the minimum dimension of the local Pauli subspace $\mathbb{H}_{\mathbb{E}_{A}}$ and $\mathbb{H}_{\mathbb{E}_{B}}$;
    \item In the case that $\gcd(N_{A},M)=\gcd(N_{B},M)=1$, $\mathbb{L}$ can be constructed according to Corollary \ref{coro::MUBs_for_Mmd_Nph}.
        The corresponding CMP of separable states is upper bounded by
        \begin{equation}
        \label{eq::CMP_sep_bnd_L}
          \cmplQ{\mathcal{F}_{\phi}}{\mathbb{L}}
          \underset{\rm{sep.}}{\le}
          \frac{|\mathbb{L}|+M-1}{|\mathbb{L}| M}.
        \end{equation}
  \end{enumerate}
\begin{proof}
  See Appendix.
\end{proof}
\end{corollary}
\noindent Note that this bound is tight and achievable for example by the separable state $\ket{11100}\ket{11000}$.

In \cite{WuHofmann2017-BiEntMltMd}, entanglement detection criterion using CMP has been derived for the specific multiphoton entangled states with $\tilde{\mu}_{l}=0$ in two complementary measurement configurations $\{\widehat{Z}\otimes\widehat{Z}, \widehat{\Lambda}_{0}\otimes\widehat{\Lambda}_{0}\}$.
Corollary \ref{coro::cmpl_MP_LONs} is a generalization of the criterion in \cite{WuHofmann2017-BiEntMltMd} for more general target entangled states and complementary measurement configurations.
For entanglement detection of the exemplary entangled state $\ket{\phi_{3_{A},2_{B}}}$ in Eq. \eqref{eq::eg_phi_3A2B},
the mutual predictability for $\ket{\phi_{3_{A},2_{B}}}$ in each $(\widehat{\alpha}\otimes\widehat{\beta})$-Pauli measurement with $\widehat{\alpha}\otimes\widehat{\beta}\in\mathbb{L}$ constructed in Eq. \eqref{eq::cmpl_msmnt_config_phi_3A2B} is given by
\begin{align}
\label{eq::CMP_phi}
  \mathcal{F}_{\phi}(\widehat{Z},\widehat{Z} | \widehat{\rho})
  & = \sum_{\mu_{A}+\mu_{B}=4} \mathrm{Pr}_{Z,Z}\left(\mu_{A},\mu_{B}|\widehat{\rho}\right),
  \\
  \mathcal{F}_{\phi}(\widehat{\Lambda}_{j},\widehat{\Lambda}_{j} | \widehat{\rho})
  & = \sum_{\mu_{A}+\mu_{B}=5-j} \mathrm{Pr}_{\Lambda_{j},\Lambda_{j}}\left(\mu_{A},\mu_{B}|\widehat{\rho}\right)
  .
\end{align}
The $\ket{\phi_{3_{A},2_{B}}}$-targeting CMP $\cmplQ{\mathcal{F}_{\phi}}{\mathbb{L}}$ of separable states is upper bounded by $(|\mathbb{L}|+4)/(5|\mathbb{L}|)$ according to Corollary \ref{coro::cmpl_MP_LONs}.
It is obvious that CMP of the target entangled state $\cmplQ{\mathcal{F}_{\phi}}{\mathbb{L}}(\ket{\phi_{3_{A},2_{B}}})$ has the maximum value, which is much larger than the separable bounds,
\begin{equation}
  \cmplQ{\mathcal{F}_{\phi}}{\mathbb{L}}(\ket{\phi_{3_{A},2_{B}}}) = 1
  >
  \frac{|\mathbb{L}|+4}{5|\mathbb{L}|}
  .
\end{equation}

\subsection{Entanglement detection under errors}

\begin{figure}
  \centering
  \subfloat[]{\includegraphics[width=0.45\textwidth]{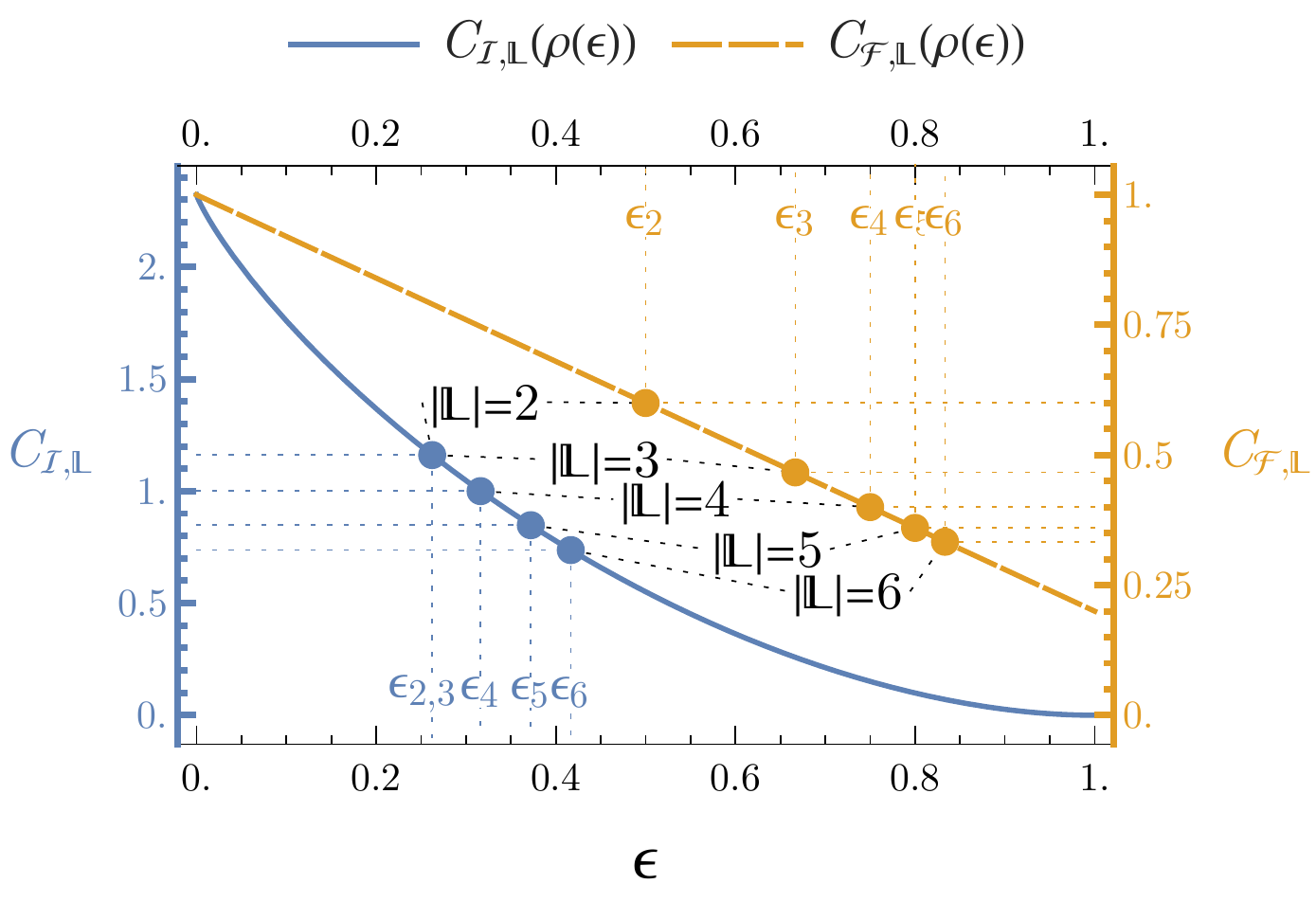}}
  \hspace{0.02\textwidth}
  \subfloat[]{\includegraphics[width=0.52\textwidth]{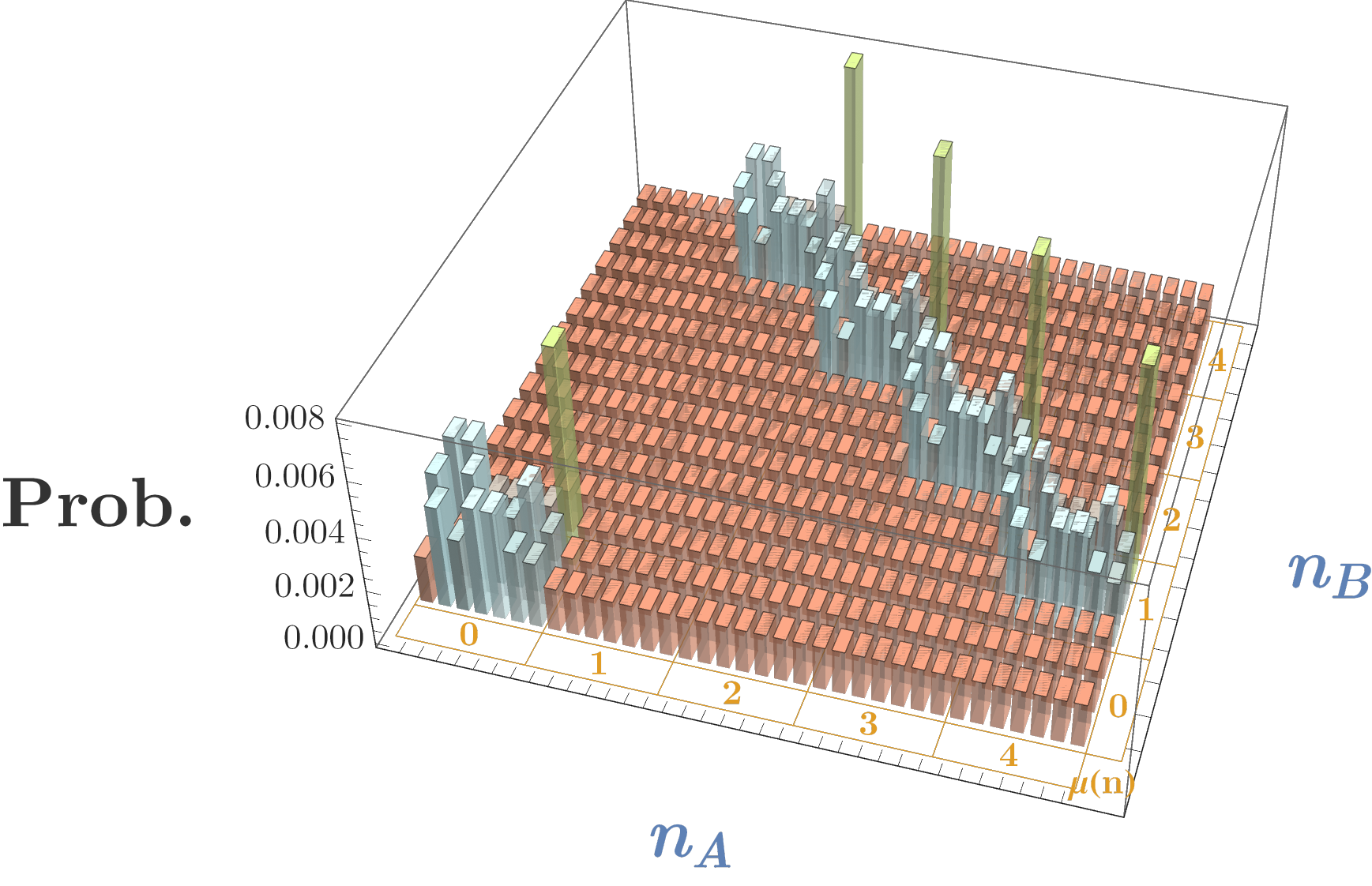}}
  \\
  \caption{
    (a) Robustness of entanglement detection against white noises. The noisy state $\widehat{\rho}_{\phi}(\epsilon)$ is given in Eq. \eqref{eq::noisy_ent_st}.
    The blue solid line and the orange dashed line plot the CMI $\cmplQ{\mathcal{I}}{\mathbb{L}}$ and the CMP $\cmplQ{\mathcal{F}_{\phi}}{\mathbb{L}}$, respectively.
    The blue points and orange points mark the upper bounds on $\cmplQ{\mathcal{I}}{\mathbb{L}}$ and $\cmplQ{\mathcal{F}_{\phi}}{\mathbb{L}}$ for separable states, respectively.
    These upper bounds are determined for complementary measurements with a number of configurations $|\mathbb{L}|=2 ,..., 6$ according to Corollary \ref{coro::cmpl_MI_LONs} and \ref{coro::cmpl_MP_LONs}.
    (b) The $\widehat{\Lambda}_{0}\otimes\widehat{\Lambda}_{0}$-measurement of $\widehat{\rho}_{\phi}(\epsilon)$ with $\epsilon = 5/6$. Neither CMI nor CMP can detect its entanglement.
  }
  \label{fig::noisy_MIMP}
\end{figure}

In either generation or measurements of a target entangled state, errors are unavoidable.
In practice, one needs to consider photon losses, which lead to faulty $N$-photon signals stimulated by irrelevant input components with a photon number higher than $N$, and a reduction of the contribution from relevant $N$-photon input components.
The latter effect can be excluded by post-selection on the $N$-photon outputs, while the former faulty signals can be avoided by employing input states that are generated from $N$ single-photon resources.
As shown in Fig. \ref{fig::ent_gen_dect}, an experiment validating the proposed entanglement detection approaches for the state $\ket{\phi_{3_{A},2_{B}}}$ can be constructed employing five good single-photon resources with high indistinguishability, well-established LONs implementing the desired Hadamard transforms, and photon number resolving detectors with detection saturation at least $N_{A,B}$ photons in the local systems $A,B$.
The photon number resolving detectors can be constructed by $N_{A,B}$-mode demultiplexers.
With post-selection on the $(3_{A},2_{B})$-photon outputs, this experimental implementation is robust against photon losses.

%
%
%
%
%
%
%

Under this experimental scheme, one can analyze the robustness of entanglement detection against noises within the Hilbert space of fixed local photon numbers $(3_{A},2_{B})$ for the state $\ket{\phi_{3_{A},2_{B}}}$.
For totally random errors, the robustness can be analyzed with the white noise model,
\begin{equation}
\label{eq::noisy_ent_st}
  \widehat{\rho}_{\phi}(\epsilon) :=
  \epsilon\frac{\id_{N}}{525} + (1-\epsilon)\projector{\phi_{3_{A},2_{B}}},
\end{equation}
where $\id_{N}=\sum_{|\boldvec{n}_{A}|=3, |\boldvec{n}_{B}|=2}\projector{\boldvec{n}_{A},\boldvec{n}_{B}}$ is the identity operator in the $(3,2)$-photon subspace.
Since the background random noise is added uniformly to every possible $(3,2)$-photon outputs, the probability distributions $\mathrm{Pr}(\mu_{A},\mu_{B})$ under the white noise are then
\begin{equation}
  \left\{
    \begin{array}{ll}
      \mathrm{Pr}(\mu_{A},\mu_{B}) = (1-\epsilon)\delta_{\mu_{A}}^{4-\mu_{B}}/5 + \epsilon/25 & \hbox{in the computational basis;} \\
      \mathrm{Pr}(\mu_{A},\mu_{B}) = (1-\epsilon)\delta_{\mu_{A}}^{5-j-\mu_{B}}/5 + \epsilon/25 & \hbox{in the $\widehat{\Lambda}_{j}\otimes\widehat{\Lambda}_{j}$-Pauli measurement.}
    \end{array}
  \right.
\end{equation}
The corresponding mutual information and mutual predictability is therefore uniform in every measurement configuration.
As a consequence, different choices of the measurement configurations $\mathbb{L}\subseteq\{\widehat{Z}\otimes\widehat{Z}, \widehat{\Lambda}_{0}\otimes\widehat{\Lambda}_{0}, ..., \widehat{\Lambda}_{4}\otimes\widehat{\Lambda}_{4}\}$ do not change the CMI and CMP of $\widehat{\rho}_{\phi}(\epsilon)$, but change the upper bounds on the CMI and CMP for separable states.
According to Corollary \ref{coro::cmpl_MI_LONs} and \ref{coro::cmpl_MP_LONs}, entanglement of $\widehat{\rho}_{\phi}(\epsilon)$ is still detectable by CMI, if
\begin{equation}
  \cmplQ{\mathcal{I}}{\mathbb{L}}(\rho_{\phi}(\epsilon)) > \log\left((|\mathbb{L}|+4)/|\mathbb{L}|\right),
\end{equation}
while it is still detectable by CMP, if
\begin{equation}
\cmplQ{\mathcal{F}_{\phi}}{\mathbb{L}}(\rho_{\phi}(\epsilon)) > (|\mathbb{L}|+4)/(5|\mathbb{L}|).
\end{equation}
There exist therefore thresholds $\epsilon_{|\mathbb{L}|}$ for white-noise errors, upon which entanglement is not detectable by CMI or CMP in the complementary measurement configurations $\mathbb{L}$.
In Fig. \ref{fig::noisy_MIMP} (a), the CMI and CMP of the noisy state $\widehat{\rho}_{\phi}(\epsilon)$ are plotted with a blue solid line and an orange dashed line, respectively.
The white noise thresholds for entanglement detection using CMI and CMP are marked by blue and orange points, respectively.
One can see that the more configurations a complementary measurement setting has, the more robust an entanglement detection is against white noises.
Entanglement is not detectable for $\epsilon>5/6$ either by CMI or CMP.
As an example, the $\widehat{\Lambda}_{0}\otimes\widehat{\Lambda}_{0}$-measurment statisics of $\widehat{\rho}_{\phi}(\epsilon=5/6)$ is shown in Fig. \ref{fig::noisy_MIMP} (b).

\begin{figure}
  \centering
  \subfloat[]{\includegraphics[width=0.45\textwidth]{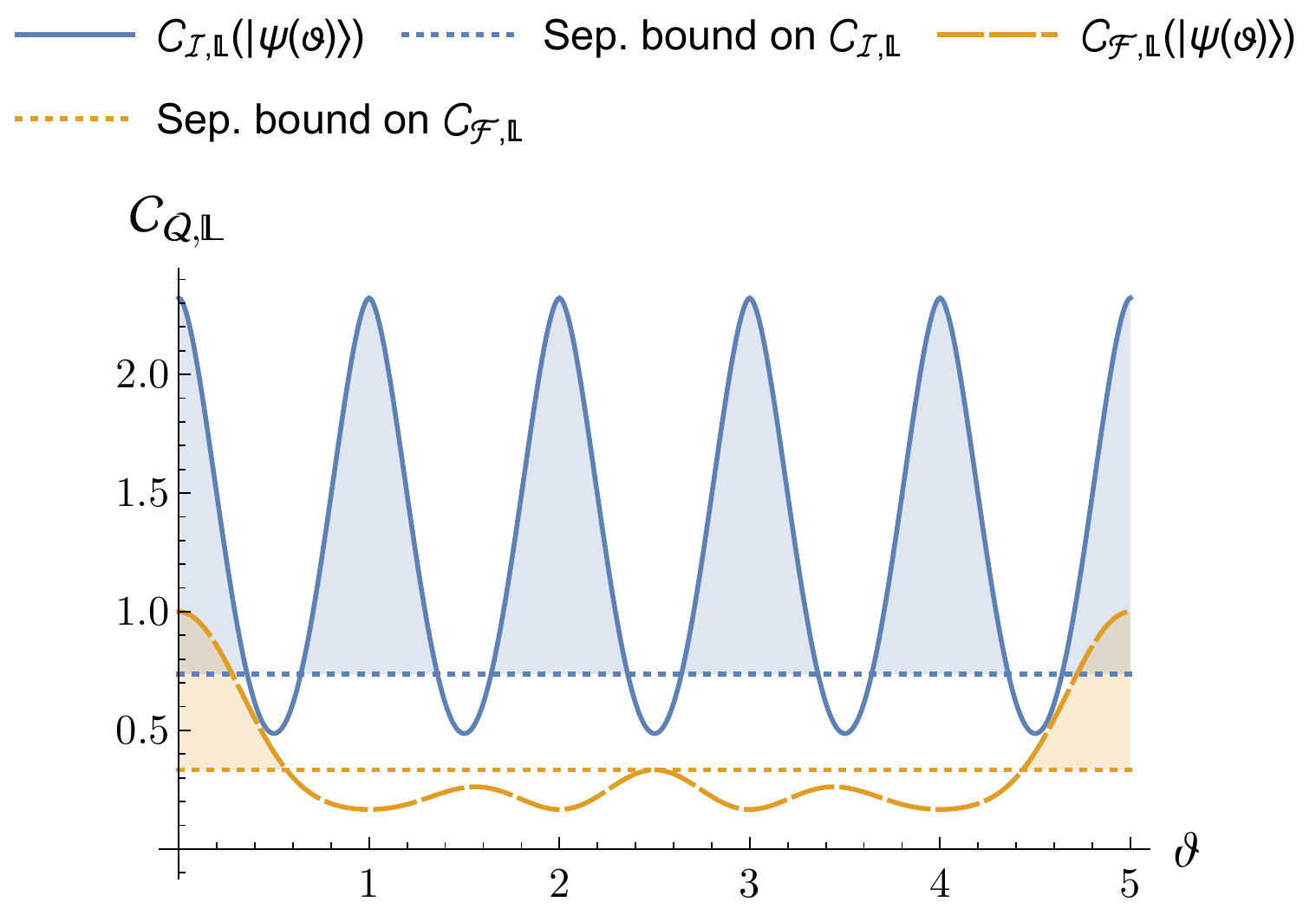}}
  \hspace{0.02\textwidth}
  \subfloat[]{\includegraphics[width=0.52\textwidth]{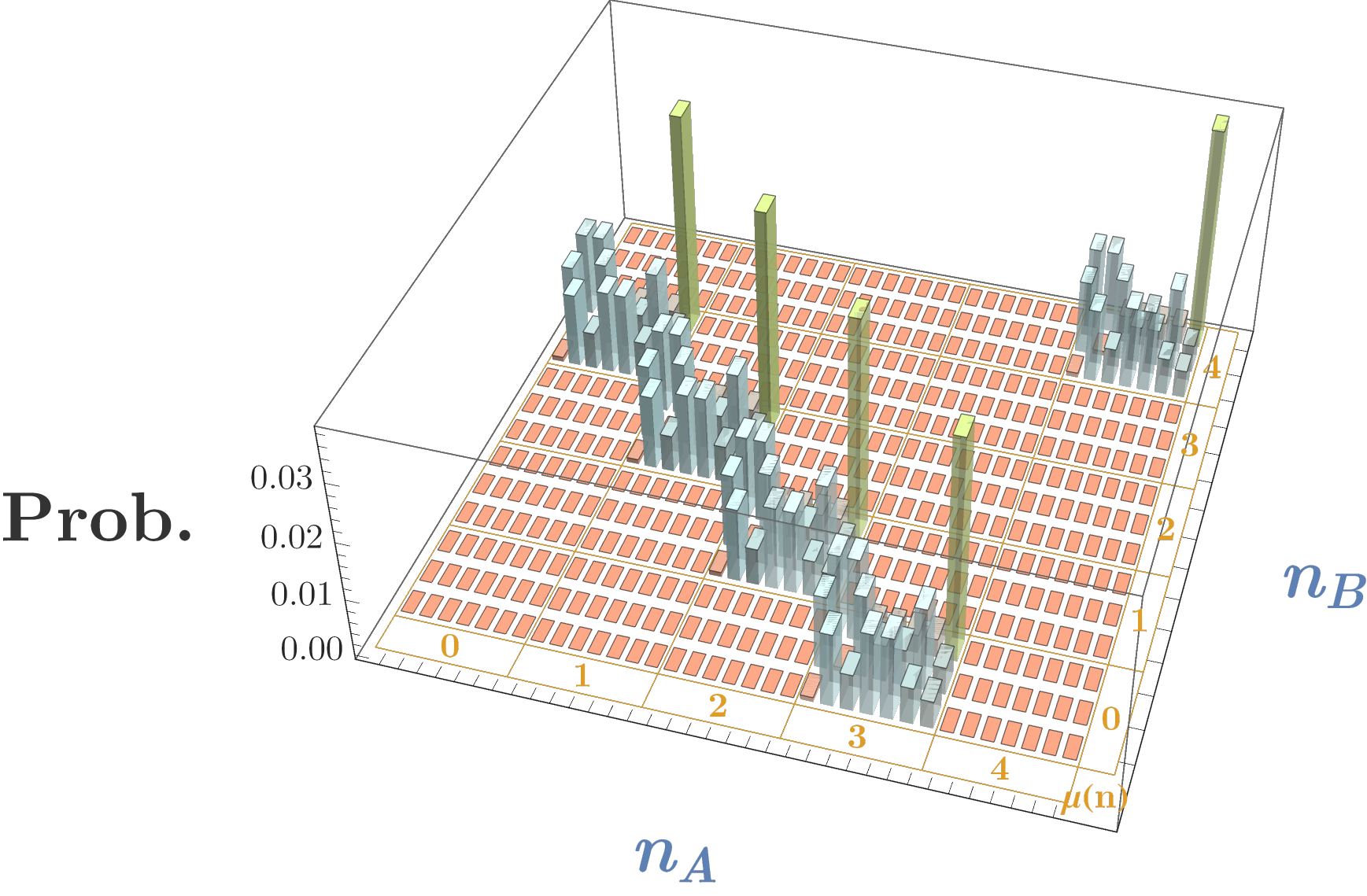}}
  \\
  \caption{
    (a) Complementary correlations of the entangled state $\ket{\psi(\theta)}$ given in Eq. \eqref{eq::phase_ent}.
    The blue solid line and the orange dashed line are its CMI $\cmplQ{\mathcal{I}}{\mathbb{L}}$ and $\ket{\phi_{3_{A},2_{B}}}$-targeting CMP $\cmplQ{\mathcal{F}_{\phi}}{\mathbb{L}}$, respectively.
    The blue and orange dotted lines mark the upper bounds on $\cmplQ{\mathcal{I}}{\mathbb{L}}$ and $\cmplQ{\mathcal{F}}{\mathbb{L}}$ for separable states, respectively.
    The color-filled areas are the interval where entanglement is detectable.
    (b) The $\widehat{\Lambda}_{0}\otimes\widehat{\Lambda}_{0}$-measurement of $\ket{\psi(\theta)}$ with $\theta = 1$. Its entanglement is detectable by CMI but not by the $\ket{\phi_{3_{A},2_{B}}}$-targeting CMP.
  }
  \label{fig::phase_MIMP}
\end{figure}

Compare these two approaches, one can see that entanglement detection using CMP is more robust against white noises than entanglement detection using CMI.
The intuition behind this is that CMP is tailor-made for the particular correlations $\mu_{A}+\mu_{B}=\tilde{\mu}_{l}$ of the target entangled state $\ket{\phi_{3_{A},2_{B}}}$, while CMI can also detect other entanglement correlations.
This intuition can be confirmed as follows.
If we introduce a phase shift $\widehat{Z}^{\theta}$ in the local system $B$ to the target entangled state $\ket{\phi_{3_{A},2_{B}}}$,
\begin{equation}
\label{eq::phase_ent}
  \ket{\psi(\theta)}
  :=
  \id\otimes\widehat{Z}^{\theta} \ket{\phi_{3_{A},2_{B}}},
\end{equation}
the modified state $\ket{\psi(\theta)}$ is still maximally entangled within the Pauli subspaces $\mathbb{H}_{\mathbb{E}_{11100}}\otimes\mathbb{H}_{\mathbb{E}_{11000}}$ and $\mathbb{H}_{\mathbb{E}_{11010}}\otimes\mathbb{H}_{\mathbb{E}_{01001}}$, but its correlations are changed.
To detect entanglement of $\ket{\psi(\theta)}$, we choose $\mathbb{L}=\{\widehat{Z}\otimes\widehat{Z},\widehat{\Lambda}_{0}\otimes\widehat{\Lambda}_{0},...,\widehat{\Lambda}_{4}\otimes\widehat{\Lambda}_{4}\}$.
The CMI and the $\ket{\phi_{3_{A},2_{B}}}$-targeting CMP of $\ket{\psi(\theta)}$ are plotted in a blue solid line and an orange dashed line, respectively, in Fig. \ref{fig::phase_MIMP} (a).
Compare these two approaches, one can see that CMI is sensitive to entanglement of the state $\ket{\psi(\theta)}$ with $\theta$ close to the values $\{0,1,2,3,4\}$, while the $\ket{\phi_{3_{A},2_{B}}}$-targeting CMP can only detect entanglement close to $\theta=0$.
The correlations of $\ket{\phi_{3_{A},2_{B}}}$ with $\mu_{A}+\mu_{B}=0$ as shown in Fig. \ref{fig::WHSt_cmpl_msmnt} (b) are transformed into the other type of correlations, e.g. $\mu_{A}+\mu_{B}=3$ for $\theta = 1$ as shown in Fig. \ref{fig::phase_MIMP} (b).
In this case, the perfect correlations of the entangled state $\ket{\psi(\theta)}$ can be detected by CMI, but not by the $\ket{\phi_{3_{A},2_{B}}}$-targeting CMP.
From the comparison between entanglement detection using CMI and CMP in Fig. \ref{fig::noisy_MIMP} (a) and Fig. \ref{fig::phase_MIMP} (a), one can see that CMI can detect entangled states of different types of correlations, while CMP is more robust against white noises than CMI.

\section{Conclusion and discussion}
\label{sec::conclusion}

In this paper, we have studied the complementary structures of generalized Pauli operators in multiphoton LONs, and found that their MUBs are constituted within Pauli subspaces that are characterized by a cyclicly translational mode shift (Theorem \ref{theorem:MUBs_in_Pauli_subspace}).
Accordingly, a set of complementary Pauli operators in fixed photon number LON systems has been constructed  (Corollary \ref{coro::MUBs_for_Mmd_Nph}).

It has been shown that, in a Pauli measurement, which is the projective measurement associated with a Pauli operator, the probability distribution over its Pauli-operator eigenspaces is given by the statistics of $\widehat{Z}$-clock labels in the outputs of its corresponding Hadamard transform (Theorem \ref{theorem::Pauli_msmnt}).
Although the explicit Hadamard transformation of multiphoton states are $\#P$-hard to calculate, this result lifts the computational complexity of Boson sampling in the simulation of Pauli measurement statistics.
It therefore allows us to predict the probability distribution of $\widehat{Z}$-clock labels in a Pauli measurement of a given state, and vice versa to access complementary properties of an unknown state from Pauli measurement statistics.

Assessment of complementary properties from complementary Pauli measurement statistics has been shown to be invariant under decoherence over Pauli subspaces (Corollary \ref{coro::Pauli_subspace_decoherence}).
As a result, we can exploit such assessed quantities, which we call \emph{complementary Pauli quantities}, to characterize the convex set of quantum states of a specific property $\mathcal{S}$ in multiphoton LONs through the convex-roof extension of its hyperplane boundaries over Pauli subspaces (Theorem \ref{theorem:convex_prop_witness}).
It therefore allows us to detect the non-$\mathcal{S}$ property of quantum states in multiphoton LON systems experimentally in complementary Pauli measurements.
Evaluation of measurement uncertainty relations in a multiphoton LON is a straightforward application of this theory (Corollary \ref{coro::un_rel}).

Exploiting this theory, we have shown that entanglement between modes in bipartite multiphoton LON systems can be physically detected by complementary correlations in complementary Pauli measurements.
We have demonstrated entanglement detection in bipartite multiphoton LON systems with the detection approaches employing complementary mutual information (Corollary \ref{coro::cmpl_MI_LONs}) and complementary mutual predictability (Corollary \ref{coro::cmpl_MP_LONs}).

\bigskip

Our results open up physical access to desired quantum coherences in the MUBs in multiphoton LONs without falling into the computational complexity in Boson samplings.
It allows us to predict and reveal the physical significance of entanglement between modes in bipartite multiphoton LONs in complementary Pauli measurements.
The developed theory provides a theoretical framework for the problems of hyperplane characterization of convex sets of multiphoton states in LON systems.
It allows us to extend well-established methods for entanglement detection in bipartite qudit systems to bipartite multiphoton LONs, if the detection approaches in bipartite qudit systems employ complementary Pauli-measurement statistics to evaluate the physical significance of entanglement.
Besides the detection of bipartite entanglement, it could be further employed in multipartite entanglement detection and entanglement dimensionality characterization in multiphoton LON systems.
Since multipartite producibility and entanglement dimensionality are convex extendible properties by definition, one can extend their detection methods in qudit systems to multiphoton LONs through convex roof extension over Pauli subspaces.
For example, in this theoretical framework, the methods in \cite{SauerweinAtElKraus2017-MltptCrrMUBs} and \cite{BavarescoEtAlHuber2018-2MUBsCrtfyHghDmEnt} can be extended for the detection of genuine multipartite entanglement and entanglement dimensionality, respectively.
The theory in this paper therefore paves a way to extend quantum information processing in multipartite single-photon LONs to the multiphoton regime.
Although our analysis is carried on in LONs, which encode paths in modes, it is general enough for any bosonic multimode system that allows generalized Hadamard transforms.

\acknowledgments
J. W. is supported by Japan Society for the Promotion of Science (JSPS) KAKENHI Grant No. 19F19817. This work is partially supported by MEXT Quantum Leap Flagship Program (MEXT Q-LEAP) Grant No. JPMXS0118069605, JSPS KAKENHI Grant No. 17H01694 and No. 18H04286.

\section*{Appendix: proofs of theorems and corollaries}

\paragraph{The Proof of Theorem \ref{theorem:MUBs_in_Pauli_subspace}}
\begin{proof}
  Let $\boldvec{n} = (\boldvec{\nu},...,\boldvec{\nu})$ be a Fock vector with repetitive components $\boldvec{\nu}$, where $\boldvec{\nu} = (\nu_{0},...,\nu_{\dimE{\boldvec{n}}-1})$ is a $\dimE{\boldvec{n}}$-dimensional non-repetitive component of $\boldvec{n}$.
  If $\boldvec{n}$ is non-repetitive, $\boldvec{n}=\boldvec{\nu}$ and $\dimE{\boldvec{n}}=M$.
  The effect of the Pauli operator $\widehat{Z}$ on a $\widehat{\Lambda}_{j}$ eigenstate is
  \begin{equation}
    \widehat{Z}\ket{\ESt{\boldvec{n}}{m}{j}} =
    w^{\mu(\boldvec{n})} \ket{ \ESt{\boldvec{n}}{m-|\boldvec{n}|}{j} },
  \end{equation}
  which leads to
  \begin{equation}
  \label{eq::proof_theorem_MUBs_1}
    \widehat{\Lambda}_{l}\ket{\ESt{\boldvec{n}}{m}{j}}
    =
    \widehat{\Lambda}_{j}\widehat{Z}^{l-j}\ket{\ESt{\boldvec{n}}{m}{j}}
    =
    w^{(l-j)\mu(\boldvec{n})+ m} \ket{\ESt{\boldvec{n}}{m+(j-l)|\boldvec{n}|}{j} }.
  \end{equation}
  Since the photon number is given by $|\boldvec{n}|=M/\dimE{\boldvec{n}}|\boldvec{\nu}|$,
  the operator $\widehat{\Lambda}_{l}$ will transform the $\widehat{\Lambda}_{j}$ eigenstate $\ket{\ESt{\boldvec{n}}{m}{j}}$ to $\ket{\ESt{\boldvec{n}}{m + (j-l)|\boldvec{\nu}| M/d_{\mathbb{E}}}{j}}$.
  One can then specify an eigensubspace of $\widehat{\Lambda}_{j}$ by a projector that is generated via $\widehat{\Lambda}_{l}$,
  \begin{equation}
    \widehat{\Pi}_{g,j} :=
    \sum_{k=0}^{\frac{\dimE{\boldvec{n}}}{\gamma}}
    \widehat{\Lambda}^{k}\projector{\ESt{\boldvec{n}}{g}{j}}\widehat{\Lambda}^{-k}
    =
    \sum_{k=0}^{\frac{\dimE{\boldvec{n}}}{\gamma}} \projector{\ESt{\boldvec{n}}{g+k(j-l)|\boldvec{\nu}|\frac{M}{\dimE{\boldvec{n}}}}{j}},
  \end{equation}
  where $\gamma := \gcd((j-l)|\boldvec{\nu}|, \dimE{{\boldvec{n}}})$ is equal to the number of different projectors $\widehat{\Pi}_{g,j}$.
  This construction leads to the invariance of $\widehat{\Pi}_{g,j}$ under the $\widehat{\Lambda}_{j}$ operation
  \begin{equation}
    \widehat{\Lambda}_{l}\widehat{\Pi}_{g,j}\widehat{\Lambda}_{l}^{\dagger}
    =
    \widehat{\Pi}_{g,j}.
  \end{equation}
  As a result, the operator $\widehat{\Lambda}_{l}$ is block-diagonal with respect to the projector $\widehat{\Pi}_{g,j}$
  \begin{equation}
    \widehat{\Lambda}_{l}
    =
    \sum_{g=0}^{\gamma-1} \widehat{\Pi}_{g,j}\widehat{\Lambda}_{l}\widehat{\Pi}_{g,j}.
  \end{equation}
  It means that for a $\widehat{\Lambda}_{l}$ eigenstate $\ket{\ESt{\boldvec{n}}{m}{l}}$, its $\widehat{\Pi}_{g,j}$ projection is either $1$ or $0$
  \begin{equation}
  \label{eq::proof_theorem_MUBs_2}
    \braket{\ESt{\boldvec{n}}{m}{l} | \widehat{\Pi}_{g,j} | \ESt{\boldvec{n}}{m}{l}}
    =1 \text{ or } 0.
  \end{equation}
  As a result of Eq. \eqref{eq::proof_theorem_MUBs_1}
  \begin{equation}
    \left|
      \braket{\ESt{\boldvec{n}}{m}{l}|\ESt{\boldvec{n}}{g + k(j-l)|\boldvec{\nu}|\frac{M}{\dimE{\boldvec{n}}}}{j}}
    \right|
    =
    \sqrt{\frac{\gamma}{\dimE{\boldvec{n}}}}
    \text{ or } 0
  \end{equation}
  for all $k=0, ..., \dimE{\boldvec{n}}/\gamma-1$.
  If $\gamma = 1$, then $\widehat{\Pi}_{g,j}=\id_{\mathbb{E}_{\boldvec{n}}}$ spans the whole Hilbert space $\mathbb{H}_{\mathbb{E}_{\boldvec{n}}}$, and Eq. \eqref{eq::proof_theorem_MUBs_2} is equal to $1$ for all $\widehat{\Lambda}_{l}$ eigenstates.
  As a result, $\{\ESt{\boldvec{n}}{m}{l}\}_{m}$ and $\{\ESt{\boldvec{n}}{m'}{j}\}_{m'}$ are MUBs, i.e.
  \begin{equation}
    \left|
      \braket{\ESt{\boldvec{n}}{m}{l}|\ESt{\boldvec{n}}{m'}{j}}
    \right|
    =
    \sqrt{\frac{1}{\dimE{\boldvec{n}}}} \text{ for all }(m,m'),
  \end{equation}
  if and only if $\gcd((l-j)|\boldvec{\nu}|,d_{\mathbb{E}})=1$.

  The equality $\gcd((l-j)|\boldvec{\nu}|,d_{\mathbb{E}})=1$ is equivalent to $\gcd((l-j)|\boldvec{n}|d_{\mathbb{E}}/M,d_{\mathbb{E}})=1$.
  This condition holds, if and only if $w^{(l-j)\mu(\boldvec{n'})}$ are not degenerated for $\ket{\boldvec{n'}}\in\mathbb{E}_{\boldvec{n}}$, which is equivalent to the non-degeneracy of $\widehat{Z}^{l-j}$ in $\mathbb{E}_{\boldvec{n}}$.
\end{proof}

\bigskip

\paragraph{The Proof of Corollary \ref{coro::un_rel}}
\begin{proof}
  According to the uncertainty relationship in qudit systems derived in \cite{WehnerWinter2010-EntrUncRel}, in a Pauli subspace $\mathbb{H}_{\mathbb{E}}$, which is a $d_{\mathbb{E}}$-dimensional qudit system, the complementary Shannon entropy $\cmplQ{\mathcal{H}}{\mathbb{L}}$ is lower bounded by
  \begin{equation}
  \label{eq::CSE_proof_1}
    \mathcal{B}_{\mathrm{quan.}}(\mathbb{E}) =
    \left\{
      \begin{array}{ll}
        \frac{1}{2}\log(d_{\mathbb{E}}),
          & |\mathbb{L}|\le\sqrt{d_{\mathbb{E}}}+1;
        \\
        -\log\frac{|\mathbb{L}|+d_{\mathbb{E}}-1}{|\mathbb{L}|d_{\mathbb{E}}},
          & |\mathbb{L}|>\sqrt{d_{\mathbb{E}}}+1.
      \end{array}
    \right.
  \end{equation}
  According to Corollary \ref{coro::MUBs_for_Mmd_Nph}, one can construct complementary measurement configurations $\mathbb{L}$ as follows.
  \begin{enumerate}
    \item For a photon number with $\gcd(N,M)\neq1$, one can only construct two complementary Pauli operators $\mathbb{L} = \{\widehat{\Xi},\widehat{\Lambda}_{j}\}$ that are complementary in all Pauli subspaces.
        In this case, the lower bound on $\cmplQ{\mathcal{H}}{\mathbb{L}}$ given in Eq. \eqref{eq::CSE_1} is determined by the convex-roof extension of Eq. \eqref{eq::CSE_proof_1} over all Pauli subspaces according to Theorem \ref{theorem:convex_prop_witness}.
    \item For $\gcd(N,M)=1$, one can construct complementary Pauli operators $\mathbb{L} \subseteq \{\widehat{\Xi},\widehat{\Lambda}_{0},...,\widehat{\Lambda}_{M-1}\}$ according to Corollary \ref{coro::MUBs_for_Mmd_Nph}.
        As result of Theorem \ref{theorem:convex_prop_witness}, the lower bound in Eq. \eqref{eq::CSE_2} is derived by the convex-roof extension of Eq. \eqref{eq::CSE_proof_1}.
  \end{enumerate}
\end{proof}

\bigskip

\paragraph{The Proof of Corollary \ref{coro::cmpl_MI_LONs}}
\begin{proof}
  The complementary mutual information $\cmplQ{\mathcal{I}}{\mathbb{L}}$ is always smaller or equal to a convex Pauli quantity $\mathcal{Q}_{\mathbb{L}}$,
  \begin{equation}
  \label{eq::proof_Cmpl_MI_1}
    \cmplQ{\mathcal{I}}{\mathbb{L}}\leq \mathcal{Q}_{\mathbb{L}}:=
    \log(M)-\frac{1}{|\mathbb{L}|}\min\left(\sum_{l}\mathcal{H}(\widehat{\alpha}_{l}|\widehat{\beta}_{l}), \sum_{l}\mathcal{H}(\widehat{\beta}_{l}|\widehat{\alpha}_{l})\right)
  \end{equation}
  where $\mathcal{H}(\widehat{\alpha}_{l}|\widehat{\beta}_{l})$ and $\mathcal{H}(\widehat{\beta}_{l}|\widehat{\alpha}_{l})$ are relative Shannon entropies of measurement statistics.
  As a result of Theorem \ref{theorem:convex_prop_witness} and Corollary \ref{coro::un_rel}, the relative Shannon entropy are lower bounded as follows.
  \begin{enumerate}
    \item  for $\gcd(N_{A},M)\neq1$ or $\gcd(N_{B},M)\neq1$, $\mathbb{L}$ is a pair of two complementary Pauli operators with $\widehat{\alpha}_{l}\in\{\widehat{\Xi},\widehat{\Lambda}_{j_{A}}\}$ and $\widehat{\beta}_{l}\in\{\widehat{\Xi},\widehat{\Lambda}_{j_{B}}\}$,
        \begin{align}
        \label{eq::proof_Cmpl_MI_2}
          \min\left(\sum_{l}\mathcal{H}(\widehat{\alpha}_{l}|\widehat{\beta}_{l}), \sum_{l}\mathcal{H}(\widehat{\beta}_{l}|\widehat{\alpha}_{l})\right)
          \ge
          \frac{1}{2}\sum p_{\mathbb{E}_{A},\mathbb{E}_{B}}\log(d_{\mathbb{E}_{A},\mathbb{E}_{B}}),
        \end{align}
        where $d_{\mathbb{E}_{A},\mathbb{E}_{B}}:=\min(\mathbb{E}_{A},\mathbb{E}_{B})$.
    \item For $\gcd(N_{A},M)=\gcd(N_{A},M)=1$ and $\widehat{\alpha}_{l}, \widehat{\beta}_{l}\subseteq\{\widehat{\Xi},\widehat{\Lambda}_{0},...\widehat{\Lambda}_{M-1}\}$, the dimension of all local $N_{A}$-photon ($N_{B}$-photon) Pauli subspace are uniform $d_{\mathbb{E}_{A}} = d_{\mathbb{E}_{B}} = M$.
        \begin{align}
        \label{eq::proof_Cmpl_MI_3}
          \min\left(\sum_{l}\mathcal{H}(\widehat{\alpha}_{l}|\widehat{\beta}_{l}), \sum_{l}\mathcal{H}(\widehat{\beta}_{l}|\widehat{\alpha}_{l})\right)
          \ge
          \max\left( - \log(\frac{|\mathbb{L}|+M-1}{|\mathbb{L}|M}), \frac{1}{2}\log(M)\right).
        \end{align}
  \end{enumerate}
  The upper bounds on $\cmplQ{\mathcal{I}}{\mathbb{L}}$ given in Eq. \eqref{eq::CMI_1} and \eqref{eq::CMI_2} follow Eq. \eqref{eq::proof_Cmpl_MI_1} - \eqref{eq::proof_Cmpl_MI_3}.
\end{proof}

\bigskip

\paragraph{The Proof of Corollary \ref{coro::cmpl_MP_LONs}}
\begin{proof}
  It is shown in \cite{SpenglerHuberEtAlHiesmayr2012-EntWitViaMUB} that the upper bounds on $\cmplQ{\mathcal{F}_{\phi}}{\mathbb{L}}$ for separable states $\widehat{\sigma}_{\mathbb{E}}$ in a Pauli subspace $\mathbb{H}_{\mathbb{E}}$ is determined  by
  \begin{equation}
    \mathcal{B}_{sep}(\mathbb{E}_{A},\mathbb{E}_{B}) =
    \frac{|\mathbb{L}|+d_{\mathbb{E}_{A},\mathbb{E}_{B}}-1}{|\mathbb{L}|d_{\mathbb{E}_{A},\mathbb{E}_{B}}},
  \end{equation}
  where $d_{\mathbb{E}_{A},\mathbb{E}_{B}}:=\min(\mathbb{E}_{A},\mathbb{E}_{B})$.
  Since CMP is linear, one can extend these upper bounds to  $(N_{A},N_{B})$-photon LON systems through convex-roof extension according to Theorem \ref{theorem:convex_prop_witness}.
  \begin{enumerate}
    \item For the first case that $\gcd(N_{A},M)\neq1$ or $\gcd(N_{B},M)\neq1$, the possible complementary measurements have two configurations $|\mathbb{L}|=2$.
        The upper bound given in Eq. \eqref{eq::CMP_sep_bnd_LEq2} is then determined by taking the average of $\mathcal{B}_{sep}(\mathbb{E}_{A},\mathbb{E}_{B})$ over all $(N_{A},N_{B})$-photon Pauli subspaces.
    \item For the second case that $\gcd{N_{A},M}=\gcd{N_{B},M}=1$, all local $N_{A}$-photon Pauli subspaces $\mathbb{H}_{\mathbb{E}_{A}}$ and $N_{B}$-photon Pauli subspaces $\mathbb{H}_{\mathbb{E}_{B}}$ have the same dimension $d_{\mathbb{E}_{A}}=d_{\mathbb{E}_{B}}=M$.
        As a result, the separable bound in each $(N_{A}, N_{B})$-Pauli subspace is uniform given by $(|\mathbb{L}|+M-1)/(|\mathbb{L}|M)$.
        After the convex-roof extension over Pauli subspaces one arrives at the upper bound given in Eq. \eqref{eq::CMP_sep_bnd_L}.
  \end{enumerate}%
\end{proof}


\myprintglossary

\myprintbibliography

\end{document}